\documentclass[prd,aps,letterpaper,twocolumn,superscriptaddress,preprintnumbers,nofootinbib,floatfix]{revtex4}
\pdfoutput=1
\usepackage[dvipsnames]{xcolor}
\usepackage{graphicx}
\usepackage{amsmath}
\usepackage{amsfonts}
\usepackage{amssymb}
\usepackage{rotating}
\usepackage{subfigure}
\usepackage{paralist} 
\usepackage{verbatim}
\usepackage{float}
\usepackage{soul} 
\usepackage{appendix}
\usepackage{natbib}
\usepackage{epsfig}
\usepackage{hyperref}
\hypersetup{colorlinks,linkcolor=red,urlcolor=blue,citecolor=blue}
\usepackage[utf8]{inputenc}
\usepackage[english]{babel}
\usepackage{tabularx}
\newcolumntype{C}{>{\centering\arraybackslash}X}
\usepackage{mathpazo} 
\usepackage{tgpagella} 
\usepackage{float}
\usepackage{listings} 

\newcommand{\mnras}{Monthly Notices of the Royal Astronomical Society}

\newcommand{\aap}{Astronomy \& Astrophysics}
\newcommand{\jcap}{J. Cosmol. Astropart. Phys.}

\normalsize

\usepackage{booktabs}
\usepackage{array, multirow}

\usepackage{lipsum}

\definecolor{backcolour}{rgb}{0.95,0.95,0.92}
\lstdefinestyle{mystyle}{
    basicstyle=\ttfamily\small\color{black},
    keywordstyle=\ttfamily\small\color{blue},
    stringstyle=\ttfamily\small\color{purple},
    breakatwhitespace=true,         
    breaklines=true,  
    backgroundcolor=\color{backcolour},
    numbers=left,
    numbersep=3pt,
    showstringspaces=false,
    numberstyle=\ttfamily\color{gray}\scriptsize,
    }
\definecolor{numcolor}{RGB}{185,0,255}
\newcommand{\num}[1]{#1}

\newcommand{\stonybrook}{Physics and Astronomy Department, Stony Brook University, Stony Brook, NY  11794, USA}
\newcommand{\cambridge}{Cavendish Astrophysics, University of Cambridge, Madingley Road, Cambridge CB3 0HA, UK}
\newcommand{\vanderbilt}{Department of Physics and Astronomy, Vanderbilt University, Nashville, TN 37240, USA}

\begin{document}

\title{CMB-HD Foregrounds: Simulations, Source Detection, and Foreground Removal}

\author{Amanda MacInnis}
\affiliation{\stonybrook}

\author{Joshua Ange}
\affiliation{\cambridge}

\author{Neelima Sehgal}
\affiliation{\stonybrook}

\author{Joshua A. Kable}
\affiliation{\stonybrook}

\author{Isabelle Blackstad}
\affiliation{\vanderbilt}

\begin{abstract}
We present simulations of the microwave sky at 2.5 arcsecond resolution over 100 square degrees, generated from existing full-sky, lower-resolution simulations, and use them to demonstrate extragalactic foreground removal for a CMB-HD survey. Our cleaning method detects and removes the cosmic infrared background and radio galaxies, yielding source catalogs that are 95\% complete down to flux limits of \num{0.008} and \num{0.04}~mJy at 90 and 148~GHz, respectively. It also identifies and removes galaxy clusters via the thermal Sunyaev-Zel'dovich effect, producing a cluster sample that is nearly complete above \num{$M_{500c} = 5\times10^{13}\,M_\odot$}. After cleaning, the residual foreground-plus-noise power spectrum of the coadded 90 and 148~GHz temperature map is \num{50\%} higher than previous idealized estimates, increasing cosmological parameter uncertainties for an 11-parameter $\Lambda\mathrm{CDM} + N_\mathrm{eff} + \sum m_\nu + T_\mathrm{AGN} + A_\mathrm{kSZ} + n_\mathrm{kSZ}$ model by less than \num{7\%}. The small impact on cosmological parameters reflects the strong constraining power of CMB polarization, the temperature-polarization cross spectra, and CMB lensing spectra reconstructed from polarization-only estimators, all of which are minimally affected by extragalactic foregrounds. In particular, the survey remains a sensitive probe of light thermal relic particles, achieving \num{$\sigma(N_\mathrm{eff}) = 0.0154$}, which can exclude any new species ($\Delta N_\mathrm{eff} \ge 0.027$) with at least \num{$90\%$} confidence.  Our simulation and foreground-removal codes are publicly available and should aid the development of analysis pipelines for ultradeep, ultrahigh-resolution microwave surveys.
\end{abstract}

\maketitle

\section{Introduction}
\label{sec:intro}

Measurements of the Cosmic Microwave Background (CMB) have played a critical role in establishing the standard model of cosmology.  Arcminute-scale measurements of the CMB, such as those from the Atacama Cosmology Telescope (ACT)~\cite{AtacamaCosmologyTelescope:2025blo, AtacamaCosmologyTelescope:2025nti}, the South Pole Telescope (SPT)~\cite{SPT3G2025}, and the {\it{Planck}} satellite~\cite{Planck:2018vyg}, have allowed rigorous stress testing of this standard model and precise parameter determinations. Current CMB experiments, such as the Simons Observatory~\cite{SimonsObservatory:2025wwn}, are poised to extend this progress further.  Beyond that, a next-generation ultradeep, subarcminute-resolution CMB experiment surveying half the sky, such as CMB-HD~\cite{HDsnowmass}, would open a new window on the microwave universe at small scales.  This would yield unprecedented constraints on the primordial CMB, constraining, in particular, light particles and inflation (e.g., via $N_{\rm{eff}}$ and $n_{\rm{s}}$)~\cite{HDparams}.  This would also revolutionize measurements of the late-time CMB anisotropies, including CMB lensing and the thermal and kinetic Sunyaev-Zel'dovich effects (tSZ and kSZ, respectively).  In addition, the five-fold increase in depth and resolution over the current class of wide-area CMB experiments would enable pioneering searches for variable and transient millimeter-wave phenomena.  

A significant challenge in reaching the key science goals of CMB-HD will be the separation of astrophysical foregrounds from the primordial CMB.  The foregrounds of greatest concern are extragalactic, as opposed to Galactic, since extragalactic foregrounds have a larger impact on small-scale measurements. These foregrounds also predominantly impact CMB temperature measurements, which is a potential concern since much of the signal-to-noise gain on cosmological parameter constraints from small-scale CMB measurements arises from temperature observations~\cite{subgalacticDM}.  In this work, we focus on the three extragalactic astrophysical foregrounds that need to be removed from microwave maps down to low residual levels to take advantage of the low instrument noise of a CMB-HD survey.  We note that this foreground removal would not be possible without the subarcminute resolution of CMB-HD, which allows faint discrete sources to be resolved and subtracted; thus, ultrahigh resolution and ultradeep sensitivity go hand in hand and require each other.  

The foregrounds we focus on in this work are the microwave emission from cosmic infrared background (CIB) and radio galaxies, as well as the tSZ effect from hot, ionized gas in galaxy clusters.  These foregrounds have a frequency dependence that differs from that of the CMB, and they are all relatively localized in real-space maps (subtending scales of about an arcminute or smaller).  We exploit the multiple frequency channels, ultrahigh resolution, and ultradeep sensitivity of a CMB-HD survey in our methods to remove these foregrounds.  We do not attempt to remove the frequency-independent kSZ signal, which is due to the bulk velocity of hot ionized gas; instead, we include the kSZ signal in all our forecasting results.

In Section~\ref{sec:key-result}, we briefly summarize the key results of this work.  In Section~\ref{sec:sims}, we describe the creation of simulations with high enough resolution to explore foreground cleaning and data analysis techniques for a CMB-HD-like survey. In Section~\ref{sec:method-points}, we describe our method for removing CIB and radio sources, and in Section~\ref{sec:method-clusters}, we describe our method for removing galaxy clusters.  We present our results in Section~\ref{sec:results}, and discuss and conclude in Section~\ref{sec:discussion}.

\section{Summary of Key Results}
\label{sec:key-result}

Below, we summarize the key results of this work. 

\begin{itemize}

    \item We create ultrahigh-resolution simulations of the microwave sky, building off a lower-resolution counterpart.  The sky area we present covers 100 square degrees, with a pixel size of 0.04 arcminutes (2.5 arcseconds), and we provide this at six frequency channels (30, 90, 148, 219, 277, and 350 GHz). While we present a particular sky region in this work and build off a particular lower-resolution simulation set, the code we make public allows the selection of any patch of sky, and in principle, can build off of any lower-resolution counterpart, with relatively minor modifications.

    \item We convolve the microwave-sky simulations with the optics and detector noise of a CMB-HD experiment~\cite{HDsnowmass} and apply the foreground-cleaning procedure of Sections~\ref{sec:method-points} and~\ref{sec:method-clusters}.  The procedure successfully detects and removes CIB and radio galaxies: when sources are detected directly in the 90 and 148~GHz maps with a signal-to-noise ratio above four, the resulting catalogs are nearly complete down to a flux limit of \num{0.1} mJy.  Extrapolating sources from other frequencies pushes this further, yielding catalogs that are 95\% complete to flux limits of \num{0.008} and \num{0.04} mJy at 90 and 148~GHz, respectively (top panels of Fig~\ref{fig:fluxhist}). 

    \item We find, as an intermediate step of our foreground cleaning procedure, that we are able to measure mean 277-to-90 and 277-to-148 CIB spectral indices to within \num{1\%} and \num{4\%} of the true mean spectral indices, respectively. We measure the 90-to-148 radio galaxy spectral index to within \num{7\%} of the true mean spectral index. See left panel of Fig~\ref{fig:TrueVsMeasuredFluxIndex} for the 277-to-148 CIB spectral index comparison.

    \item We use a multi-frequency matched-filter to detect galaxy clusters via the tSZ effect. For a signal-to-noise ratio threshold of \num{four or five}, we find clusters above \num{$M_{\rm{500c}} \approx 5 \times 10^{13} M_\odot$} or \num{$\approx 7 \times 10^{13} M_\odot$} with about \num{99\%} completeness, respectively (see Fig~\ref{fig:clustermass}); for these thresholds, the samples of detected clusters are more than \num{65\%} or \num{80\%} pure, respectively. For our foreground cleaning procedure, we detect and remove clusters with a signal-to-noise ratio above \num{four} since we detect many clusters below \num{$M_{\rm{500c}} \approx 5 \times 10^{13} M_\odot$} as well with the lower threshold cut (see Fig~\ref{fig:clustermass}).

    \begin{figure}[t]
    \centering
    \includegraphics[width=\columnwidth]{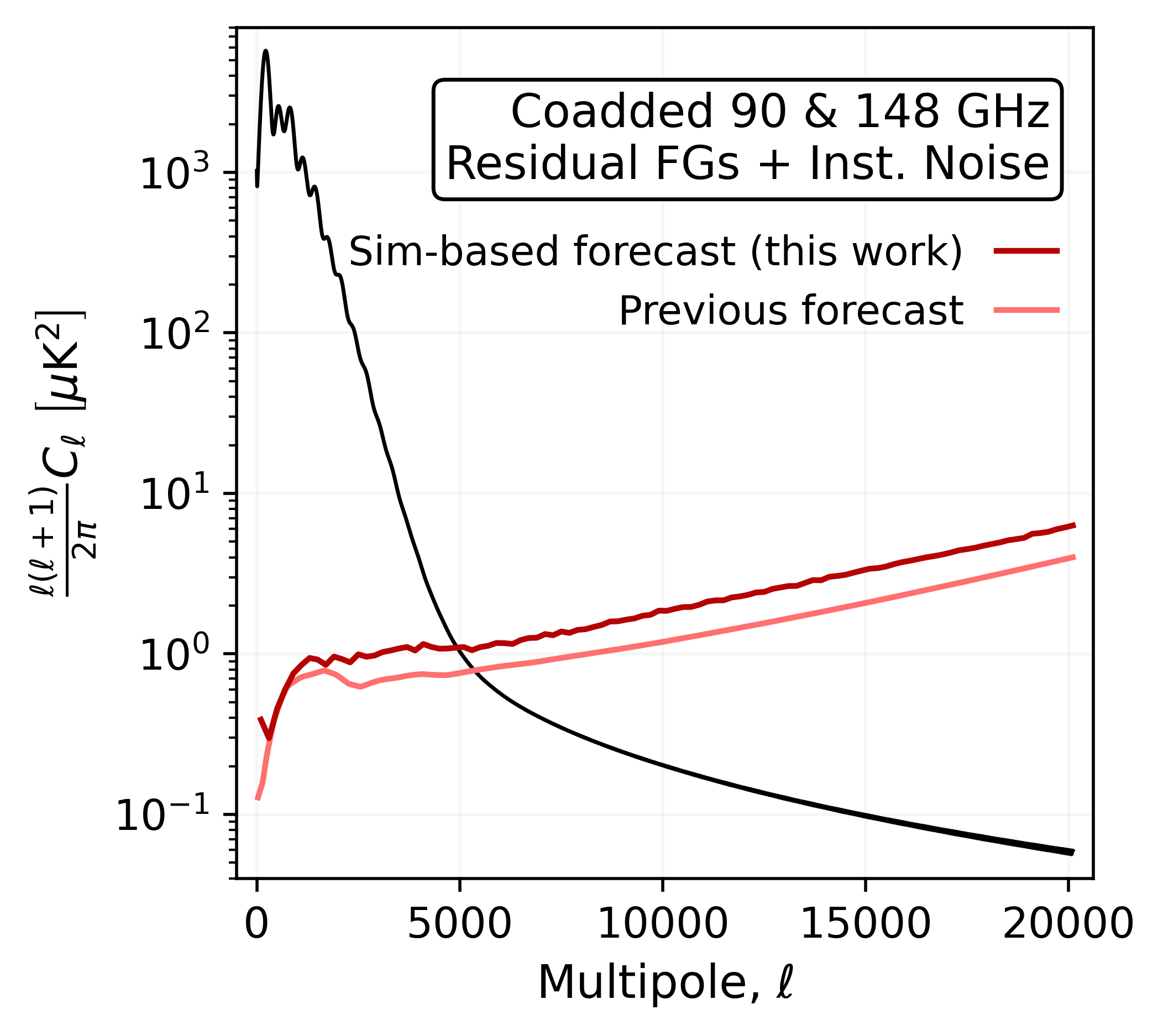}
    \caption{We show in dark red the power spectrum of the beam-deconvolved instrument noise plus residual extragalactic temperature foregrounds obtained from our simulation-based analysis. In light red, we show the previous estimate from~\protect{\cite{han22,HDparams,subgalacticDM}}. We show here the spectrum for the coadded 90 and 148 GHz maps, and show in Figure~\ref{fig:spectra} the power spectra for the 90~and 148~GHz maps separately. We find that the simulation-based forecast is about 50\% higher than the previous estimates.  Note that these red curves show the noise per $\ell$-mode, so each curve should be divided by $\sqrt{2\ell+1}$ to get the error bar per $\ell$-mode. This 50\% increase in $TT$ noise power, when combined with the significant constraining power of the $TE, EE, BB$, and polarization-only $\kappa\kappa$ spectra (all of which are minimally impacted by extragalactic foregrounds), results in at most a 7\% increase in cosmological parameter uncertainties for an 11-parameter $\Lambda$CDM+$N_{\rm{eff}}$+$\sum m_{\nu}$+$T_{\rm{AGN}}+A_{\rm{kSZ}}+n_{\rm{kSZ}}$ model. }
    \label{fig:coaddedspectra}
    \end{figure}

    \begin{figure*}[t]
    \centering
    \includegraphics[width=\textwidth]{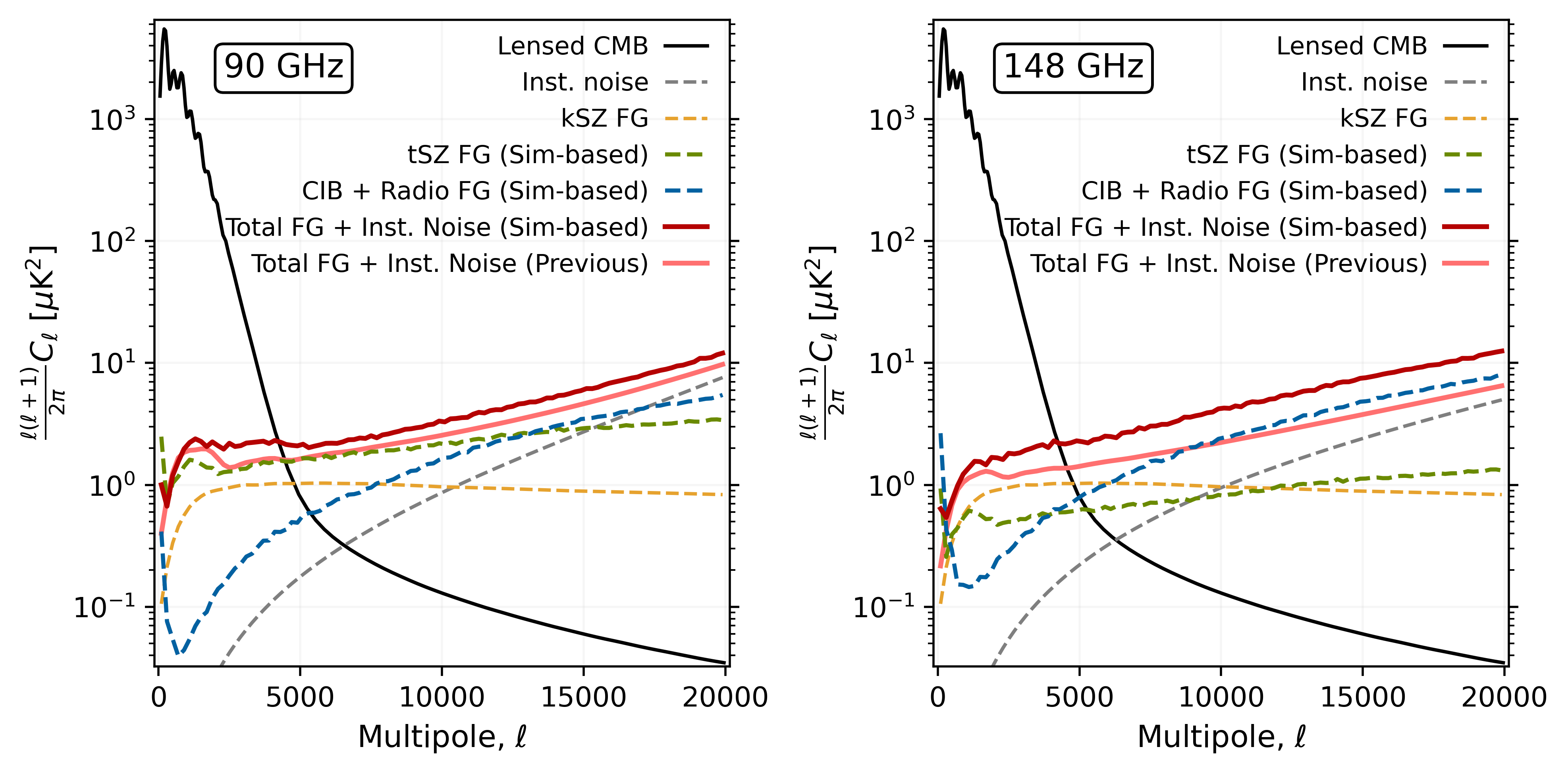}
    \caption{The 90~and 148~GHz residual power spectra of the CIB and radio galaxies (blue dashed) and the tSZ (green dashed) obtained after foreground cleaning simulations of the microwave sky. We also show the lensed CMB power spectrum (black), the full kSZ signal (orange dashed), and the beam-deconvolved instrument noise for a CMB-HD survey (gray dashed). We find that the simulation-based total spectrum of residual foregrounds plus instrument noise (dark red) is slightly higher than that from previous theoretical estimates~\protect{\cite{han22,HDparams,subgalacticDM}} (light red): with an increase of about 30\% for 90 GHz, 80\% for 148 GHz, and 50\% for the 90 and 150 GHz coadded temperature map (see Figure~\ref{fig:coaddedspectra} for the latter).  We also find that we lose less than 1.5\% of the total sky area with the foreground cleaning procedure.  Forecasting constraints on cosmological parameters for an 11-parameter $\Lambda$CDM+$N_{\rm{eff}}$+$\sum m_{\nu}$+$T_{\rm{AGN}}+A_{\rm{kSZ}}+n_{\rm{kSZ}}$ model, including the new simulation-based residual foregrounds and loss of sky area, increases parameter uncertainties by less than~7\% relative to the forecasts in~\protect{\cite{subgalacticDM}}. }  
    \label{fig:spectra}
    \end{figure*}
    
    \item After applying the foreground cleaning procedure on the simulations, the total power spectrum of residual foregrounds plus instrument noise matches that from previous idealized estimates~\cite{han22,HDparams,subgalacticDM} to within about \num{50}\% for the 90 and 150 GHz coadded map (see Fig~\ref{fig:coaddedspectra}), and to within about \num{30}\% for 90 GHz and \num{80}\% for 148 GHz, separately (see Fig~\ref{fig:spectra}). We also find that we lose only \num{1.5}\% of the total sky area with this foreground cleaning procedure.

    \item For the parameter forecasts, we use the delensed power spectra from CMB temperature ($TT$) and E-mode and B-mode polarization ($EE$ and $BB$), as well as the delensed temperature and E-mode cross spectra ($TE$) out to $\ell = 20,000$. We also include the lensing convergence power spectrum ($C_L^{\kappa\kappa}$) to $L=20,000$. We adopt two versions of $C_L^{\kappa\kappa}$ for the parameter forecasts in Table~\ref{tab:params}. The first assumes only polarization quadratic estimators ($EE$ and $EB$) are used in the lensing reconstruction, which removes a potential source of lensing bias from residual non-Gaussian extragalactic foregrounds, as explored in~\cite{vanEngelen:2013rla}. The second adds the $TE$ and $TB$ estimators, restricting the temperature data to CMB multipoles below $\ell_\mathrm{max}^T = 5{,}000$ to avoid high-$\ell$ foreground contamination in these four-point estimators. It further adds the $TT$ estimator: for lensing multipoles $L < 5000$ we again impose $\ell_\mathrm{max}^T = 5{,}000$, while for $L > 5000$ we adopt a previous simulation-based estimate of the $TT$ lensing noise~\cite{han22}, scaled to match the higher $TT$ noise found in this work relative to idealized estimates. In both cases we assume the CMB spectra are delensed with the corresponding lensing maps, and we refer to these as the ``pol-only $\kappa\kappa$'' and ``MV $\kappa\kappa$'' cases.
    
    \item Forecasting constraints on cosmological parameters for an 11-parameter $\Lambda$CDM+$N_{\rm{eff}}$+$\sum m_{\nu}$+$T_{\rm{AGN}}+A_{\rm{kSZ}}+n_{\rm{kSZ}}$ model, we find changes in cosmological parameter errors of less than \num{7\%} relative to the idealized forecasts of~\cite{subgalacticDM}.  This small change reflects the strong constraining power of CMB polarization, the temperature-polarization cross-spectra, and CMB lensing spectra reconstructed from polarization-only estimators, all of which are minimally affected by extragalactic foregrounds.  For the MV $\kappa\kappa$ case we find, in particular, a constraint of \num{$\sigma(N_{\rm{eff}}) = 0.0154$} for a survey covering 60\% of the sky, whereas for the pol-only $\kappa\kappa$ case, we find \num{$\sigma(N_{\rm{eff}}) = 0.0167$} (see Table~\ref{tab:params}); these constraints can exclude any new species ($\Delta N_\mathrm{eff} \ge 0.027$) with at least \num{$92\%$} or \num{$89\%$} confidence, respectively.  We only find a difference in cosmological parameter constraints between the MV $\kappa\kappa$ and pol-only $\kappa\kappa$ cases when simultaneously freeing the amplitude and shape of the kSZ power spectrum ($A_{\rm{kSZ}}$ and $n_{\rm{kSZ}}$); the lower small-scale lensing noise when including temperature data helps break the lensing-kSZ degeneracy in the CMB $TT$ power spectrum.
    
    \item The ultrahigh-resolution simulations and code to generate them are publicly available at \url{https://lambda.gsfc.nasa.gov/simulation/ultrahigh_resolution_sims.html} and \url{https://github.com/CMB-HD/hdsims}, respectively. The foreground cleaning code presented here is publicly available at \url{https://github.com/CMB-HD/hdfgclean}. We release the simulation-based CMB temperature and CMB lensing noise curves obtained in this work at \url{https://github.com/CMB-HD/hdMockData} as version \texttt{v1.2}, which also contains the CMB polarization noise spectra and the covariance matrices derived from the full set of noise curves.  The code used to calculate the covariance matrices will be available at \url{https://github.com/CMB-HD/hdcov}.
    
\end{itemize}

\section{Creation of Simulations}
\label{sec:sims}

The highest resolution of wide-area CMB surveys to date is about 1 to 1.5 arcminutes at 150 GHz~\cite{dr6maps,enhancedSOforecasts2025,SPT3Gmaps2026}.  These CMB experiments have required simulations with about 0.5 arcminute resolution to test analysis pipelines and foreground cleaning procedures.\footnote{CMB maps are usually pixelized with pixels about three times smaller than the telescope resolution~\cite{Sullivan2024}.}  Since CMB-HD will have a dish size about five times larger than precursor wide-area CMB surveys (and thus five times higher resolution), we need to create higher-resolution CMB simulations to investigate CMB-HD foreground removal techniques.  Resolution is also a function of frequency, with higher frequency channels having higher resolution for a fixed dish size.  The highest frequency channel of CMB-HD critical to foreground removal is 277 GHz, with a resolution of 0.13 arcminutes.  Thus, we make CMB simulations with a resolution of 0.04 arcminutes (2.5 arcseconds), which is three times smaller than the 277 GHz beam.

We take as a starting point the microwave sky simulations and source catalogs from~\cite{microwaveSims} (hereafter S10 simulations), which are publicly available.\footnote{\url{lambda.gsfc.nasa.gov/simulation/full_sky_sims_ov.html}}  Although there are other microwave sky simulations publicly available (e.g.~Websky~\cite{websky}, Agora~\cite{agora}, FLAMINGO~\cite{flamingo}, HalfDome~\cite{halfdome}, mmDL~\cite{mmDL}, BACKLIGHT), we use these largely because of familiarity. However, we note that our procedure for generating ultrahigh-resolution simulations can be applied to any lower-resolution simulation set. A useful feature of the S10 simulations is that the CIB and radio galaxies are correlated with the kSZ and tSZ signals, as well as the lensing convergence map.  The tSZ clusters are also modeled with realistic profiles, as opposed to spherical halos.  The above adds realism to the simulations and, in principle, increases the difficulty of separating the foregrounds from the primordial CMB.  

To extrapolate the S10 simulations to higher resolution, we upsample the diffuse extragalactic foreground maps of the S10 simulations (i.e.~the tSZ, kSZ, and lensing convergence maps) to 0.04 arcminute pixels.  Since the S10 simulations did not aim to model these signals on such small scales, we extend, in particular, the kSZ and lensing convergence signals to $\ell > 8000$ and $\ell > 4000$, respectively, by modeling them as described in Section~\ref{sec:smallscalesims}. For discrete sources, we use the S10 catalogs of CIB and radio galaxies to generate CIB and radio source maps natively at 0.04 arcminute, with some modifications to the CIB model discussed in detail in Appendix~\ref{sec:CIBmodel}.  

A new realization of the unlensed primordial CMB is created for both temperature and polarization maps. The $\Lambda$CDM cosmology adopted is ($\Omega_\mathrm{b}h^2,\Omega_\mathrm{c}h^2,h,n_\mathrm{s},A_\mathrm{s}, \tau$) = (0.0222, 0.111, 0.71, 0.961, $2.41\times 10^{-9}$, 0.089), which is the cosmology of the S10 simulations.\footnote{This corresponds to ($\Omega_\mathrm{b},\Omega_\mathrm{m},\Omega_{\Lambda},h,n_\mathrm{s},\sigma_8$) = (0.044, 0.264, 0.736, 0.71, 0.96, 0.80), which is consistent with {\it{WMAP-5}} results~\cite{WMAP5}, and with {\it{Planck}} 2018 results in terms of $\sigma_8$~\cite{planck18params}, which determines the amount of foreground structure.} We also assume three massive neutrinos with $\sum m_\nu$ = 0.06~eV. The unlensed CMB maps are then lensed with the newly created ultrahigh-resolution lensing convergence map. The CIB, radio sources, tSZ, and kSZ are added only to the CMB temperature map; we expect that these foregrounds are much less of a concern for CMB polarization~\cite{Datta:2018oae,SPT:2019wyt}. Below we summarize basic properties of the simulation products that we make public at \url{https://lambda.gsfc.nasa.gov/simulation/ultrahigh_resolution_sims.html}, and in the following subsections we describe each step of their creation in detail.  \\

{\it{Publicly Released Products:}} 
\begin{itemize}
    \item The ultrahigh-resolution simulations that we create and release cover a sky region of \num{121 square degrees ($11^\circ \times 11^\circ$)}, however, the usable area after \num{$0.5^\circ$} apodization around the map borders when convolving with an instrument beam is 100 square degrees. The sky region is centered at a right ascension (RA) of $6^\circ$ and a declination (DEC) of $6^\circ$, and the pixel size is 0.04 arcminute (2.5 arcseconds). While this particular sky region is what we present in this work, the code we make public allows for the selection of any patch of sky. This sky region is large enough to be representative of the foregrounds, and small enough to be computationally efficient given the pixel size.\footnote{It takes about \num{2 hours} to generate the full set of 100 square degree ultrahigh-resolution simulations from the lower-resolution counterparts.}
    
    \item The simulation set has six frequency channels: 30, 90, 148, 219, 277, and 350 GHz, which match the original S10 frequencies and six of the seven frequency channels of CMB-HD~\cite{HDsnowmass} (which will also have a 40~GHz channel).  
    
    \item We provide maps of the unlensed and lensed $T$, $Q$, and $U$ CMB, with power up to \num{$\ell=40,000$} using the CAMB accuracy settings given in Appendix~\ref{sec:CAMBaccuracy}.\footnote{We find that using a maximum $\ell$ of 40,000 to generate an unlensed CMB realization results in less than a 1\% difference between the simulation and theory power spectra given our accuracy settings. This maximum $\ell$ value was also used in~\cite{subgalacticDM}. For comparison, using a maximum $\ell$ of 24,000 gives a 2\% difference in simulation vs theory spectra.} We also provide maps at each frequency of each foreground component (tSZ, kSZ, CIB, radio, and lensing convergence), and a combined temperature map that includes the lensed CMB and tSZ, kSZ, CIB, and radio foreground components.
    
    \item The maps are in the plate carr\'ee (CAR) format~\cite{Calabretta2002}\footnote{We use the pixell package (\url{pixell.readthedocs.io}) when working with CAR maps, unless otherwise stated.}, and all have been convolved with the CAR pixel window function, except the lensing convergence map. All maps are in units of $\mu$K, except the lensing convergence map, which is dimensionless.  
    
    \item We provide catalogs of the CIB and radio sources within the 100 square degree maps, giving the location of each source and its flux (in units of mJy) at each frequency.  We also provide catalogs of the galaxy clusters in these maps, giving the cluster positions, masses, sizes, redshifts, and tSZ and kSZ properties.
\end{itemize}

\begin{figure*}[t]
    \centering
    \includegraphics[height=\textheight]{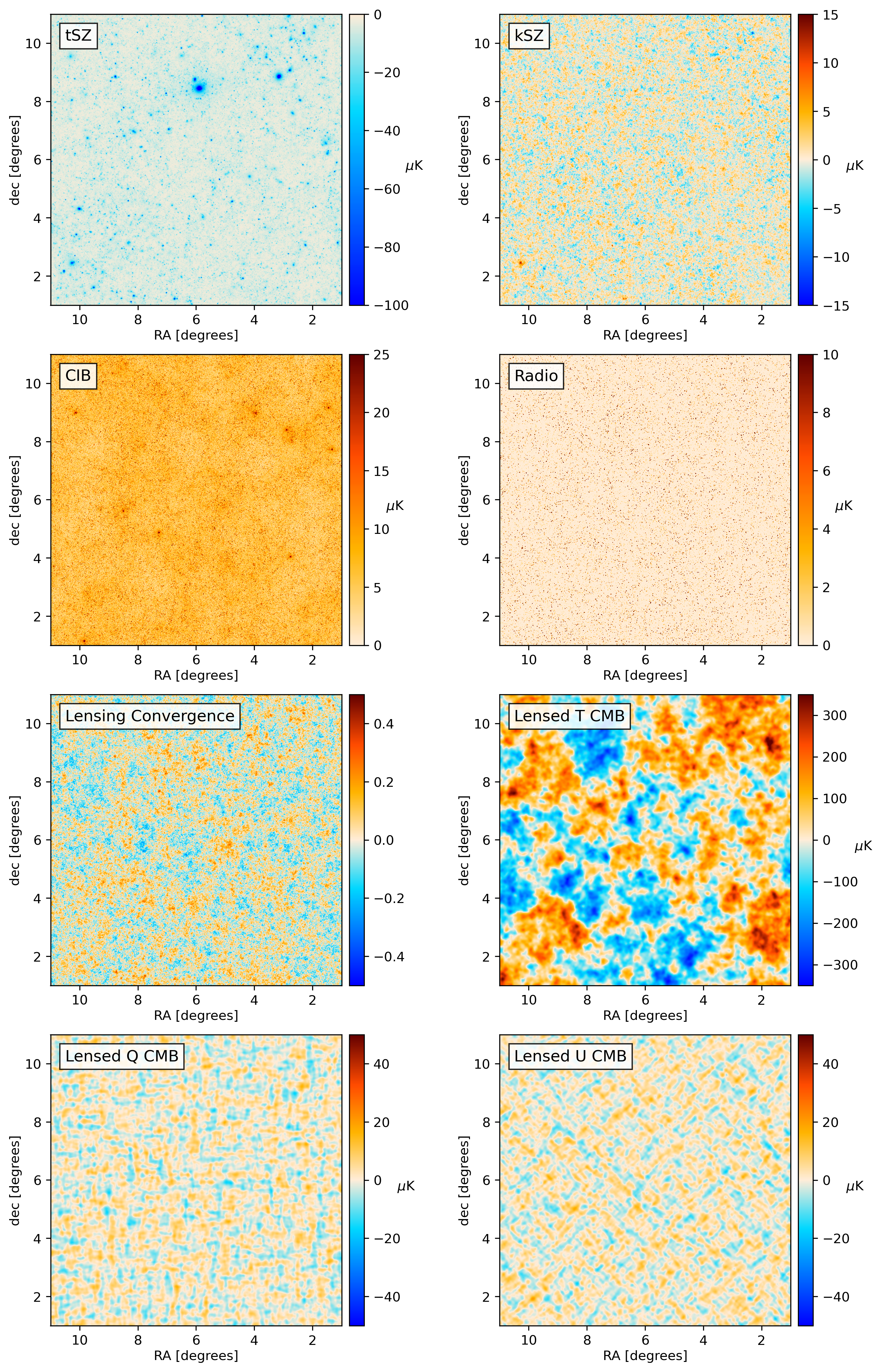}
    \caption{100 square degree microwave sky simulations at 90 GHz with 0.04 arcminute pixelization.  All, except the lensing convergence map, are convolved with the pixel window function and the CMB-HD beam.}
    \label{fig:maps}
\end{figure*}

\begin{figure*}[t]
    \centering
    \includegraphics[width=\textwidth]{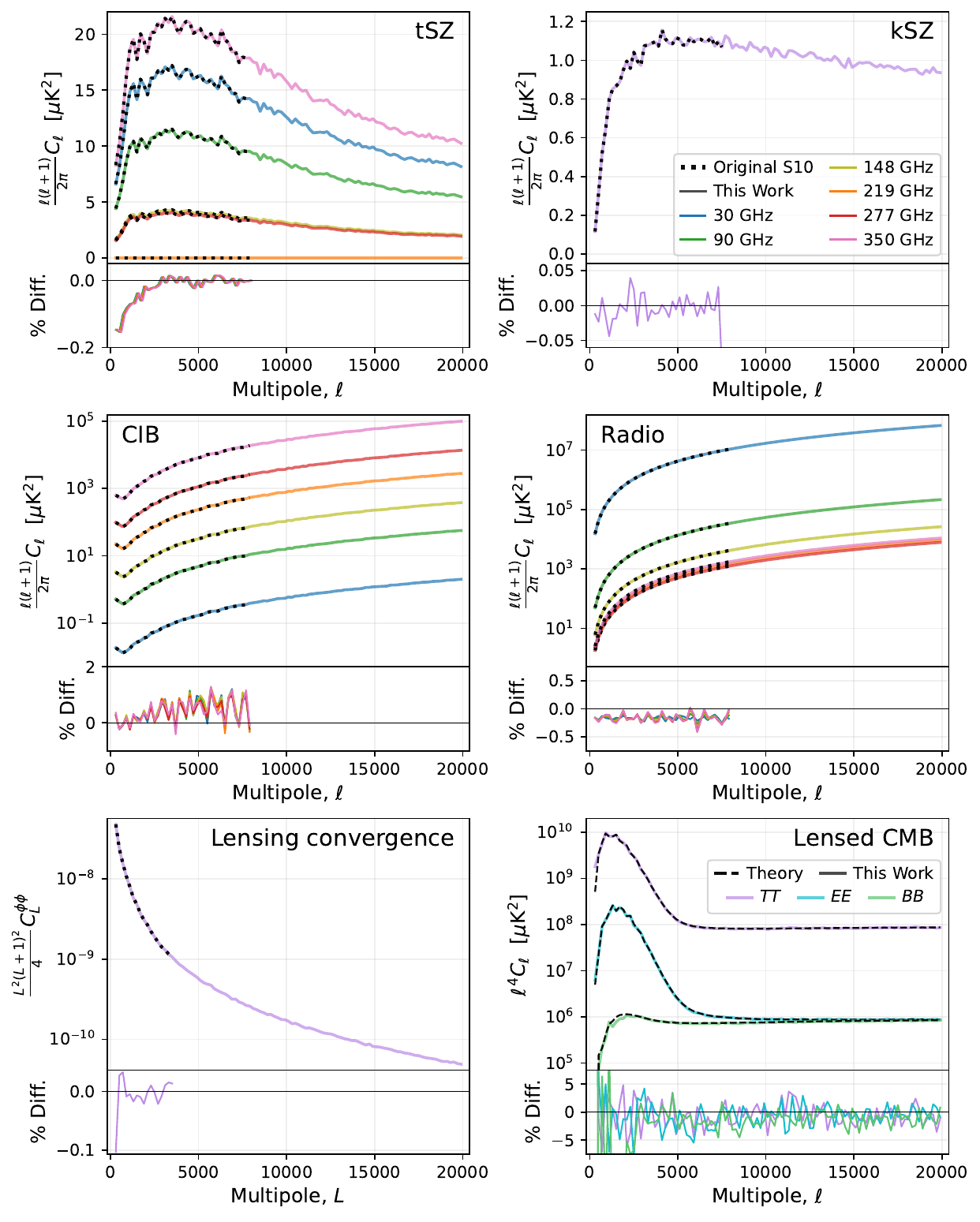}
    \caption{We show the power spectra of the simulation maps shown in Figure~\ref{fig:maps} (solid curves), and the spectra of the S10 simulations in the same patch of sky up to $\ell=8000$ ($\ell=4000$ for the lensing convergence) (dotted curves). The percent difference between the S10 simulations and this work is indicated in the bottom panels. (Different frequency curves are offset in $\ell$ for clarity.) For the CIB sources, the roughly 2\% deviation is due to adding extra scatter in the source positions for the higher resolution simulations (see Appendix~\ref{sec:CIBmodel}). For the lensed $TT, EE,$ and $BB$ spectra, the difference plot is with respect to the theory (dashed).} 
    \label{fig:sim-spectra}
\end{figure*}

In Section~\ref{sec:diffuse}, we discuss how we generate maps of the diffuse components (tSZ, kSZ, and lensing convergence).  We describe the creation of the lensed CMB in Section~\ref{sec:lensedCMB}.  In Section~\ref{sec:discrete}, we describe how we make maps of the discrete components, namely the CIB and radio galaxies, and in Section~\ref{sec:instrument}, we discuss the application of instrument properties.  Figure~\ref{fig:maps} shows images of these simulated maps at 90 GHz. Figure~\ref{fig:sim-spectra} shows the power spectra of these maps up to $\ell=20,000$ and the match to the S10 power spectra over multipoles where they overlap.

\begin{figure*}[t]
    \centering
    \includegraphics[width=\textwidth]{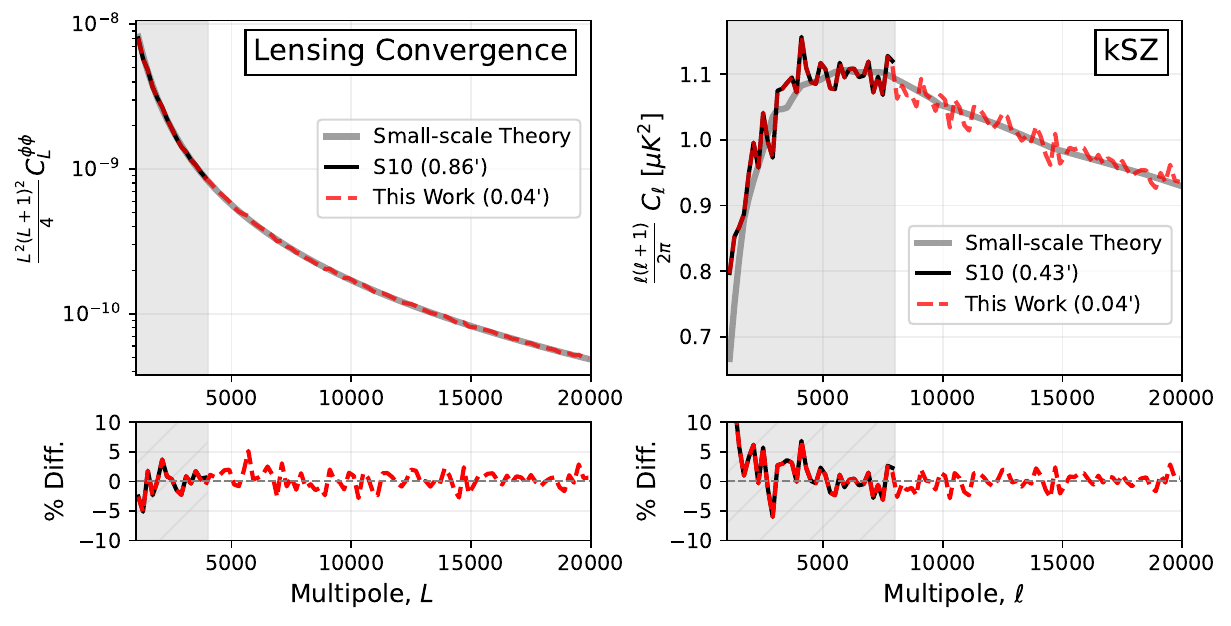}
    \caption{Power spectra of the kSZ and lensing convergence maps in this work after extending to small scales (red curves). We overplot the power spectra of the S10 simulations for this region of sky (black curves). We compare these simulation spectra to theory spectra (gray curves) that are matched to the S10 simulations for this sky region below $\ell=4000$ for the lensing convergence and $\ell=8000$ for the kSZ (gray shaded regions), as described in Section~\ref{sec:smallscalesims}. In the bottom panels, we show the percent difference between the simulation power spectra and the theory spectra.}
    \label{fig:sim-stitching}
\end{figure*}

\subsection{Diffuse Foregrounds: kinetic SZ, thermal SZ, and lensing convergence} \label{sec:diffuse}

Here we describe how we create ultrahigh-resolution maps of the diffuse extragalactic foregrounds (tSZ, kSZ, and lensing convergence) from lower-resolution S10 counterparts.

The S10 simulations are provided as full-sky HEALPix maps.\footnote{\url{healpix.jpl.nasa.gov}; we use the healpy package (\url{healpy.readthedocs.io}) when working with HEALPix maps.} The S10 tSZ and kSZ maps have HEALPix \texttt{nside} = 8192 (corresponding to 0.43 arcminute resolution), while the lensing convergence ($\kappa$) map has \texttt{nside} = 4096 (corresponding to 0.86 arcminute resolution). The tSZ and kSZ maps are available in units of flux per solid angle (Jy/sr). We convert them to CMB temperature units by dividing by a factor of $\left.\left(\partial B_\nu(T) / \partial T\right)\right|_{T_\mathrm{CMB}}$ (which has units of Jy/sr/K), where $B_\nu$ is the Planck intensity function at frequency $\nu$ and we use $T_\mathrm{CMB}$ = 2.7255 K for the CMB temperature~\cite{Fixsen1996}. 

We multiply the S10 tSZ maps at each frequency by 0.75, as was done in~\cite{SOforecasts}, to better match current observations.  We also multiply the S10 kSZ map by a factor of $1/\sqrt{2}$, which reduces the power by a factor of 2, in order to better match recent data \cite{Beringue2025}; this reduced kSZ amplitude was also assumed in previous CMB-HD forecasts \cite{HDparams,subgalacticDM}.

To obtain higher-resolution versions of these maps with a usable area of 100 square degrees, we:
\begin{itemize}
    \item Deconvolve the pixel window function from the full-sky S10 tSZ and kSZ maps. (Note that the S10 lensing convergence map does not contain the pixel window.)
    
    \item Reproject from HEALPix to CAR, cutting out a \num{$12^\circ \times 12^\circ$} patch of sky. 
    
    \item Apodize a \num{$0.5^\circ$} region around the edges of each map; this will taper the edges so that they smoothly go to zero, ensuring periodic boundary conditions.
    
    \item Upsample the map resolution to 0.04 arcminutes by zero-padding them in Fourier space, using the \texttt{pixell.enmap.resample} function. 
    
    \item Extend the kSZ and lensing convergence maps to smaller scales following the procedure described in Section~\ref{sec:smallscalesims} below. 
    
    \item Convolve the kSZ and tSZ maps with the CAR pixel window function and cut out the inner \num{$11^\circ \times 11^\circ$} region.\footnote{We use \texttt{pixell.enmap.project} for cutting out smaller patches in CAR maps, which we find to be more accurate than \texttt{pixell.enmap.submap} for maps of arbitrary size.} Note that the lensing convergence map is not convolved with a pixel window, and the full \num{$12^\circ \times 12^\circ$} lensing convergence map is needed to lens the unlensed CMB map (see Section~\ref{sec:lensedCMB}); we later cut this convergence map down to \num{$11^\circ \times 11^\circ$} to be consistent with the other maps.
\end{itemize}

To use these simulations to test CMB-HD-like data analyses, we apodize the \num{$11^\circ \times 11^\circ$} maps with a \num{$0.5^\circ$} border, and convolve with the appropriate beam for that frequency, leaving an inner 100 square degree un-apodized region. In Figure~\ref{fig:maps}, we show the 100 square degree tSZ, kSZ, and lensing convergence maps generated from the full procedure (including as described in the sub-sections below) convolved with a CMB-HD-like beam; the tSZ map shown is at 90 GHz, while the kSZ and lensing convergence maps are frequency independent. In Figure~\ref{fig:sim-spectra}, we show the match between the original S10 power spectra for this particular patch of sky (black dotted curves) and the power spectra of the ultrahigh-resolution counterparts (solid curves). In the bottom panels of Figure~\ref{fig:sim-spectra}, we show a fractional difference plot between the S10 simulations and this work; the fractional difference between these spectra for the tSZ, kSZ, and lensing convergence is better than 0.2\%.

\subsubsection{Extending to Small Scales: kinetic SZ and lensing convergence} \label{sec:smallscalesims}

To make simulations suitable for CMB-HD, we need the tSZ, kSZ, and lensing convergence maps to contain accurate power up to at least \num{$\ell=20,000$}.  The S10 simulations did not aim to model the kSZ and tSZ signals beyond $\ell \sim 8000$, and the lensing convergence beyond $\ell \sim 4000$. However, since the dominant contribution to the tSZ power spectrum is from localized halos of roughly arcminute size, the S10 simulations have a reasonable power spectrum for the tSZ signal up to $\ell=20,000$, as shown in Figure~\ref{fig:sim-spectra}.  Thus, we do not further modify the tSZ, other than upsampling the pixelization of the map. This also preserves the correlation between the tSZ and the other foreground components.

In contrast, the S10 power spectra of the kSZ signal and lensing convergence deviate significantly from theoretical expectations at small scales ($\ell > 10,000$). (For the kSZ, this deviation reaches as high as a factor of two, and for the lensing convergence, there is no power in the S10 simulations above $\ell = 12,500$.) Thus, to generate $0.04$ arcminute maps with physical accuracy for these two signals, we retain the match to the S10 simulations on large scales and replace the small-scale signals with Gaussian random fields matched to theory. Note that retaining the match to the S10 simulations for $\ell < 4,000$ for the lensing convergence and $\ell < 8,000$ for the kSZ preserves the correlation among the kSZ signal, the lensing convergence, the tSZ signal, and the underlying galaxy clusters and groups, which are on roughly arcminute scales (about $\ell = 3,000$); for the distribution of dark matter and the kSZ on smaller scales than this, a Gaussian random field is a reasonable approximation.

To obtain theoretical spectra on small scales that match the power in the S10 simulations on scales $\ell \lesssim 8000$ for the kSZ and $\ell \lesssim 4000$ for the lensing convergence, we:
\begin{itemize}
    \item Measure the power spectrum of the kSZ and lensing convergence maps from the S10 simulation for our patch of sky (i.e., we cut the \num{$12^\circ \times 12^\circ$} S10 CAR patches to $10^\circ \times 10^\circ$ and take the power spectra). We will denote these power spectra by $C_{\ell}^{\mathrm{S10,kSZ}}$ and $C_{\ell}^{\mathrm{S10},\kappa}$, respectively.
    
    \item Obtain a template theory power spectrum out to \num{$\ell = 40,000$} for the kSZ and lensing convergence, denoted by $C_{\ell}^{0\text{,kSZ}}$ and $C_{\ell}^{0,\kappa}$, respectively. (This value of $\ell$ was chosen to be consistent with the CMB theory.) The kSZ template spectrum is from~\cite{Battaglia2010} (for the late-time kSZ signal) and~\cite{Smith2018,Park2013} (for the reionization contribution). The lensing convergence template is generated with CAMB~\cite{CAMBLewis:1999bs,CAMBHowlett:2012mh}\footnote{\url{camb.readthedocs.io}}, using the S10 cosmology described above and the CAMB accuracy settings given in Appendix~\ref{sec:CAMBaccuracy}. 
    
    \item Since the template spectra will not exactly match the spectra in our chosen patch of sky, we adjust the theory spectra to fit those in our patch.  We do this by defining a model for the theory power spectrum based on the template, given by 
    \begin{equation} 
        C_\ell^{\mathrm{theo},X}  = A_X \left(\frac{\ell}{\ell_0}\right)^{n_X}  C_\ell^{0,X}, \label{eq:smallscaletheo}
    \end{equation}
    for $X \in \{\mathrm{kSZ},~\kappa\}$, where $\ell_0 = 3100$ is a fixed pivot scale,\footnote{The measured power spectrum of the simulations is binned, so we choose this value of $\ell_0$, instead of $\ell_0=3000$, because it approximately corresponds to the center of a bin.} and $A_X$ and $n_X$ are free parameters for an amplitude and slope, respectively. 
    
    \item We fit the amplitude and slope parameters above to obtain theory spectra that are consistent with the power spectra of the S10 simulations within our patch of sky. 
    \begin{itemize}
        \item We determine the amplitude by requiring that the theory spectrum equal the simulation spectrum at the pivot scale. (To obtain the spectrum of the simulation at the pivot scale, we average over three bins centered around $\ell_0$ to reduce scatter.)  
        \item We determine the slope by minimizing the squared difference $\left(C_{\ell_X}^{\mathrm{theo},X} - C_{\ell_X}^{\mathrm{S10},X} \right)^2$ between the theory spectrum and the simulation spectrum at $\ell_X$, which we take to be 8100 for kSZ and 3900 for $\kappa$.\footnote{We use \texttt{scipy.optimize.minimize} to perform the minimization, which by default utilizes the Broyden–Fletcher–Goldfarb–Shanno algorithm.} (Again, we average the simulation power in three bins centered around $\ell_X$ to reduce scatter.)   
    \end{itemize}
\end{itemize}

For our patch of sky centered at RA = $6^\circ$, Dec = $6^\circ$, this fitting procedure results in $A_\mathrm{kSZ} \approx \num{1.05}$, $n_\mathrm{kSZ} \approx \num{0.036}$, $A_{\kappa} \approx \num{0.984}$, and $n_{\kappa} \approx \num{-0.105}$. Using these theory curves, we generate ultrahigh-resolution maps of the kSZ and lensing convergence signals as follows. 
\begin{itemize}
    \item We apodize the \num{$12^\circ \times 12^\circ$} upsampled 0.04 arcminute resolution kSZ and lensing convergence maps (described above) and take their harmonic transforms to obtain their $a_{\ell m}$ coefficients.\footnote{ Note that a real-space map, $M(\hat{\mathbf{n}})$, where $\hat{\mathbf{n}}$ is the position on the sky, can be expanded in  spherical harmonics as $M(\hat{\mathbf{n}}) = \sum_{\ell, m}  a_{\ell m} Y_{\ell m}(\hat{\mathbf{n}})$.} 

    \item We generate a Gaussian random realization of $a_{\ell m}$ coefficients from the theory curves described above.

    \item We then replace the upsampled map $a_{\ell m}$'s with the Gaussian random $a_{\ell m}$'s on scales $\ell > 8000$ for the kSZ and $\ell > 4000$ for the lensing convergence, and then transform back to real space to obtain final \num{$12^\circ \times 12^\circ$}, 0.04 acrminute resolution maps. 
\end{itemize}
This procedure preserves the large-scale characteristics of the S10 simulations while maintaining accuracy to theory at small scales, as can be seen in Figure~\ref{fig:sim-stitching}.

\subsection{Lensed CMB} \label{sec:lensedCMB}

To make lensed CMB maps, we first use CAMB~\cite{CAMBLewis:1999bs,CAMBHowlett:2012mh} to generate the theoretical unlensed CMB temperature and polarization power spectra up to \num{$\ell_\mathrm{max} = 40,000$}. We adopt the same cosmology as the S10 simulations and use the high-accuracy CAMB settings described in Appendix~\ref{sec:CAMBaccuracy}.   We then generate a random realization of $a_{\ell m}$'s from the unlensed theory power spectra and use these to generate \num{$12^\circ \times 12^\circ$} unlensed CMB $T$, $Q$, and $U$ maps. 

The unlensed CMB maps are then lensed with the lensing convergence map extended to small scales, described in Section~\ref{sec:diffuse}. We do this using the curved-sky lensing routine \texttt{pixell.lensing.lens\_map\_curved}\footnote{While flat-sky lensing routines are computationally less expensive, the resulting lensed CMB power spectra deviate from those of the curved-sky routine by up to \num{2\%} for our multipole range of interest for \num{$11^\circ \times 11^\circ$} maps.}, which requires the harmonic coefficients of the lensing potential map $\phi$, defined by $\phi_{\ell m} = 2\kappa_{\ell m} / \ell(\ell + 1) $ \cite{Planck:2015mym}, and of the unlensed CMB map. We apodize both the lensing convergence and unlensed CMB maps by \num{$0.5^\circ$} around the borders, harmonically transform them, lens the CMB, and then cut out the inner \num{$11^\circ \times 11^\circ$} region of the resulting lensed CMB map. We show our lensed CMB $T$, $Q$, and $U$ maps in Figure~\ref{fig:maps}.

To cross-check our map-level lensing of the CMB, we obtain lensed CMB theory spectra from CAMB as follows. We give CAMB the measured power spectrum of the lensing convergence map of our patch of sky. (For multipoles $L \leq 100$, which are not captured by our patch of sky, we give CAMB the $C_\ell^{\mathrm{theo},\kappa}$ obtained above by fitting a CAMB generated lensing power spectrum template to the S10 patch (see Equation~\ref{eq:smallscaletheo}).)  We then use CAMB to lens our unlensed theory spectra with that lensing convergence power.\footnote{We use \texttt{get\_lensed\_cls\_with\_spectrum}, which calculates the lensed spectra using the curved-sky correlation function method.} We show the agreement between the lensed CMB theory spectra from CAMB and the simulation lensed CMB spectra in Figure~\ref{fig:sim-spectra}.

\subsection{Discrete Foregrounds: Radio sources and CIB} \label{sec:discrete} 

We create ultrahigh-resolution maps of CIB and radio galaxies by using the catalogs of these sources directly.  The S10 galaxy catalogs contain the position of each source in RA and DEC, and its flux in units of mJy at each of the six frequencies considered here.  We truncate this catalog to include only those sources within our \num{$12^\circ \times 12^\circ$} patch of the sky.  This results in a catalog of about~\num{200,000} radio sources and \num{17,800,000} CIB sources.  We lower the fluxes of all CIB sources by scaling them by a factor of 0.75, as was done in~\cite{SOforecasts}, to better match current observations. The CIB model in the S10 simulations has many sources that fragment into ``a cloud'' of lower flux sources when mapped at a resolution much higher than the S10 resolution of 0.43 arcminutes.  Thus, we explore three cases: the original S10 model, a model with sources within the same 0.43 arcminute pixel summed together, and a model with sources within the same 0.25 arcminute pixel summed together. Although the original S10 model matches the observed number counts well for very low fluxes (0.01 mJy at 148 GHz), for fluxes close to our expected measurement thresholds (\num{0.05 mJy} at 148 GHz), the latter two models match the observed number counts better (see Figure~\ref{fig:cibmodels} in Appendix~\ref{sec:CIBmodel}).  We choose the model with 0.25 arcminute summation as our baseline CIB model since it has the best overall match to observations, and we explore foreground cleaning results for the other two CIB models in Appendix~\ref{sec:CIBmodel}.  The baseline CIB model results in a CIB catalog of about \num{7,260,000} sources for the \num{$12^\circ \times 12^\circ$} patch of the sky. 

We then generate empty \num{$12^\circ \times 12^\circ$} sky maps at 0.04 arcminute resolution for each frequency and populate them with the CIB and radio sources from the truncated catalogs.  For each source in the catalog, we find the pixel in the map that corresponds to its RA and DEC coordinates and add the amplitude of the source (its flux in units of Jy) to that pixel. We then divide each pixel's flux value by the pixel's solid angle to obtain a map in units of Jy/sr, and divide by $\left.\left(\partial B_\nu(T) / \partial T\right)\right|_{T_\mathrm{CMB}}$ to produce a map in CMB temperature units of $\mu$K.  We apodize these sky patches over \num{$0.5^\circ$} at the edges, convolve with the CAR pixel window function, and cut out the inner \num{$11^\circ \times 11^\circ$} region.

In Figure~\ref{fig:maps}, we show the CIB and radio source maps at 90 GHz, after apodizing with a \num{$0.5^\circ$} border, convolving with the CMB-HD beam, and cutting out the inner $10^\circ \times 10^\circ$ square degrees.   In Figure~\ref{fig:sim-spectra}, we show the match between the original S10 power spectra for this particular patch of sky (black dotted curves) and the power spectra of the ultrahigh-resolution counterparts (solid curves). The fractional difference between these spectra is better than \num{2\%}; the deviation at small scales for the CIB is due to added extra scatter in the source positions for the higher resolution simulations, as described in Appendix~\ref{sec:CIBmodel}.

\subsection{Instrument Properties}
\label{sec:instrument}

\begin{table}[t]
    \begin{center}
    \begin{tabular}{c@{\hskip 2em} c@{\hskip 2em} c}
      \toprule
      \toprule
      Frequency  & White Noise level  & Beam size \\
      (GHz) & ($\mu$K-arcmin) & (arcmin) \\
      \midrule
      30 & 6.5 & 1.25 \\
      90 & 0.7 & 0.42 \\
      148 & 0.8 & 0.25 \\
      219 & 2.0 & 0.17 \\
      277 & 2.7 & 0.13 \\
      350 & 100 & 0.11 \\
      \bottomrule
    \end{tabular}    
    \caption{Listed are the instrumental specifications for the CMB-HD experiment as described in~\protect{\cite{HDsnowmass}}. The columns are observing frequency in GHz, the white noise level for temperature in $\mu$K-arcmin, and the beam full-width at half-maximum (FWHM) in arcmin. Noise levels for polarization are a factor of $\sqrt{2}$ higher. (CMB-HD will also have a 40~GHz channel with 3.4 $\mu$K-arcmin noise and 0.94~arcmin resolution, which is not modeled here.)} 
    \label{tab:frequencies}
    \end{center}
\end{table}

From the maps described above, we take the sum of the \num{$11^\circ \times 11^\circ$} pixel-window-convolved kSZ, tSZ, CIB, radio, and lensed CMB temperature maps to produce a single temperature map at each frequency.  
We apodize this combined $T$ map, as well as the $Q$ and $U$ maps, at each frequency by \num{$0.5^\circ$} and convolve with the appropriate instrument beam for that frequency.\footnote{We model the beam with a Gaussian profile, and use the \texttt{pixell.enmap.smooth\_gauss} function to convolve the beam in flat-sky Fourier space. We found that convolving the beam in curved-sky harmonic space produced ringing around bright sources and is more computationally expensive.} Table~\ref{tab:frequencies} gives the beam size and white noise level for each frequency of CMB-HD.  Given these levels of white noise, we make a map of the instrumental white noise for each frequency,\footnote{We use the \texttt{pspy.so\_map.white\_noise} function of the pspy package (\url{pspy.readthedocs.io}) to generate the instrumental noise maps.} assuming independent noise realizations for the $T$, $Q$, and $U$ maps.  We add these instrument noise maps to the beam convolved simulations and use this set of CMB-HD-like simulations in the following foreground removal analysis.\footnote{Since the first step of our foreground cleaning procedure is to filter the maps in Fourier space, we begin with the \num{$11^\circ \times 11^\circ$} apodized maps, which allow us to remove foregrounds from the full $10^\circ \times 10^\circ$ inner region.}

\section{Foreground Cleaning: Method to remove CIB and Radio Sources}
\label{sec:method-points}

For the foreground cleaning analysis described throughout this work, our main focus will be on residual foregrounds in the 90 GHz and 148 GHz maps. Maps at these frequencies will have the lowest instrumental noise and, thus, will be primarily responsible for cosmological parameter constraints.  

One traditional technique to remove CIB and radio sources is to exploit their frequency dependence and create a new map that is some linear combination of CMB maps at different frequencies. The linear combination is chosen to eliminate sources with a given spectral dependence.  This is called ``constrained ILC'' or ``deprojection'' in some works and can come at the cost of increased noise in the final map since the constraint uses a degree of freedom, i.e., one of the frequency maps~\cite{Remazeilles2011, Surrao2025, SOforecasts, Kwok:2025npw}. A constrained ILC approach also does not exploit the fact that these sources are localized in the maps.  CMB-HD will have seven frequency channels, and a constrained ILC method may be helpful, potentially after sources have already been removed to a low flux level. Similarly, galaxy-tracer-assisted ILC cleaning methods may provide further gains~\cite{Chen:2026cad}. We also note that frequency-dependent foregrounds will also be constrained by the multi-frequency likelihood traditionally adopted for parameter forecasts~\cite{Dunkley:2013vu, Kokron:2024ioy,Beringue2025}.  We leave the exploration of these avenues for future work. 

Instead, we focus on another traditional technique, which is to use a matched filter to exploit the localization and shape of CIB and radio sources.  In CMB maps, these sources are mostly ``point-like'' and have the shape of the telescope beam.  Thus, filtering CMB maps for objects that match the shape of the beam is an effective way of finding and removing them.  This technique gains effectiveness as the telescope resolution increases, provided that the resolution is such that most of the sources remain point-like; this is the case for CMB-HD, which has a resolution of about 10 arcseconds.  The point-like nature of the extragalactic CIB and radio sources on these scales means that their signal increases faster than the noise in a given map pixel, as the telescope resolution increases and the map pixel size decreases for a fixed instrument noise level. Therefore, each source's signal-to-noise ratio increases as the resolution increases.  

Another advantage of the matched-filter technique is that by measuring the flux of each detected source and knowing that the source's shape matches that of the telescope beam, each detected source can be subtracted from the original map without losing any sky area or increasing the noise of the map. A final advantage of the matched-filter technique is that the subarcminute resolution of CMB-HD allows for relatively clean separation of CIB and radio galaxies from tSZ clusters, the latter of which primarily subtend scales greater than an arcminute.  Thus, we utilize this technique throughout this work and expect potentially further gains from constrained ILC methods applied in addition.  

Since we would like to remove CIB and radio sources down to very low flux levels for a CMB-HD survey, we require a number of modifications to the traditional matched-filter method. For example, we employ an iterative approach, making many passes through the maps and removing successively dimmer sources at each pass.  When we can no longer detect sources directly in the 90 and 148 GHz maps, we use the 277 GHz map (where the CIB sources are brighter and the resolution is higher) to identify dim CIB sources at 90 and 148 GHz.  Similarly, we use the 90 GHz map to identify dim radio sources at 148 GHz (since radio sources are brighter at 90 GHz).  We note that the 30, 40, and 350 GHz channels have noise levels too high to be useful for these purposes. Another modification is that we identify any spurious \num{4$\sigma$} sources in our near-final source-subtracted maps, add their full flux back to the maps, and repeat our procedure at those locations; thus, we remeasure and resubtract these sources with a cleaner background environment to resolve any initial mis-subtraction, potentially due to nearby sources. 

In the following, we itemize the steps of our procedure and provide further details in subsequent sections.

\subsection{Overview of Procedure}
\label{sec:overview}

Here we give an overview of the procedure we use to remove CIB and radio sources from the 90~and 148~GHz maps. Where we refer to {\it{filter map}}, {\it{detect sources}}, and {\it{measure the fluxes}}, the reader can refer to Sections~\ref{sec:filter},~\ref{sec:detect-src}, and~\ref{sec:measure-flux}, respectively, for details on these procedures.  We {\it{subtract sources}} by ``reconstructing'' a map of sources at a given frequency and then subtracting this source map from the corresponding original map.  Below, ``bright sources'' refer to sources detected with a signal-to-noise ratio (SNR) greater than \num{four}, and ``dim sources'' refer to sources with an SNR less than \num{four}. 

\begin{enumerate}
    \item {\ul{Make a catalog of 277~GHz sources:}}  We {\it{filter}} the 277~GHz map, {\it{detect sources}} with SNR greater than \num{four}, {\it{measure the fluxes}} of these sources, and create a catalog of them. 

    \item {\ul{Remove bright CIB and radio sources from 90 GHz maps:}} We {\it{filter the map}} at 90 GHz and {\it{detect sources}} with SNR greater than \num{four}. If one of these sources is within the 90 GHz beam size of a source in the 277 GHz CIB  catalog, then we use the position of the source in the 277 GHz catalog. (Using the 277 GHz positions aids in CIB source localization since CIB sources are brighter, and the resolution is higher at 277 GHz compared to lower frequencies. It also aids in knowing which 277 GHz sources have not already been subtracted for \num{Step 5} below.) We identify sources without a brighter 277 GHz counterpart as radio sources and make a 90 GHz radio source catalog.  We {\it{measure the fluxes}} of all detected 90 GHz sources and {\it{subtract}} them from the map. 

    \item {\ul{Remove bright CIB and radio sources from 148 GHz maps:}} We {\it{filter the map}} at 148 GHz and {\it{detect sources}} with SNR greater than \num{four}. If one of these sources is within the 148 GHz beam size of a source in the  277 GHz CIB or 90 GHz radio catalog, then we use the position of the source in the 277/90 catalog.   We {\it{measure the fluxes}} of these sources and {\it{subtract}} them from the 148 GHz map. 

    \item {\ul{Get 277-to-90 and 277-to-148 CIB spectral indices, and 90-to-148 radio spectral index:}}  We collect sources that have an SNR greater than ten in the 90 and 148 GHz maps. If they have counterparts in the 277 GHz catalog that are brighter, we use them to determine the 277-to-90 GHz and 277-to-148 GHz CIB spectral indices.  If $10\sigma$ sources in the 148 GHz map have counterparts in the 90 GHz catalog that are brighter, we use them to determine a 90-to-148 GHz radio spectral index.\footnote{To give a sense of flux, sources detected at greater than $10\sigma$ have a flux greater than 0.2 mJy at 90 and 150 GHz. It is possible that this population could have a different spectral index than the population of dimmer sources; this is something we can check and potentially correct for if needed, using the 277, 220, 30, and 40 GHz frequency channels.} 

    \item {\ul{Remove dim sources from 90 and 148 GHz maps using 277 GHz CIB and 90 GHz radio source catalogs and corresponding spectral indices:}} We {\it{subtract}} from the 90 and 148 GHz maps all sources in the 277 GHz CIB catalog that have not already been subtracted, using the 277-to-90 GHz and 277-to-148 GHz CIB spectral indices to estimate the 90 and 148 GHz fluxes. We {\it{subtract}} from the 148 GHz map all sources in the 90 GHz radio catalog that have not already been subtracted, using the 90-to-148 radio spectral index to estimate the 148 GHz flux.   

    \item \ul{Remeasure any missubtracted sources in the 90~and 148~GHz maps}: After subtracting all sources using the steps above, we \textit{filter} the source-subtracted maps at 90 and 148 GHz and \textit{detect} any sources with $|$SNR$|$ greater than \num{four}, which are either over- or under-subtracted sources. We add the full flux of these missubtracted sources back to the maps and repeat  \num{steps 1 through 3} at these locations. This allows us to gain a better measurement of both the position and flux of these sources, which may have been missubtracted due to source blending.
\end{enumerate}

After following the procedure summarized above to remove point sources, we remove the tSZ signal due to galaxy clusters, which we describe in Section~\ref{sec:method-clusters}. While we do not utilize the 219~GHz map to subtract point sources, this map is used when removing the tSZ signal. Therefore, we also remove point sources from the 219~GHz map following the same procedure that is applied to the 148~GHz map in order to use it for cluster removal. 

Below, we describe the steps summarized above in more detail.

\subsection{Filter Map}
\label{sec:filter}

CIB and radio galaxies are localized in CMB maps and have a small enough extent that they generally look like point sources that have the shape of the instrument beam.\footnote{Nearby sources at low redshifts can have a larger spatial extent.  We anticipate that this will be \num{0.002\%} and \num{0.6\%} of the total number of galaxies at \num{148 GHz} and \num{277 GHz}, respectively, given the CMB-HD beam sizes. This is because most of the galaxies are CIB galaxies, which are predominantly at higher redshifts ($z\approx 2$). In this work, we treat all CIB and radio galaxies as point sources.}  Thus, to isolate these sources, we filter the microwave sky map with a filter matched to the shape of the instrument beam.  We \textit{detect sources} (see Section~\ref{sec:detect-src}) by finding regions in the filtered map with an SNR above a given threshold, and \textit{measure the flux} of each source (see Section~\ref{sec:measure-flux}) by measuring its amplitude in the filtered map at the source position. 

We apply a matched filter to the 90, 148, 219, and 277~GHz simulated maps that we created and described in Section~\ref{sec:sims}, which contain the tSZ and kSZ effects (Section~\ref{sec:diffuse}), the CIB and radio galaxies (Section~\ref{sec:discrete}), the lensed CMB (Section~\ref{sec:lensedCMB}), and the instrumental noise and beam (Section~\ref{sec:instrument}).\footnote{We apply a mask to any bright sources with flux above 1000 mJy in any of the maps prior to running our foreground cleaning procedure; there is only one such source in our maps.} \\

{\it{Construct and apply the filter:}}  To construct a matched filter, we first model the unfiltered map as a sum of the sources we are trying to subtract (in this case, CIB and radio galaxies) and the other components of the map (in this case, CMB, kSZ, tSZ, and instrumental noise), so that 
\begin{equation} \label{eq:TsrcTother}
    T(\pmb{x}) = T_\mathrm{src}(\pmb{x}) + T_\mathrm{other}(\pmb{x}).
\end{equation}
All signal components are assumed to be convolved with the beam and pixel window function, as discussed in Section~\ref{sec:instrument}.  A matched filter $\Phi(\pmb{k})$ is calculated using the Fourier transform of the source profile convolved with the beam profile; for the case of ``point sources'', we just calculate the Fourier transform of the beam profile, $B(\pmb{k})$, at a given map frequency. We also need the power spectrum of the other components in the map, $P_\mathrm{other}(\pmb{k}) = |T_\mathrm{other}(\pmb{k})|^2$. Here both the beam profile and the power spectrum are two-dimensional in Fourier space. While we apodize the $T_\mathrm{other}(\pmb{x})$ map before taking its Fourier transform, we do not apply any other correction for the mode-coupling that is introduced when cutting out a patch of the full sky (see Appendix~\ref{sec:ps}), since no other correction is being applied to the map being filtered.\footnote{We tested applying a mode-decoupling matrix to the power spectrum of the $T_\mathrm{other}(\pmb{x})$ map and found that it made no difference.} The filter is given by
\begin{equation} \label{eq:beamfilter}
        \Phi(\pmb{k}) = \frac{B(\pmb{k})}{P_\mathrm{other}(\pmb{k})} \left[\int d^2k~\frac{|B(\pmb{k})|^2}{P_\mathrm{other}(\pmb{k})}\right]^{-1},
\end{equation}
as in, e.g.,~\cite{Haehnelt1995,VargasACT2023srcs}, where we assume a flat-sky approximation.  

We model the instrument beam with a Gaussian profile, given in real space by 
\begin{equation} \label{eq:beamprofile}
    B(\pmb{x}) = e^{-|\pmb{x}|^2 / 2 \sigma_\mathrm{B}^2},
\end{equation}
where $|\pmb{x}|$ is the angular distance from the origin, $\sigma_\mathrm{B} = \theta_\mathrm{FWHM} / \sqrt{8 \ln(2)}$, and $\theta_\mathrm{FWHM}$ is the beam full-width at half-maximum (FWHM). The CMB-HD beam size, $\theta_\mathrm{FWHM}$, is given in Table~\ref{tab:frequencies} for each frequency.  We make a real-space map of the beam profile $B(\pmb{x})$ by placing a single non-zero pixel at the center of a blank map with the same sky coordinates as the map to be filtered, convolving it with the beam, and then normalizing it so that $B(0) = 1$ at the map center.

We simulate the ``noise map'' $T_\mathrm{other}(\pmb{x})$ to obtain the noise power spectrum $P_\mathrm{other}(\pmb{k})$ in Equation~\ref{eq:beamfilter}.  This simulation includes the lensed CMB, kSZ, tSZ, and instrument noise, generated as described in Section~\ref{sec:sims} on a different patch of the sky (same size, different RA center) so that we have a realization that differs from the input simulation.\footnote{Note that this deviates from the standard practice of  using the input $T(\pmb{x})$ map itself, which includes both the signal to be removed and all noise sources, as a proxy for the noise map, e.g.~\cite{VargasACT2023srcs}.} We smooth the two-dimensional noise power spectrum by convolving it with a Gaussian function with a standard deviation equal to \num{3 pixels} in Fourier space, similar to what is done in~\cite{VargasACT2023srcs}. 

We apply our matched filter to the Fourier transform of the apodized input map, i.e.~$T_\mathrm{filt}(\pmb{k}) = \Phi(\pmb{k}) T(\pmb{k})$.  Then we deconvolve the pixel window (in order to get accurate flux measurements) and take the inverse Fourier transform of the result to obtain the filtered map $T_\mathrm{filt}(\pmb{x})$. \\

{\it{Calculate SNR of sources in filtered map:}} Once we have the filtered map, we want to estimate the SNR of each source within it.  Thus, we need an estimate of the noise in the filtered map.  We explored using the filtered $T_{\mathrm{other}}(\pmb{x})$ map to obtain this noise estimate. However, we found that this underestimated the noise and therefore overestimated the SNR, leading to too many false detections.\footnote{This is in part due to the absence of dim sources in the $T_{\mathrm{other}}(\pmb{x})$ map as well as features in the filtered map around bright sources.} So, instead, we use the filtered $T(\pmb{x})$ map itself, $T_\mathrm{filt}(\pmb{x})$, when creating an estimate of the noise in the filtered map. 

We do this by dividing the filtered map into a grid of \num{10 arcminute cells}, and calculating the mean and standard deviation of the map within each cell.  (We found that 10 arcminute cells are big enough to get a representative estimate of the background noise, but small enough to capture the unique noise variations around dense/bright source regions in the filtered map.) We then identify any 0.04' pixels that are more than \num{three} standard deviations away from the mean of the larger grid cell it is within, and recalculate the mean and standard deviation excluding those pixels. The value of each 0.04' pixel in a given cell is then set to the standard deviation of that cell obtained after repeating this procedure \num{10} times.  We call this the ``RMS map''.

We divide the original filtered map, $T_\mathrm{filt}(\pmb{x})$, by this RMS map to make a signal-to-noise ratio map (``SNR map'').  This approach is similar to that used by~\cite{HiltonACTclusters2020nemo} and implemented in the Nemo package\footnote{\url{nemo-sz.readthedocs.io}}, but we make the following two additions. (1) We compute the RMS map using the inner, un-apodized region (ignoring the apodized region), and then elongate the cells along the border to fill the entire  map (so we preserve continuity and minimize false detections over the apodized region); (2) We then smooth the RMS map with a Gaussian that has a standard deviation equal to \num{10\%} of the grid cell size of the RMS map, in order to remove any sharp transitions from lower-noise to higher-noise regions (which would then also appear in the SNR map).

\subsection{Detect Sources}
\label{sec:detect-src}

{\it{Iteratively detect sources:}}  We iteratively detect, measure, and remove sources down to an SNR threshold of \num{four}.  We choose this minimum threshold since it is the lowest threshold that does not alter the underlying CMB maps (see Appendix~\ref{sec:CMBtest} for details).  We do this by picking a set of SNR thresholds; we chose 250, 100, 75, 50, 40, 30, 25, 20, 15, 12.5, 10, 7.5, 5, and 4.  We start with the highest SNR threshold and work down the list as described below. We detect all sources with an SNR above each threshold in the following way:

\begin{itemize}
    \item We filter the map. This could be the original input map or a source-subtracted version of it from a previous iteration. Then we create an SNR map as described in Section~\ref{sec:filter}.
    
    \item In the SNR map, we identify regions of contiguous pixels that exceed or are equal to a given SNR threshold.  
    
    \item In each contiguous region with SNR above the threshold, we identify a {\it{single source}} in that region using the pixel with the highest SNR as its position. (We will catch other neighboring sources in the next iteration.)
    
    \item We measure the flux of each identified source (see Section~\ref{sec:measure-flux}), make a map of detected sources (as described below), and then subtract this map from the starting map for the given iteration.
    
    \item Using this source-subtracted map, we repeat the above steps.  We continue iterating through these steps until we find fewer than \num{10} sources for a given SNR threshold (except SNR~=~\num{4}); when we have fewer than \num{10} sources, we remove them and then move down to the next SNR threshold. For SNR~=~\num{4}, we stop when zero sources are found.
\end{itemize}

When we detect a source in the 90~or 148~GHz map that is located within the 90~or 148~GHz beam size of a source already found at a different frequency (277~or 90~GHz), we modify the procedure described above to use the source position measured at the other frequency. \\

{\it{Subtract sources:}} To subtract the detected sources from a map, we create a catalog of their measured positions and fluxes (see Section~\ref{sec:measure-flux} for flux measurement), place the sources on a map following the procedure described in Section~\ref{sec:discrete}, and then convolve the map with the beam. This map of detected sources is then subtracted from the original map to produce a source-subtracted map.  We subtract dim sources in the same way, using the source positions measured at another frequency and extrapolating the flux from that frequency using the measured spectral index (see Section~\ref{sec:measure-flux} for spectral index measurement).

\subsection{Measure Fluxes}
\label{sec:measure-flux}

{\it{Measure source fluxes:}} As described in Section~\ref{sec:detect-src}, we detect point sources by looking for regions of contiguous pixels above a given SNR threshold in the SNR map; we identify the position of each source as the pixel with the highest SNR.  The flux $S$ of the source is measured from the amplitude of the pixel in the filtered map, $T_\mathrm{peak}$ (in units of $\mu$K), using
\begin{equation} 
\label{eq:measuredflux}
    S = T_\mathrm{peak} \Omega_\mathrm{B} \left.\frac{\partial B_\nu(T)}{\partial T}\right|_{T_\mathrm{CMB}},
\end{equation}
where the factor $\left.\left(\partial B_\nu / \partial T\right)\right|_{T_\mathrm{CMB}}$ converts between intensity and CMB temperature units as described in Section~\ref{sec:sims}, and the beam solid angle $\Omega_\mathrm{B}$ corrects for the effect of beam smoothing~\cite{VargasACT2023srcs}. For a Gaussian beam as in Eq.~\ref{eq:beamprofile}, $\Omega_\mathrm{B} = 2\pi\sigma_\mathrm{B}^2$. However, since the map is pixelized and then smoothed with a Gaussian beam, we need to calculate $\Omega_\mathrm{B}$ more precisely, or we will under- or over-estimate the flux by up to about \num{1\%} over a 100 square degree map (see Appendix~\ref{sec:beam} for details on our exact calculation of $\Omega_\mathrm{B}$). \\

{\it{Measure spectral index and extrapolate source fluxes:}} For CIB galaxies that are dim at 90 and 148 GHz and thus not detected with SNR greater than \num{four} in those maps, we use the 277 GHz map, where they are brighter, to detect them. We then extrapolate their flux down to 90 and 148 GHz using a spectral index that we measure as described below.  Similarly, we use the 90 GHz map to find radio sources that are dim at 148 GHz, and we use a measured spectral index to estimate their 148 GHz flux. 

We assume the flux $S_2$ of a point source at frequency $\nu_2$ is related to the flux $S_1$ at frequency $\nu_1$ via
\begin{equation} \label{eq:fluxscaling}
    S_2 = S_1 \left(\frac{\nu_2}{\nu_1}\right)^\alpha,
\end{equation}
where $\alpha$ is the spectral index. Here we take $\nu_2$ to be the frequency to which we are extrapolating.  To measure the $\nu_1$-to-$\nu_2$ spectral index, we use sources measured inside the un-apodized $10^\circ \times 10^\circ$ region with SNR greater than \num{10} at $\nu_2$ that also have SNR $\geq 10$ counterparts in either the 277 or 90 GHz catalogs. If a given source at $\nu_2 \in$ \{90~GHz,~148~GHz\} has a higher flux counterpart at 277 GHz, then we assume it is a CIB galaxy; if a source at $\nu_2$ = 148~GHz has a higher flux counterpart at 90 GHz, then we assume it is a radio galaxy. We calculate a mean spectral index and standard deviation $\sigma$ using these sources and use sigma clipping to iteratively remove sources that deviate by \num{$\pm 2.5\sigma$}, recalculating the mean and standard deviation each time.  We use the measured spectral index and Eq.~\ref{eq:fluxscaling} to extrapolate the fluxes of dim sources from $\nu_1$ to $\nu_2$, and then subtract these sources from the $\nu_2$ map. \\

{\it{Remeasure missubtracted sources:}}  After performing the source subtraction procedure above, there will be some sources that are missubtracted, either because a neighboring source interfered with their flux measurement or the flux extrapolated from a different frequency deviated from the true flux.  Some of these missubtractions are large enough that their residual signals have $|$SNR$|$ greater than \num{four} in the source-subtracted map.  Thus, we identify these locations in the source-subtracted map and the corresponding missubtracted sources. We undo the source subtraction just for these sources and repeat the source subtraction procedure at these locations to remeasure the remaining sources.  Since these sources are now isolated in the map, with all neighbors removed, we have an opportunity to improve their position and flux measurement.

\section{Foreground Cleaning: Method to Remove Galaxy Clusters}
\label{sec:method-clusters}

Once we have maps that have been cleaned of CIB and radio sources down to low flux levels, as described above, we use a matched filter to remove galaxy clusters that appear in the microwave maps via the thermal Sunyaev-Zel'dovich (tSZ) effect~\cite{Melin2006}. The ultrahigh resolution of CMB-HD yields a natural separation of scales between galaxy clusters, which are on arminute scales, and the CIB and radio sources, which have the subarcminute resolution of the CMB-HD beam.  This allows us to perform the point source removal as a separate step from the cluster removal; in particular, we found that the inclusion of tSZ clusters did not impact the efficacy of point source removal since the clusters are on a large enough scale that the they are filtered out of the maps when applying the point-source matched filter. This differs from previous and current CMB surveys, which typically have 1 to 1.5 arcminute resolution, resulting in CIB and radio galaxies subtending the same angular scale as galaxy clusters.

The tSZ effect also has a unique spectral dependence~\cite{1972CoASP...4..173S}, which we exploit when removing the signal. Specifically, we use a {\it{multi-frequency}} matched filter, which searches for localized sources of a particular profile shape (which we vary), as well as for sources that have the unique tSZ frequency dependence.  To implement the multi-frequency matched filter, we also use point-source-subtracted CMB-HD maps at 219~GHz, in addition to 90, 148, and 277~GHz. We find that adding 219 GHz significantly improves tSZ cluster removal, since the tSZ signal is close to zero at 219~GHz, providing an important null map. We also find that 30~GHz does not help significantly with cluster subtraction, likely due to its larger beam and lower tSZ signal compared to 90~GHz; thus, we do not consider 30~GHz further in our method.\\

{\it{Differences with respect to point source removal:}} We remove galaxy clusters using a similar method to that described above for point sources, with the main differences as follows.
\begin{itemize}
    \item While we still use a match filter to search for clusters, the profiles of the  clusters are not known.  Thus, we apply different matched filters with different cluster profiles to the microwave maps, and identify each cluster with the profile that has the highest SNR.  In this way,  we measure {\it{the shape}} as well as the amplitude of each cluster.
    \item We also use a multi-frequency matched filter, which is applied simultaneously to all frequencies (90, 148, 219, 277~GHz); the output of applying this filter is a single Compton-$y$ map, where Compton $y$ is the frequency-independent part of the tSZ signal. 
    \item We utilize a simpler iterative procedure than for point sources, with no extrapolation from other frequencies or re-measuring.
\end{itemize}

{\it{Multi-frequency Matched Filter:}} The equation for the multi-frequency matched filter in flat-sky Fourier space~\cite{Melin2006} is given by:
\begin{equation} \label{eq:clusterfilter}
    \Psi_{\nu_i}(\pmb{k};\pmb{\theta}) = A(\pmb{\theta}) \sum_{j=1}^{N_\nu} \bigl[\mathbf{P}^{-1}(\pmb{k})\bigr]_{\nu_i \nu_j} ~ F_{\nu_j}(\pmb{k},\pmb{\theta}) ,
\end{equation}
where $A$ is the normalization, given by
\begin{equation} \label{eq:clusterfilternorm}
    A(\pmb{\theta}) =  \left[\int \frac{d^2k}{(2\pi)^2}  \sum_{i=1}^{N_\nu} \sum_{j=1}^{N_\nu} F_{\nu_i}^*(\pmb{k},\pmb{\theta}) \bigl[\mathbf{P}^{-1}(\pmb{k})\bigr]_{\nu_i \nu_j} ~ F_{\nu_j}(\pmb{k},\pmb{\theta}) \right]^{-1},
\end{equation}
with the $*$ indicating complex conjugation. Here, $N_\nu = 4$ for the number of frequencies, and $\Psi_{\nu}$ is a collection of $N_\nu$ filters, each of which is applied to the microwave map at the corresponding frequency. 

$\pmb{F}_\nu(\pmb{k},\pmb{\theta})$ is also a vector with $N_\nu$ components, where $F_\nu(\pmb{k},\pmb{\theta}) =  f_\mathrm{tSZ}(\nu) B_\nu(\pmb{k}) S(\pmb{k},\pmb{\theta})$.  $f_\mathrm{tSZ}(\nu)$ gives the frequency dependence of the tSZ effect, which is $f_\mathrm{tSZ}(\nu) = x \mathrm{coth}(x/2) - 4$, where $x = h\nu / k_\mathrm{B} T_\mathrm{CMB}$, $h$ is the Planck constant, $k_\mathrm{B}$ is the Boltzmann constant, and $T_\mathrm{CMB}$ is the mean CMB temperature today. $S(\pmb{k},\pmb{\theta})$ is the projected, radial cluster profile, described by some parameters $\pmb{\theta}$ (e.g., for a Gaussian profile, this would be a single parameter $\sigma$).  We normalize the real-space profile $S(\pmb{x},\pmb{\theta})$ to have a maximum amplitude of 1 at the origin. In practice, we calculate the normalization $A$ by applying the filter to a cluster with a known amplitude and the same profile $S(\pmb{k},\pmb{\theta})$ as in $\pmb{F}(\pmb{k},\pmb{\theta})$. 

$\mathbf{P}(\pmb{k})$ is a $N_\nu \times N_\nu$ matrix at \textit{each} $\pmb{k}$, with elements $P_{\nu_i \nu_j}(\pmb{k})$ given by $T_{\mathrm{other},\nu_i}(\pmb{k}) T_{\mathrm{other},\nu_j}^*(\pmb{k})$.  Here $T_{\mathrm{other},\nu}$ contains all components in the map being filtered that are not the tSZ, including residual point sources. We make the $T_{\mathrm{other},\nu}$ maps at each frequency in a way similar to what was done for point sources, in that we use a different patch of sky with different CMB, foreground, and white noise realizations; since we want residual point sources in the $T_{\mathrm{other},\nu}$ map, we first include the tSZ, subtract the sources as described in Section~\ref{sec:method-points}, and then subtract the tSZ. 

We use Gaussian profiles for the cluster profiles, which we found to work just as well as more complicated profiles with extra parameters. We employ \num{11} Gaussian profiles in total with \num{$\sigma$ = 0.25', 0.3', 0.35', 0.4', 0.45',} \num{0.5', 0.55', 0.6', 0.65', 0.7'}, and \num{0.75'}.\footnote{We explored using different sets of Gaussian profiles, but found that including smaller ($<$ 0.25') or much larger profiles ($>$ 1.5') increased the number of false detections; in addition, using much larger profiles resulted in over-subtracting the measured clusters.} \\

{\it{Applying the filter and getting an SNR map:}} We apply the multi-frequency matched filter in Fourier space on the point-source-subtracted maps, i.e.~$\hat{y}(\pmb{k}, \pmb{\theta}) = \sum_{j=1}^{N_\nu} \Psi_j(\pmb{k},\pmb{\theta}) T_j(\pmb{k})$, where $T_j(\pmb{k})$ is the input map at frequency $\nu_j$. We then deconvolve the pixel window from the filtered map and transform it back to real space, which results in a single map $\hat{y}$ in $y$ units.  The calculation of the SNR map is identical to that for the point sources, except we use \num{40'} grid cells instead of \num{10'} when calculating the ``RMS map'' (since clusters are larger than point sources).\\

{\it{A single iteration of cluster detection:}} For a single iteration of cluster detection, we apply the \num{11} different multi-frequency matched filters, corresponding to the \num{11} different cluster profiles, to the microwave maps.  This results in a set of \num{11} filtered maps (in $y$ units) and their corresponding SNR maps.  For each profile, we create a catalog of cluster ``candidates'' by identifying regions of contiguous pixels in the SNR map that exceed a given SNR threshold and measuring a single cluster candidate in each of these regions.  We set the location of the cluster candidate to be the location of the pixel with the maximum SNR, and the amplitude of the cluster candidate to be the value of that pixel in the filtered map. 

We identify the highest SNR detection across \textit{all} profiles, and use this detection as the measurement for this cluster, removing all other detections (e.g., from other profiles) within \num{1'} of this cluster from the candidate lists.  We repeat this using the next-highest SNR detection until no candidates remain.  From the measured positions, amplitudes, and profiles of the detected clusters, we make a map of the measured clusters (in $y$ units).  For each frequency, we convert the Compton-y maps to $\mu$K, and then convolve these maps with the pixel window and appropriate beam.  We then subtract the map of detected clusters for each frequency from the corresponding input map.\\

{\it{Iteratively detecting, measuring, and subtracting clusters:}} We follow the procedure above using SNR thresholds from the following list: \num{50, 25, 15, 12.5, 10, 7.5, 5, 4}. We start with the highest SNR threshold, and subtract the clusters found after a single iteration (described above). Then we use that cluster-subtracted map as input for a subsequent iteration using the same SNR threshold.  We repeat this procedure until no further clusters are found above this SNR threshold.  Then we switch to the next lower SNR threshold in the list.  This iterative detection method allows us to find dimmer clusters more easily after the brighter ones are removed.\\

{\it{Masking out leftover missubtracted point sources and clusters:}}  After the full CIB/radio point source and galaxy cluster subtraction procedures, there may still be significantly missubtracted locations in the maps.  To find missubtracted clusters, we apply the set of \num{11} filters once more, but this time identify \textit{all} regions in the \num{11} SNR maps with \num{$|$SNR$| \geq 5$}. We make a binary mask where pixels with \num{$|$SNR$| \geq 5$} in any of the SNR maps are set to zero, and other pixels set to one.  To find missubtracted point sources at 90 and 148 GHz, we calculate a new point-source filter, where now $T_\mathrm{other}$ includes only the \textit{residual} tSZ plus CMB, kSZ, and white noise. We apply this filter to the 90 and 148 GHz maps and make a point-source mask where pixels with \num{$|$SNR$| \geq 5$} in the SNR map are set to zero, and other pixels set to one.

We make a final mask for each frequency by multiplying the cluster and point source source masks. We then apodize the holes using a \num{1'} apodization width.\footnote{We additionally mask clusters that are simultaneously large (angular size $> 5'$), nearby ($z~<~0.2$), and massive ($M_{500\mathrm{c}} > 2 \times 10^{14} \, M_\odot$), which are not detected by our masking procedure due to their size; there are four such clusters in our 100 square degree maps.  We apodize the holes in this mask using a 5' apodization width.}  Before applying this final mask to our source/cluster subtracted maps, we apodize them and deconvolve the pixel window.  When taking power spectra of the masked maps, we recompute the mode-decoupling matrix to account for the mask.  We find that this final mask results in only a \num{1.5\%} loss of sky area.

\section{Results}\label{sec:results}

We run the foreground cleaning procedure discussed above on \num{4 square degree ($2^\circ\times2^\circ$)} regions of the ultrahigh-resolution simulations described above.  We then stitch together \num{25} of these regions to make a \num{100 square degree ($10^\circ\times10^\circ$)} patch of sky.  The advantages of performing the foreground cleaning procedure on smaller regions of sky are that:
\begin{itemize}
    \item The source flux extrapolation from one frequency to another is potentially more accurate when using CIB and radio spectral indices measured from the same local region. 
    \item The profiles for the beam and clusters in the matched filter are more accurate when computed in the local region in which they are applied. (See Figure~\ref{fig:beam} in Appendix~\ref{sec:beam} for the variation of the beam solid angle across a 100 square degree patch of sky.) 
    \item Running on smaller patches of sky is faster and can be parallelized.  For reference, foreground removal with the procedure described above takes \num{5} hours on one 4 square degree patch (see Table~\ref{tab:FGtimes} in Appendix~\ref{sec:computationalNeeds}).
\end{itemize}
In the following, we present the results of the foreground removal method and the subsequent cosmological parameter constraints.

\begin{figure*}[t]
    \centering
    \includegraphics[width=\textwidth]{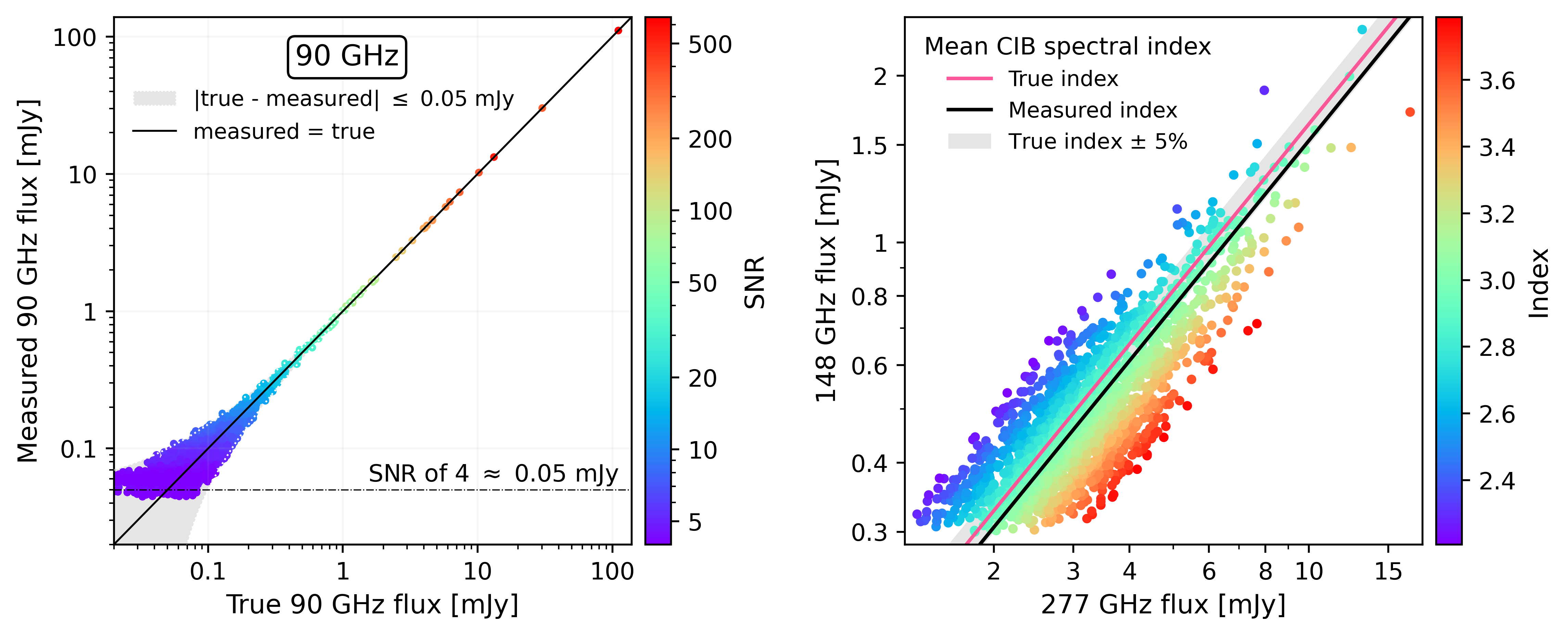}
    \caption{\textit{Left panel}: Comparison between true and measured fluxes of CIB and radio sources in the 90~GHz map within a four square degree region of sky. Each point corresponds to a source that is measured with SNR $\geq$ 4 at 90~GHz and matched to a true source from the simulation catalog; the color of the points corresponds to their measured SNR. The diagonal black line shows where measured flux equals true flux.  The gray shaded region shows where the difference between true and measured flux is less than our SNR threshold of four; this threshold corresponds to about 0.05~mJy (indicated by the dashed black horizontal line). We see good agreement between measured and true flux. \textit{Right panel}: Measured CIB spectral index between 277 and 148~GHz for point sources measured with SNR $\geq$ 10 at both 277~and 148~GHz within four square degrees; the color indicates the measured CIB spectral index of each source.  The average measured 277-to-148~GHz CIB spectral index of the detected sources is indicated by the black line, and the average true spectral index is shown as the pink line; these agree to within 5\%, indicated by the gray shaded region.}
    \label{fig:TrueVsMeasuredFluxIndex}
\end{figure*}

\subsection{Source and Cluster Recovery}
\label{sec:SourceClusterRecovery}

{\it{Measured source fluxes:}} We show in the left panel of Figure~\ref{fig:TrueVsMeasuredFluxIndex} the match between the true and measured fluxes of  sources measured at a single frequency.  Here we show results for sources measured at 90~GHz and the match to their counterparts in the simulation catalog.  We identify a source match using both the position and flux of the sources.  We use the standard deviation of the gaussian beam, $\sigma_\mathrm{B}$, at each frequency as the match radius.  Within this match radius, we match sources by flux.  

To match by source flux, we first sort the measured sources by SNR in order to match the brightest sources first.  For each measured source, we exclude all true sources within the match radius that have $|$true flux - measured flux$|$ / (measured flux error) $>$ 3. Here, the measured flux error is given by measured flux / SNR for bright sources; for dim sources, it is given by the average flux error for sources with SNR between 4 and 5 (since dim sources have no flux measurement at that frequency).  From the true sources within the match radius that are not excluded, we match to the true source with the closest flux, i.e.~smallest value of $|$true flux - measured flux$|$ / (measured flux error).  Once we have a match, we remove the true source from its catalog so that it cannot be matched more than once. If there are no true sources within the match radius after removing the excluded sources, then we label that a false detection.  

The left panel of Figure~\ref{fig:TrueVsMeasuredFluxIndex} compares the measured versus true flux of all sources measured in the 90~GHz map with SNR $\geq$ \num{4}, corresponding to measured fluxes down to $\sim 0.05$~mJy (indicated by the black horizontal dashed line).  The color of each point shows the SNR of the measured flux. The bright sources fall on the black diagonal line with little scatter, indicating a good match between true and measured flux.  At lower fluxes, there is more scatter, and the measured flux in this region tends to be higher than the true flux due to Eddington bias/flux boosting~\cite{Marriage2011,VargasACT2023srcs}.  The gray shaded region indicates where the difference between true and measured flux is less than our SNR = \num{4} threshold.

{\it{Measured spectral indices:}} Part of the procedure to remove CIB and radio galaxies involves measuring the CIB and radio spectral indices to extrapolate their flux from another frequency.  In Table~\ref{tab:spectralindex}, we show the match between measured and true spectral indices for CIB and radio galaxies within a \num{$2^\circ \times 2^\circ$} region.  We find the agreement to be better than \num{$7\%$} across all the frequency combinations we use.  

In the right panel of Figure~\ref{fig:TrueVsMeasuredFluxIndex}, we show our most discrepant spectral index measurement for CIB galaxies between 277~and 148~GHz. We plot the fluxes of sources at 277~and 148~GHz that were used to calculate the 277-to-148 CIB spectral index following the procedure described in Section~\ref{sec:measure-flux}. The color of each point indicates the spectral index of each individual source, calculated from the pair of flux measurements at both frequencies (see the colorbar for numerical values). The black line indicates the measured index, and the pink line indicates the true index; these values agree to within less than \num{5\%}, indicated by the shaded gray region.

\begin{table}[t]
    \begin{center}
    \begin{tabular}{c@{\hskip 1.5em}  c@{\hskip 1.5em}  c@{\hskip 1.5em} c}
      \toprule
      \toprule
      Spectral Index & Measured & True & \% Difference  
      \\
      \midrule
      277-to-90 CIB & 3.02 $\pm$ 0.21 & 3.02 & 0.04 \\
      277-to-148 CIB & 3.00 $\pm$ 0.32 & 2.88 & 4.1 \\
      277-to-219 CIB & 2.79 $\pm$ 0.79 & 2.71 & 3.0 \\
      \midrule
      90-to-148 Radio & -0.75 $\pm$ 0.29 & -0.80 & -6.6  \\
      90-to-219 Radio & -0.82 $\pm$ 0.09 & -0.80 & 2.9 \\
      \bottomrule
    \end{tabular}
    \caption{The average measured and true CIB and radio spectral indices within a four square degree region. The first column lists the two frequencies used to calculate the index, and whether the index corresponds to CIB or radio sources; the first frequency given is where the source is brighter. The second column lists the mean and standard deviation of the measured spectral indices, calculated from sources measured at both frequencies with SNR $\geq$ 10. The third column lists the true average spectral index, and the last column lists the fractional difference between them. We find that measured and true spectral indices agree to better than 7\%.}
    \label{tab:spectralindex}
    \end{center}
\end{table}

We also find that our results are rather insensitive to the exact spectral index.  For example, using the true mean indices from the catalogs for both radio and CIB galaxies, instead of the measured mean indices, results in a 0.15\% difference in the total power spectrum of residual foregrounds plus instrument noise for both 90~and 148~GHz. \\

\begin{figure*}[t]
    \centering
    \includegraphics[width=\textwidth]{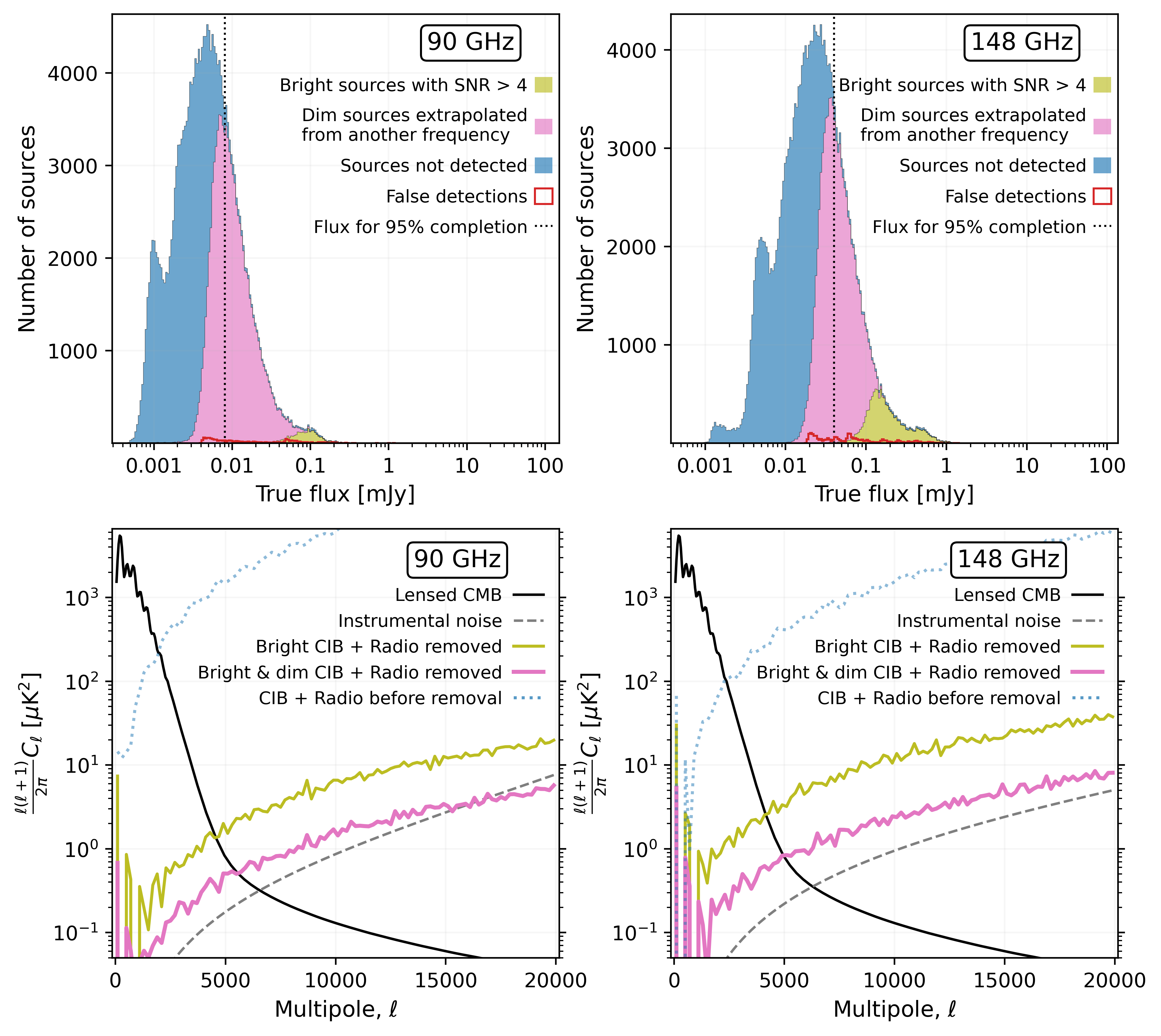}
    \caption{{\it{Upper panels}} show histograms of CIB and radio sources in four square degree maps at 90~GHz (left) and 148~GHz (right).  Bright sources detected with SNR $\geq$ 4 in the 90~or 148~GHz map are shown in yellow.  Dim sources that were measured in a map at a different frequency with SNR $\geq$ 4 and subtracted from the 90~or 148~GHz map after extrapolating their fluxes are shown in pink. Sources that were not detected and subtracted are shown in blue.  The true flux of detected and subtracted sources (pink and yellow) is determined by matching to the catalog of all sources in the map, as described in Section~\ref{sec:results}.  In red, we show the false detections (plotting their measured flux on the $x$-axis), i.e., sources that have been detected and subtracted, but do not correspond to a true source in the map.  The histograms show significantly more sources are removed when taking advantage of measurements at a different frequency.  The resulting catalogs are 95\% complete down to flux limits of 0.008 and 0.04~mJy for 90 and 148 GHz, respectively (black dotted lines). {\it{Lower panels}} show power spectra of CIB and radio sources in the 90~GHz (left) and 148~GHz (right) maps at three stages of our source-subtraction procedure: 1)~before any subtraction (blue dotted), 2)~after subtracting only the bright sources detected with SNR $\geq$ 4 in the 90~or 148~GHz maps (yellow), and 3)~after also subtracting dimmer sources (pink). For comparison we also show the instrumental noise (dashed gray) and lensed CMB (solid black) power spectra. These power spectra show the significant impact of removing both bright and dim sources in lowering the total power.}
    \label{fig:fluxhist}
\end{figure*}

{\it{Statistics of detected sources:}} In the upper panels of Figure~\ref{fig:fluxhist}, we show histograms of detected sources in the 90~GHz (left) and 148~GHz (right) maps. In yellow are bright sources detected with SNR $\geq$ \num{4} in the 90~or 148~GHz maps (for the left and right panels, respectively). In pink are shown sources measured with SNR $\geq$ \num{4} at a different frequency, whose fluxes are then extrapolated to 90~or 148~GHz.  We show in blue the sources not detected and subtracted from the maps.  False detections are indicated in red using the measured, as opposed to the true, flux.

We see that we remove significantly more sources by using other frequencies where the sources are brighter and extrapolating to 90 and 148 GHz; in particular, roughly \num{95\%} and \num{85\%} of the sources removed from the 90~and 148~GHz maps, respectively, were detected at a different frequency. Overall, we remove about \num{45\%} of all sources in the 90~and 148~GHz maps and produce catalogs that are 95\% complete down to flux limits of \num{0.008} and \num{0.04}~mJy for 90 and 148 GHz, respectively (black dotted lines).  The remaining sources are mostly CIB sources that are too dim to be detected at 277~GHz. 

In the lower panels of Figure~\ref{fig:fluxhist}, we show power spectra of CIB and radio sources in the 90~GHz (left) and 148~GHz (right) maps.  In dotted blue are the power spectra before any point source subtraction.  In yellow are the power spectra after subtracting only the bright sources detected with SNR $\geq$ \num{4} in the 90~or 148~GHz maps.  In pink are the power spectra after also subtracting dimmer sources.  We see that removing both bright and dim sources significantly lowers the total point source power.\\

\begin{figure}
    \centering
    \includegraphics[width=\linewidth]{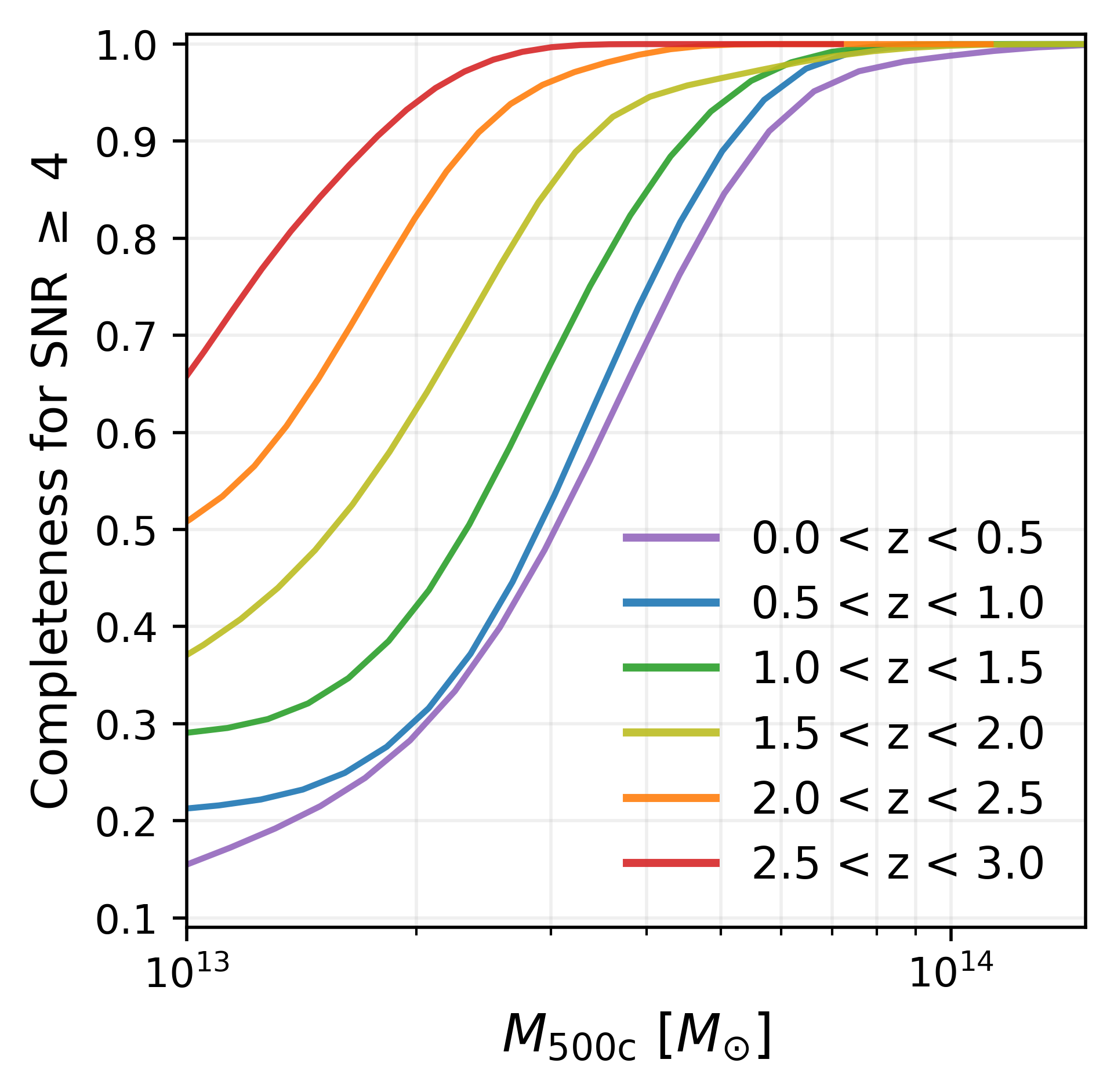}
    \caption{Completeness per mass and redshift bin of the detected clusters with SNR $\geq$ 4. We see that for a fixed mass, the completeness improves for clusters at higher redshift, as also noted in~\protect{\cite{SPT-3G:2025rxd}}.}
    \label{fig:clustercompleteness}
\end{figure}

\begin{figure}
    \centering
    \includegraphics[width=\linewidth]{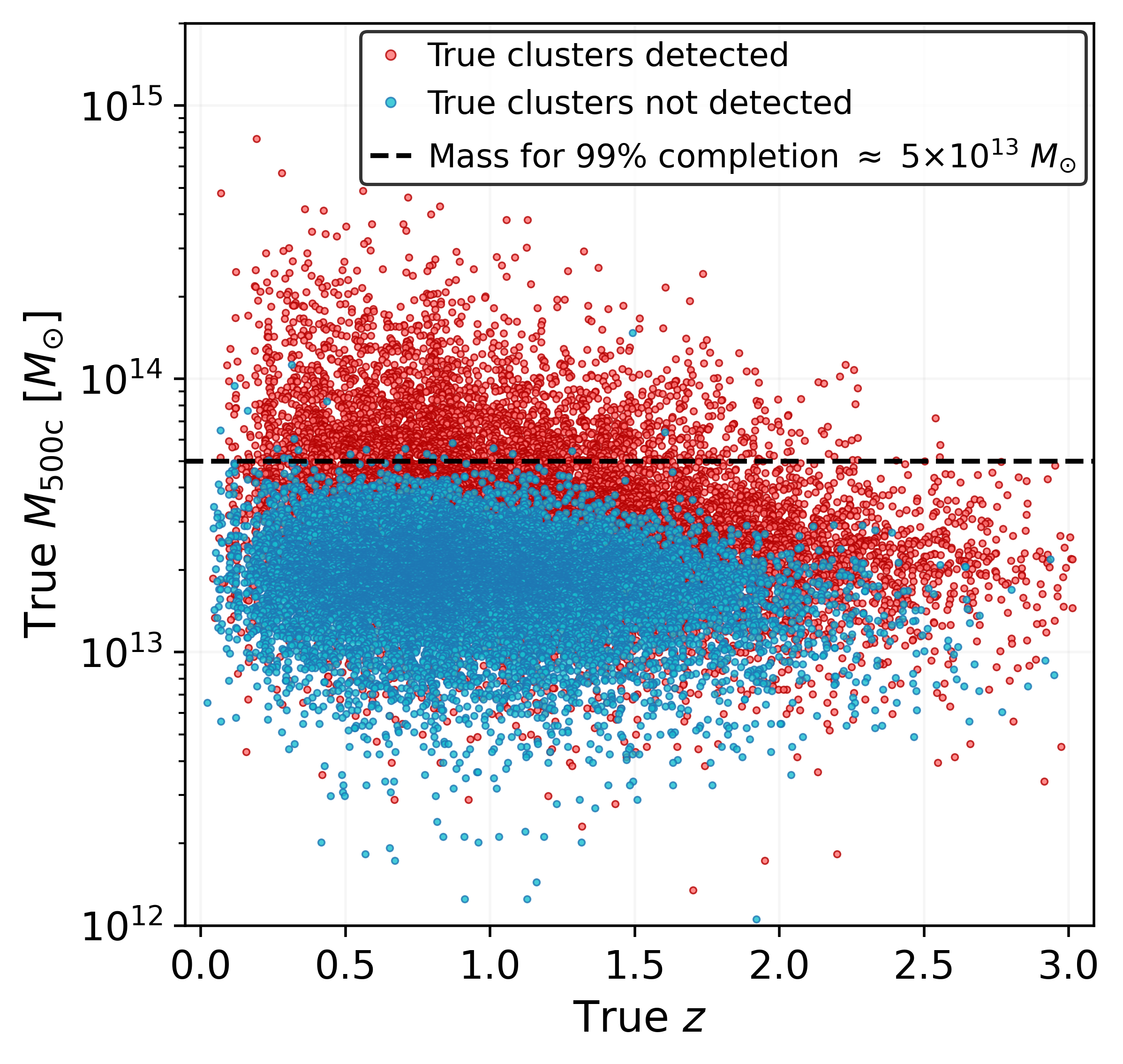}
    \caption{The true mass and redshift of all SZ clusters within a 100 square degree region of the maps. Clusters that have been detected with SNR $\geq$ 4 and subtracted from the maps are shown in red, while the remaining clusters that have not been detected are shown in blue. We find 99\% of the true clusters above a mass of about $5 \times 10^{13} M_\odot$ (black dashed line).}
    \label{fig:clustermass}
\end{figure}

{\it{Statistics of detected clusters:}} To match a detected cluster with a true cluster, we adopt a matching radius of \num{1 arcmin}.  We identify a true cluster as detected if we have at least one measured cluster within \num{1 arcmin} of the true position. In Figure~\ref{fig:clustercompleteness}, we show the completeness per mass and redshift bin, i.e.~the number of detected clusters as a fraction of all true clusters within a given mass and redshift range. We find that, for a given mass range, the completeness improves as the redshift increases, as was also seen in other work~\cite{SPT-3G:2025rxd}.  In Figure~\ref{fig:clustermass}, we show the true mass and redshift of detected clusters in red and undetected clusters in blue.  We find \num{99\%} of the true clusters above \num{$M_{\rm{500c}} \approx 5 \times 10^{13} M_\odot$} using an SNR threshold of \num{four}. Regarding the purity of the cluster samples, for an SNR threshold of \num{five}, our detected sample of clusters is more than \num{80\%} pure; this drops to \num{65\%} pure for an SNR threshold of \num{four}.  Overall, we detect about \num{30\%} and \num{50\%} of the true clusters in the maps with an SNR threshold of \num{five} and \num{four}, respectively. 

\begin{figure*}[t]
    \centering
    \includegraphics[width=\textwidth]{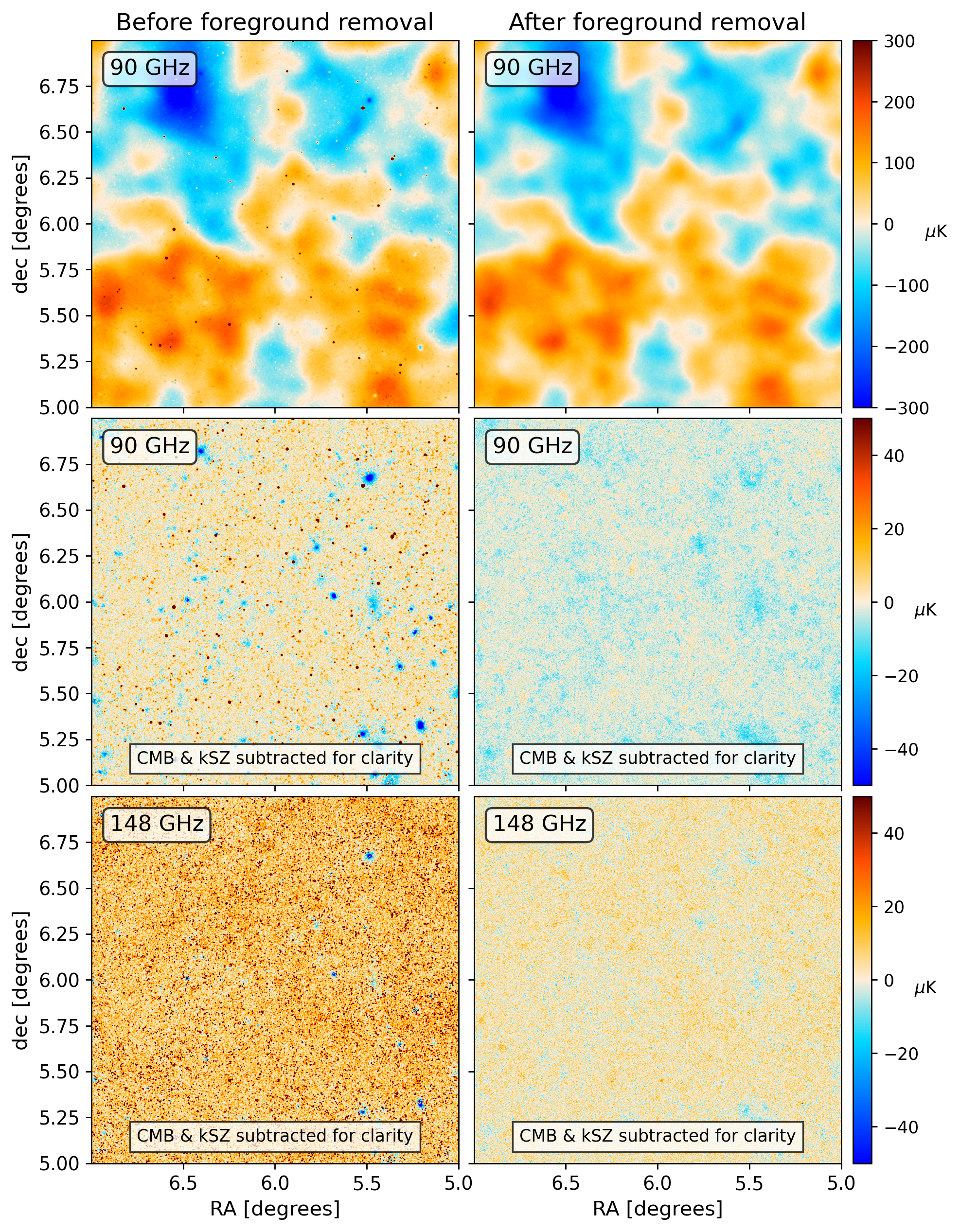}
    \caption{A four square degree region of sky at 90~GHz (first two rows) and 148~GHz (last row), before (left) and after (right) foreground cleaning. In the last two rows, the CMB and kSZ signals have been removed from each map for clarity.} 
    \label{fig:maps_before_after}
\end{figure*}

\begin{figure*}[t]
    \centering
    \includegraphics[width=\textwidth]{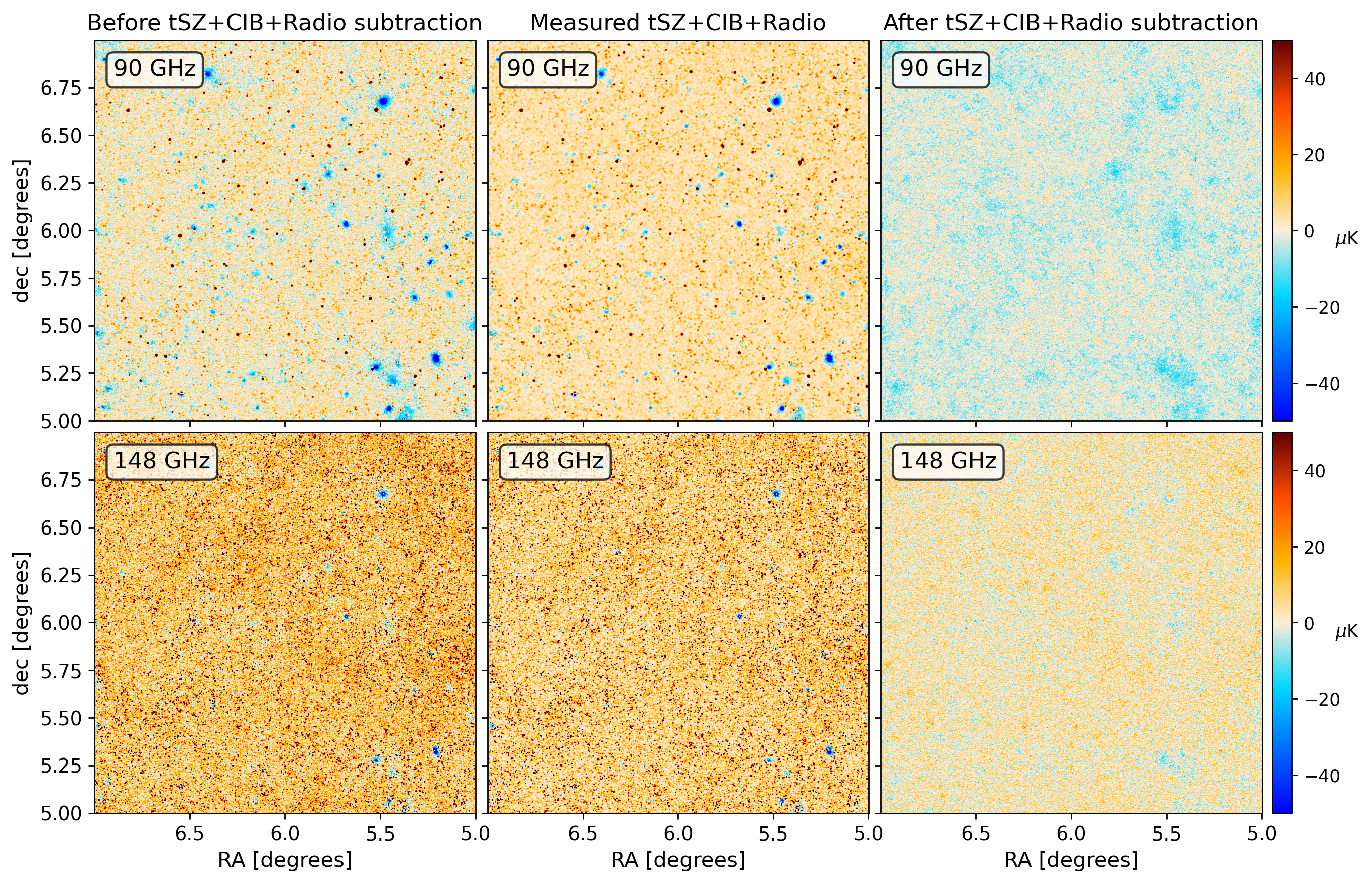}
    \caption{A four square degree sky region at 90~GHz (top) and 148~GHz (bottom), before (leftmost) and after (rightmost) cleaning the CIB, radio, and tSZ signals as described in Sections~\ref{sec:method-points} and~\ref{sec:method-clusters}. The frequency-independent CMB and kSZ signals have been removed from each map for clarity in order to see the source subtraction more easily.  The middle panel shows maps of the measured CIB, radio, and tSZ signals that are subtracted from the maps on the left to produce the maps on the right.} 
    \label{fig:maps_before_after_nocmb}
\end{figure*}

\subsection{Foreground-Cleaned Maps and Power Spectra} \label{sec:MapsAndSpectra}

{\it{Foreground-cleaned maps:}} Figures~\ref{fig:maps_before_after} and~\ref{fig:maps_before_after_nocmb} show microwave sky maps before and after foreground removal.  The top panels of Figure~\ref{fig:maps_before_after} show the 90~GHz map before (left) and after (right) foreground cleaning. The middle panels show the same set of maps, but with the CMB and kSZ removed in order to visualize the point source and cluster subtraction more clearly. (Note that we do not attempt to remove the kSZ in our foreground removal procedure, as we only focus on frequency-dependent and localized sources.) The bottom panels show the corresponding maps at 148~GHz (again with the CMB and kSZ removed to facilitate the observation of the source and cluster subtraction). 

Figure~\ref{fig:maps_before_after_nocmb} shows the same maps as in the last two rows of Figure~\ref{fig:maps_before_after}, in the leftmost and rightmost columns.  Additionally, in the middle column, we show the CIB, radio, and tSZ signals that we measure and subtract from the maps at 90~GHz (top) and 148~GHz (bottom); we see that our model for the sources and clusters in the middle column matches well with the true source/cluster distribution shown in the leftmost column.  From the rightmost panels, we see that the residual 90~GHz map is dominated by dim, diffuse tSZ that remains unsubtracted, which makes the overall map lower than the mean CMB temperature (i.e.~blue).  In contrast, the residual 148~GHz map is dominated by dim unsubtracted CIB sources, which make the overall map higher than the mean CMB temperature (i.e.~orange).  This is reasonable since the tSZ is brighter at 90~GHz than at 148~GHz, and the situation is reversed for the CIB. \\

{\it{Foreground-cleaned power spectra:}} Figures~\ref{fig:coaddedspectra} and~\ref{fig:spectra} show the power spectra of our final 100 square degree region after foreground removal. Figure~\ref{fig:spectra} shows, in dark red, the total residual noise power spectra of the 90~GHz (left) and 148~GHz (right) maps after foreground cleaning, which includes the beam-deconvolved instrument noise, the kSZ (late-time and reionization components), and the residual tSZ, CIB, and radio sources after foreground removal.  Figure~\ref{fig:coaddedspectra} shows the same for the coadded 90~and 148~GHz spectra, also in dark red. These noise curves show the noise per $\ell$-mode, so each curve should be divided by $\sqrt{2\ell+1}$ to get the error bar per $\ell$-mode.

In Figure~\ref{fig:spectra}, we separate the different noise components, showing the power spectra of the residual tSZ in green and the residual CIB plus radio point sources in blue. We also show spectra for the lensed CMB (black), instrumental noise (dashed gray), and kSZ (dashed orange).  We see that residual point sources dominate the foreground spectra at 148~GHz, whereas residual clusters dominate for all but the highest multipoles at 90~GHz.   

In light red, in both Figures~\ref{fig:coaddedspectra} and~\ref{fig:spectra}, we show the previous idealized estimate of the total residual noise power spectra from~\cite{han22, HDparams, subgalacticDM}.  From Figure~\ref{fig:coaddedspectra}, we see that the simulation-based coadded residual noise power spectra (this work) is about \num{50\%} higher than the previous estimate.  Separating by frequency, Figure~\ref{fig:spectra} shows that the simulation-based residual noise power spectra is higher than the previous estimate by \num{30\%} at 90~GHz and \num{80\%} at 148~GHz.

\subsection{Parameter Constraints} \label{sec:params}

We also forecast how well a CMB-HD-like survey would measure cosmological parameters, using the simulation-based noise curves from this work. We do this using a Fisher forecast, which was shown in~\cite{HDparams} to give the same results as a Markov Chain Monte Carlo method.  In Table~\ref{tab:fisher}, we give the fiducial parameter values and step sizes used for the Fisher forecast.  We forecast parameters constrained by CMB-HD delensed CMB $TT$, $TE$, $EE$, $BB$ and CMB lensing $\kappa\kappa$ power spectra, combined with mock DESI BAO (as done in in~\cite{HDparams,subgalacticDM}), assuming $f_\mathrm{sky} = 0.6$.  We use the same method for Fisher forecasts as in~\cite{HDparams,subgalacticDM}, except for the following changes:
\begin{itemize}
    \item We calculate the covariance matrix for the delensed CMB $TT$, $TE$, $EE$, $BB$ and CMB lensing $\kappa\kappa$ power spectra in the same way as in~\cite{subgalacticDM}, except we: 
    \begin{enumerate}
        \item Update the $TT$ noise curve used for the covariance matrix calculation with the simulation-based coadded 90 plus 148~GHz total residual noise curve (dark red curve shown in Figure~\ref{fig:coaddedspectra}). 
        \item Update the CMB lensing noise curve for the following two cases:
        \begin{itemize}
        \item pol-only $\kappa\kappa$: assume that only polarization estimators (i.e.~$EE$ and $EB$) are used to measure the CMB lensing power spectrum and a diagonal auto-covariance matrix for the lensing spectrum, with the assumption that RDN0 subtraction will effectively remove the off-diagonal terms, as discussed in~\cite{Peloton:2016kbw,Nguyen2017}.
        \item MV $\kappa\kappa$: add to the pol-only $\kappa\kappa$ the $TE$ and $TB$ estimators, restricting the temperature data to CMB multipoles below $\ell_\mathrm{max}^T = 5{,}000$ to avoid high-$\ell$ foreground contamination in these four-point estimators. We further add the $TT$ estimator: for lensing multipoles $L < 5000$ we again impose $\ell_\mathrm{max}^T = 5{,}000$, while for $L > 5000$ we adopt a previous simulation-based estimate of the $TT$ lensing noise~\cite{han22}, scaling the diagonal elements of this covariance matrix by the ratio of $TT$ lensing noise (when allowing $\ell_\mathrm{max}^T = 20{,}000$) resulting from a quadratic estimator applied to the $TT$ residual noise curves shown as the light and dark red curves in Figure~\ref{fig:coaddedspectra}.
        \end{itemize}
    \end{enumerate}
    
    \item We use the CAMB accuracy settings listed in Appendix~\ref{sec:CAMBaccuracy} instead of those given in~\cite{HDparams,subgalacticDM}. (We increased the accuracy of the unlensed CMB at high-$\ell$; in contrast to~\cite{HDparams,subgalacticDM}, here we first generate the unlensed CMB and then lens it with a convergence map to make the lensed CMB.)  

    \item We decrease our Gaussian prior on $\tau$ to $\sigma(\tau) = 0.005$, consistent with the results found from a combination of Planck, ACT, SPT, and DESI DR2 data~\cite{SPT3G2025}.
\end{itemize}

\begin{table}[t]
    \begin{center}
    \begin{tabular}{l@{\hskip 1.5em} c@{\hskip 1.5em} c}
      \toprule
      \toprule
      Parameter & Fiducial  & Step Size 
      \\
      \midrule
      $\Omega_\mathrm{b} h^2$\dotfill & $0.02237$ & 1\% 
      \\
      $\Omega_\mathrm{c} h^2$\dotfill & $0.1200$ & 1\% 
      \\
      $\ln(10^{10} A_\mathrm{s})$\dotfill & $3.044$ & 0.3\%\footnote{The step size of 0.3\% $\ln(10^{10} A_\mathrm{s})$ corresponds to an approximate step size of 1\% on $A_\mathrm{s}$.} 
      \\
      $n_\mathrm{s}$\dotfill & $0.9649$ & 1\% 
      \\
      $\tau$\dotfill & $0.0544$ & 5\% 
      \\
      $100 \theta_\mathrm{MC}$\dotfill & $1.04071$ & 1\% 
      \\
      $N_\mathrm{eff}$\dotfill & $3.044$ & 5\% 
      \\
      $\sum m_\nu$ [eV]\dotfill & $0.06$ & 10\% 
      \\
      $\log_{10}\left(T_\mathrm{AGN}/\mathrm{K}\right)$\dotfill & 7.8 & 0.05
      \\
      $A_\mathrm{kSZ}$\dotfill & 1 & 0.1
      \\
      $n_\mathrm{kSZ}$\dotfill & 0 & 0.01
      \\
      \bottomrule
    \end{tabular}
    \caption{The parameters considered in the Fisher forecasts, with their fiducial values (middle column) and the step sizes (last column) by which they are varied when calculating derivatives. We apply a Gaussian prior of $\sigma(\tau) = 0.005$ on $\tau$ based on the constraints from a combination of Planck, ACT, SPT, and DESI DR2 data~\protect{\cite{SPT3G2025}}, and a 0.06\% prior on $\log_{10}\left(T_\mathrm{AGN}/\mathrm{K}\right)$ as was done in~\protect{\cite{HDparams,subgalacticDM}}.} 
    \label{tab:fisher}
    \end{center}
\end{table}

\begin{table*}
    \centering
    \begin{tabular}{l@{\hskip 2em} l@{\hskip 1em} l@{\hskip 2em}  l@{\hskip 1em} l@{\hskip 2em}  l@{\hskip 1em} l }
        \toprule
        \toprule
        \multicolumn{1}{l}{} &  \multicolumn{6}{c}{ $1\sigma$ Errors using Simulation-based Forecast} \\
        \cmidrule(){2-7}
        & \multicolumn{2}{c}{$\Lambda$CDM+$N_\mathrm{eff}$+$\sum m_\nu$} &  \multicolumn{2}{c}{+$\log_{10}\left(T_\mathrm{AGN}/\mathrm{K}\right)$} &  \multicolumn{2}{c}{+$A_\mathrm{kSZ}$+$n_\mathrm{kSZ}$}  \\
        \cmidrule(lr){2-3} \cmidrule(lr){4-5} \cmidrule(lr){6-7}
        Parameter & MV $\kappa\kappa$ & pol.-only $\kappa\kappa$ & MV $\kappa\kappa$ & pol.-only $\kappa\kappa$  & MV $\kappa\kappa$ & pol.-only $\kappa\kappa$  \\
        \midrule
        $\Omega_\mathrm{b} h^2$\dotfill                           & 0.0000254  & 0.0000255  & 0.0000253  & 0.0000255  & 0.0000256  & 0.0000259  \\
        $\Omega_c h^2$\dotfill                                    & 0.000366   & 0.000368   & 0.000363   & 0.000365   & 0.000370   & 0.000378   \\
        $\ln \left(10^{10} A_\mathrm{s}\right)$\dotfill           & 0.00787    & 0.00794    & 0.00844    & 0.00846    & 0.00853    & 0.00857    \\
        $n_\mathrm{s}$\dotfill                                    & 0.00153    & 0.00155    & 0.00148    & 0.00151    & 0.00162    & 0.00179    \\
        $\tau$\dotfill                                            & 0.00421    & 0.00425    & 0.00447    & 0.00449    & 0.00452    & 0.00454    \\
        $100\theta_\mathrm{MC}$\dotfill                           & 0.0000595  & 0.0000598  & 0.0000603  & 0.0000605  & 0.0000605  & 0.0000608  \\
        $N_\mathrm{eff}$\dotfill                                  & 0.0142     & 0.0143     & 0.0140     & 0.0141     & 0.0154     & 0.0167     \\
        $\sum m_\nu$~[eV]\dotfill                                 & 0.0262     & 0.0266     & 0.0287     & 0.0288     & 0.0289     & 0.0290     \\
        $\log_{10}\left(T_\mathrm{AGN}/\mathrm{K}\right)$\dotfill & ---        & ---        & 0.00464    & 0.00464    & 0.00466    & 0.00466    \\
        $A_\mathrm{kSZ}$\dotfill                                  & ---        & ---        & ---        & ---        & 0.001172   & 0.001574   \\
        $n_\mathrm{kSZ}$\dotfill                                  & ---        & ---        & ---        & ---        & 0.000821   & 0.000984   \\
        \bottomrule
    \end{tabular}
    \caption{Forecasted $1\sigma$ parameter errors for the combination of CMB-HD delensed CMB $TT$, $TE$, $EE$, $BB$ and CMB lensing $\kappa\kappa$ power spectra plus DESI BAO data, using the simulation-based power spectra obtained after foreground cleaning from this work. These forecasts assume the CMB survey covers 60\% of the sky and the BAO survey covers 14,000 square degrees.  We present parameter forecasts for three models: a model that varies the six $\Lambda$CDM parameters plus $N_\mathrm{eff}$ and $\sum m_\nu$ (second column), a model that additionally varies $\log_{10}\left(T_\mathrm{AGN}/\mathrm{K}\right)$ to include baryonic feedback effects (third column), and a model that also varies the amplitude and slope of the kSZ power spectrum ($A_{\rm{kSZ}}$ and $n_{\rm{kSZ}}$) (last column).  We also include a Gaussian prior on $\tau$ of $\sigma{(\tau)} = 0.005$ for all models, and a 0.06\% prior on $\log_{10}\left(T_\mathrm{AGN}/\mathrm{K}\right)$ as was done in~\protect{\cite{HDparams,subgalacticDM}}. We show these forecasts for two cases of the CMB lensing spectra: one where the lensing reconstruction is done with polarization-only estimators (pol.-only $\kappa\kappa$) and one where temperature data is also included (MV $\kappa\kappa$); we show the comparison since the former is immune to lensing bias from extragalactic foregrounds, even though the latter is expected to have biases below statistical errors~\protect{\cite{han22}}.  We see little change in cosmological parameters when freeing $\log_{10}\left(T_\mathrm{AGN}/\mathrm{K}\right)$ or when switching between pol.-only and MV $\kappa\kappa$.  The exception to the latter is when additionally freeing $A_{\rm{kSZ}}$ and $n_{\rm{kSZ}}$, which we attribute to the lower high-$L$ lensing noise from the MV $\kappa\kappa$ breaking the lensing-kSZ degeneracy in the CMB $TT$ power spectrum.  }
    \label{tab:params}
\end{table*}

In Table~\ref{tab:params}, we show the simulation-based forecasts obtained using the simulation-based coadded residual noise power spectra from this work (dark red curve in Figure~\ref{fig:coaddedspectra}).  We show these constraints for three models: (1)~a $\Lambda$CDM+$N_\mathrm{eff}$+$\sum m_\nu$ model, (2)~a model that also includes baryonic feedback~\cite{Mead2020} using a single parameter $\log_{10}(T_\mathrm{AGN}/\mathrm{K})$ (we apply a 0.06\% prior on this parameter, as was done in~\cite{HDparams,subgalacticDM}), and (3)~a model that additionally includes parameters for the amplitude and slope of the kSZ power spectrum~\cite{subgalacticDM}, $A_\mathrm{kSZ}$ and $n_\mathrm{kSZ}$, respectively.\footnote{These amplitude and slope parameters have the same meaning as described in Section~\ref{sec:smallscalesims}.}  We find that marginalizing over the baryonic physics and kSZ parameters only slightly increases cosmological parameter errors.

\begin{table}[t]
    \centering
    \begin{tabular}{l@{\hskip 1em} l@{\hskip 1em}  l l@{\hskip 1em} c}
        \toprule
        \toprule
        \multicolumn{1}{l}{} &  \multicolumn{2}{c}{Forecasted  $1\sigma$ Errors} &  & \multicolumn{1}{c}{Ratio of $1\sigma$ Errors} \\
        \cmidrule(){2-3} \cmidrule(){5-5}
        Parameter & Previous & Sim-based & & Sim-based/Previous \\
        \midrule
        $\Omega_\mathrm{b} h^2$\dotfill                 & 0.0000249  & 0.0000256  &  & 1.03 \\
        $\Omega_c h^2$\dotfill                          & 0.000374   & 0.000370   &  & 0.99 \\
        $\ln \left(10^{10} A_\mathrm{s}\right)$\dotfill & 0.00850    & 0.00853    &  & 1.00 \\
        $n_\mathrm{s}$\dotfill                          & 0.00152    & 0.00162    &  & 1.07 \\
        $\tau$\dotfill                                  & 0.00451    & 0.00452    &  & 1.00 \\
        $100\theta_\mathrm{MC}$\dotfill                 & 0.0000600  & 0.0000605  &  & 1.01 \\
        $N_\mathrm{eff}$\dotfill                        & 0.0150     & 0.0154     &  & 1.03 \\
        $\sum m_\nu$~[eV]\dotfill                       & 0.0289     & 0.0289     &  & 1.00 \vspace{1mm} \\
        \hline
        $\log T_\mathrm{AGN}$\dotfill                   & 0.00465    & 0.00466    &  & 1.00 \\
        $A_\mathrm{kSZ}$\dotfill                        & 0.000808   & 0.001172   &  & 1.45 \\
        $n_\mathrm{kSZ}$\dotfill                        & 0.000564   & 0.000821   &  & 1.46 \\
        \bottomrule
    \end{tabular}
    \caption{Parameter forecasts for the combination of CMB-HD delensed CMB $TT$, $TE$, $EE$, $BB$ and CMB lensing MV $\kappa\kappa$ power spectra plus DESI BAO data. The ''previous'' column lists the $1\sigma$ error on each parameter calculated using the previous idealized estimate of the CMB-HD residual foreground power spectra at 90~and 148~GHz from~\protect{\cite{subgalacticDM}}, while the forecasts in the ''sim-based" column use the corresponding simulation-based power spectra obtained in this work; the previous and simulation-based spectra are shown in Figures~\ref{fig:coaddedspectra} and~\ref{fig:spectra} as the light and dark red curves, respectively. The last column lists the ratio of the simulation-based forecasts to the previous estimates. The model fit is an 11-parameter $\Lambda$CDM+$N_{\rm{eff}}$+$\sum m_{\nu}$+$T_{\rm{AGN}}+A_{\rm{kSZ}}+n_{\rm{kSZ}}$ model, which varies baryonic physics ($T_{\rm{AGN}}$) and kSZ parameters ($A_{\rm{kSZ}}$ and $n_{\rm{kSZ}}$) in addition to the cosmological parameters.  The eight cosmological parameters are separated from the three additional nuisance parameters by the horizontal line. We see that the two sets of cosmological parameter forecasts agree to within 7\% or better. The uncertainty on the kSZ parameters increases by up to 46\%, suggesting that they are absorbing the excess foreground uncertainty of the simulation-based residual foregrounds.}
    \label{tab:params_comparison}
\end{table}

\begin{figure*}[t]
    \centering
    \includegraphics[width=\textwidth]{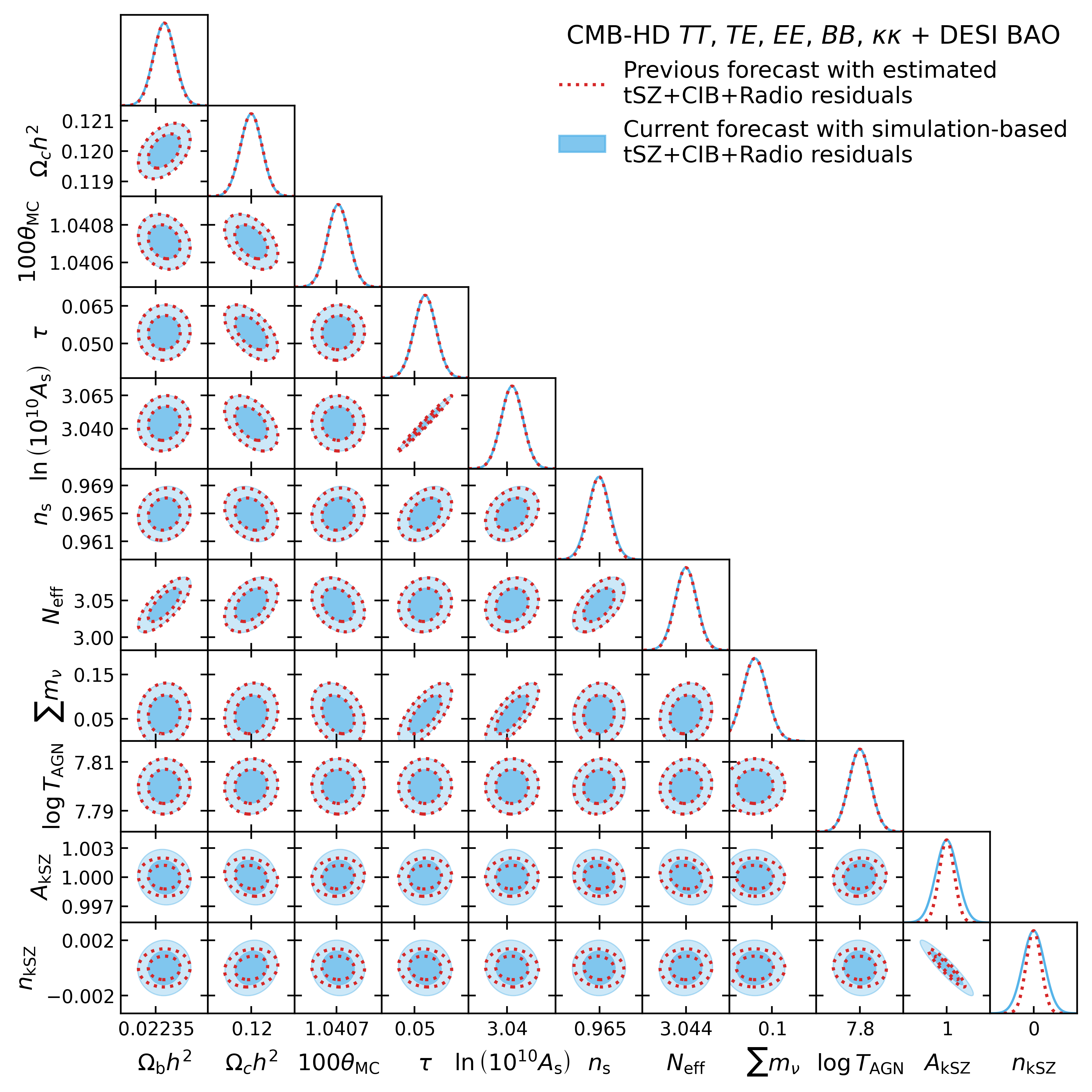}
    \caption{Comparison of the current simulation-based cosmological parameter forecasts derived in this work (blue contours) to the previous estimates of~\protect{\cite{HDparams, subgalacticDM}} (red dotted contours), for the combination of CMB-HD delensed CMB $TT$, $TE$, $EE$, $BB$ and CMB lensing $\kappa\kappa$ power spectra plus DESI BAO data. The model fit is an 11-parameter $\Lambda$CDM+$N_{\rm{eff}}$+$\sum m_{\nu}$+$T_{\rm{AGN}}+A_{\rm{kSZ}}+n_{\rm{kSZ}}$ model, which varies baryonic physics ($T_{\rm{AGN}}$) and kSZ parameters ($A_{\rm{kSZ}}$ and $n_{\rm{kSZ}}$) in addition to the cosmological parameters.   The previous forecasts are based on idealized estimates of the residual foreground and instrumental noise power spectra, while the current forecasts use the corresponding simulation-based power spectra obtained after foreground cleaning (shown as the lighter and darker red curves, respectively, in Figures~\ref{fig:coaddedspectra} and~\ref{fig:spectra}). We find that the two sets of forecasts are consistent to within 7\% for the cosmological parameters considered, as quantified in Table~\ref{tab:params_comparison}. We find that the uncertainty on the kSZ parameters increases up to 46\% (see Table~\ref{tab:params_comparison}), absorbing the excess foreground uncertainty of the simulation-based residual foregrounds.}
    \label{fig:params}
\end{figure*}

\begin{figure}[t]
    \centering
    \includegraphics[width=\columnwidth]{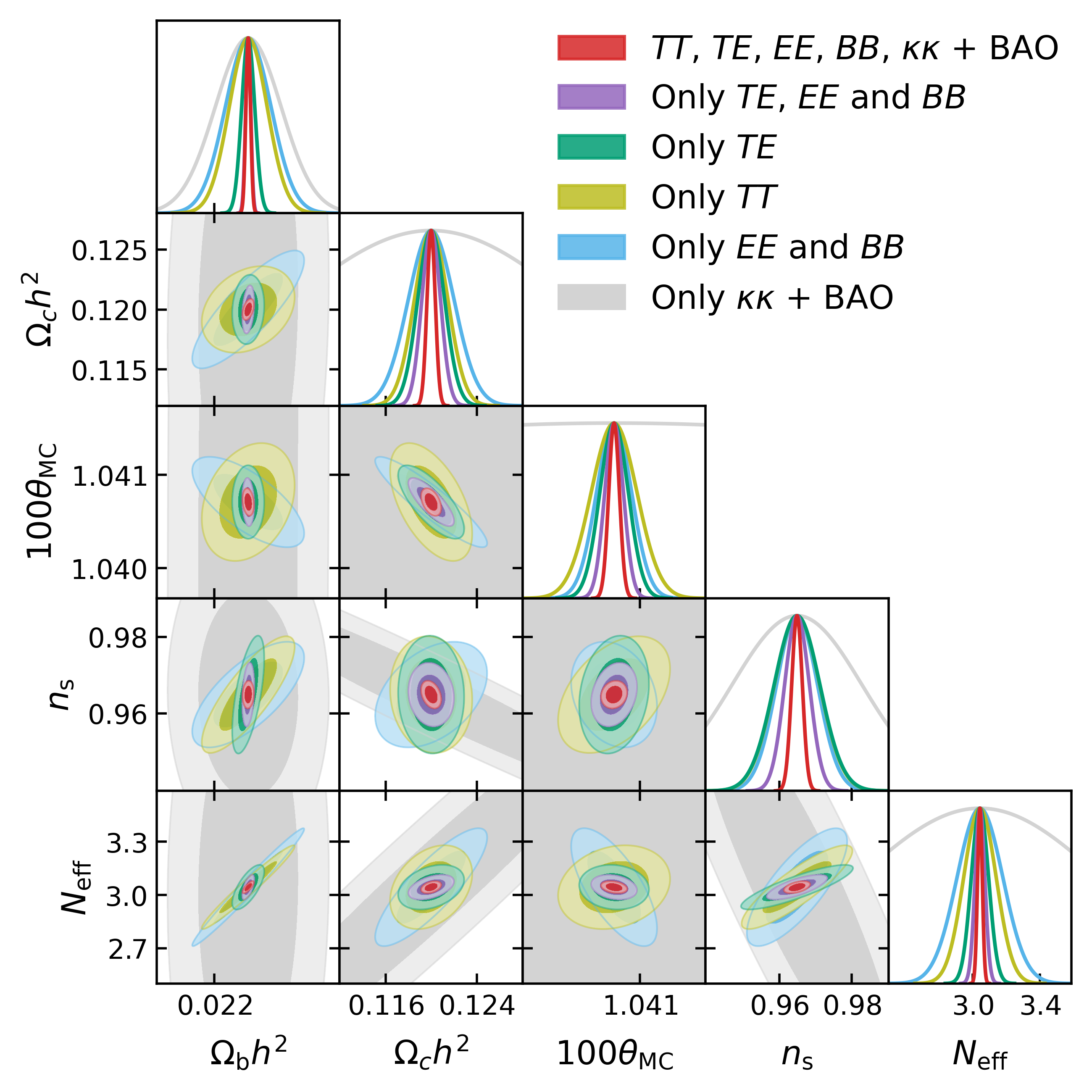}
    \caption{Forecasted parameter constraints for a 9-parameter $\Lambda$CDM+$N_{\rm{eff}}$+$\sum m_{\nu}$+$T_{\rm{AGN}}$ model which varies baryonic physics ($T_{\rm{AGN}}$) in addition to the cosmological parameters.  Here we focus on a subset of key parameters and separate the constraints from different spectra; we do not vary the kSZ parameters in order to ensure a fair comparison since we cannot vary them without including the $TT$ power spectrum. We show constraints from the combination of CMB-HD and DESI BAO when including all delensed CMB and CMB lensing power spectra (red).  We also show constraints from only CMB-HD $TE$, $EE$, and $BB$ (purple), $TE$ (green), $EE$ and $BB$ (blue), $TT$ (yellow), and $\kappa\kappa$ and BAO (gray); for the first three cases, we also use pol-only $\kappa\kappa$ to delens the spectra. For the $\kappa\kappa$ plus BAO case, we include Gaussian priors on $\Omega_\mathrm{b}h^2$ and $n_\mathrm{s}$ with $\sigma(\Omega_\mathrm{b}h^2) = 0.00036$ and $\sigma(n_\mathrm{s}) = 0.02$, as done in~\protect{\cite{Madhavacheril2023}}.   We see that the $TE$, $EE$ and $BB$ spectra combined are very constraining, and have different degeneracy directions from the $TT$ spectra. This mitigates the impact of extragalactic foreground noise in the $TT$ spectra. In Table~\ref{tab:ParamsDataCombinations} in Appendix~\ref{sec:CIBmodel}, we give the parameter constraints for this model with and without the $TT$ spectra; the inclusion of $TT$ improves constraints, in particular on $n_\mathrm{s}$ and $N_{\rm{eff}}$ by about 25\%.}
    \label{fig:ParamsFromTempPol}
\end{figure}

In the ``sim-based'' column of Table~\ref{tab:params_comparison}, we list the forecasted $1\sigma$ errors for the 11-parameter $\Lambda$CDM+$N_{\rm{eff}}$+$\sum m_{\nu}$+$T_{\rm{AGN}}$+$A_{\rm{kSZ}}$+$n_{\rm{kSZ}}$ model (same as the last column of Table~\ref{tab:params}). We also plot these parameter constraints as the blue contours in Figure~\ref{fig:params}.  We show the corresponding set of previous forecasts from~\cite{subgalacticDM} in the ``previous'' column of Table~\ref{tab:params_comparison}\footnote{Note that here we use a $\tau$ prior of 0.005 and three degenerate massive neutrinos, as opposed to a $\tau$ prior of 0.007 and one massive neutrino, as was done in~\cite{HDparams,subgalacticDM}.}, and as the red dotted contours in Figure~\ref{fig:params}; these were obtained using previous idealized estimates of the residual noise spectra from~\cite{HDparams,subgalacticDM} (light red curve in Figure~\ref{fig:coaddedspectra}). We find that the uncertainties on the cosmological parameters increase by less than \num{7\%} relative to previous forecasts, as shown in the last column of Table~\ref{tab:params_comparison}. In particular, $\sigma(N_{\rm{eff}})$ increases from \num{0.0150} to \num{0.0154} for this 11-parameter model, with neutrino mass and baryonic physics simultaneously varied.\footnote{$N_{\rm{eff}}$ constraints are sensitive to the sky area surveyed.  We assume a 60\% sky fraction in this work and obtain \num{$\sigma(N_{\rm{eff}})=0.0154$}; a CMB-HD survey covering 70\% of the sky would yield \num{$\sigma(N_{\rm{eff}})=0.0143$}. For comparison, adding {\it{Planck}} data over 20\% of the sky non-overlapping with CMB-HD improves $\sigma(N_{\rm{eff}})$ by 1\% and $\sigma(n_{\rm{s}})$ by 5\%.} In the case of a non-detection of $\Delta N_{\rm{eff}}$, this would rule out the existence of new light particle species at a \num{92\%} confidence level~\cite{Baumann:2015rya,Baumann:2017gkg}. 

Part of the reason that the parameter constraints do not increase that much, even though the noise increased by \num{50\%} for $TT$ compared to previous estimates, is that the polarization spectra play a significant role in determining the cosmological parameters. In Figure~\ref{fig:ParamsFromTempPol}, we show the constraints for a 9-parameter model ($\Lambda$CDM+$N_{\rm{eff}}$+$\sum m_{\nu}$+$T_{\rm{AGN}}$), focusing on a subset of key parameters and separating the constraints from different spectra; we do not vary the kSZ parameters here to ensure a fair comparison since we cannot vary them without including the $TT$ power spectrum. We show the constraints from only $TE$ (green), only $TT$ (yellow), the combination of $EE$ and $BB$ (blue), and the combination of $\kappa\kappa$ and BAO (gray); for the $\kappa\kappa$ plus BAO case, we include Gaussian priors on $\Omega_\mathrm{b}h^2$ and $n_\mathrm{s}$ with $\sigma(\Omega_\mathrm{b}h^2) = 0.00036$ and $\sigma(n_\mathrm{s}) = 0.02$ (as in~\cite{Madhavacheril2023}). We also show the constraints from the combination of $TE$, $EE$, and $BB$ (purple), which spectra are insensitive to the foreground noise in $TT$.  We see that the $TE$, $EE$, and $BB$ spectra combined are very constraining and have different degeneracy directions from the $TT$ spectra. In Table~\ref{tab:ParamsDataCombinations} in Appendix~\ref{sec:CIBmodel}, we provide the parameter constraints for this model with and without the $TT$ spectra; we find that while the non-$TT$ spectra are very constraining, the inclusion of $TT$ improves constraints, in particular on $n_\mathrm{s}$ and $N_{\rm{eff}}$ by about \num{25\%}.  

In addition to the polarization spectra mitigating the impact of extragalactic foreground noise in temperature spectra, varying the kSZ parameters absorbs some of this uncertainty as well, as shown in Table~\ref{tab:params_comparison} and Figure~\ref{fig:params}.  These parameters  increase in uncertainty by \num{45\%} and \num{46\%} for $A_{\rm{kSZ}}$ and $n_{\rm{kSZ}}$, respectively, when using the simulation-based noise curves. In contrast, errors on the cosmological parameters remain roughly unchanged. Nonetheless, the parameters that have the highest increase in uncertainty when switching to simulation-based noise curves are $n_\mathrm{s}$ (\num{7\%}) and $N_\mathrm{eff}$ (\num{3\%}), which are key scientific targets for a CMB-HD-like experiment, highlighting the importance of foreground removal.

\section{Discussion and Conclusion}
\label{sec:discussion}

In this work, we have explored whether extragalactic foregrounds in temperature maps can be removed to low enough levels for a CMB-HD-like survey such that improvements in instrumental sensitivity of such a survey can be capitalized upon. In particular, foreground removal needs to be sufficient to preserve the CMB-HD light relic target of $\sigma(N_\mathrm{eff}) \approx 0.015$, which would rule out or detect with more than 90\% confidence any new light particle species that was in thermal equilibrium with standard model particles at any time after the Big Bang~\cite{Baumann:2015rya,Baumann:2017gkg}.  Reaching specifications to achieve this key science target also enables a wealth of other powerful constraints on inflation, dark matter, dark energy, and astrophysics more broadly.  

To answer this question, we created a new set of microwave sky simulations with more than ten times higher resolution than previous millimeter-wave simulations.  This was done by developing a general procedure to turn lower-resolution simulations ($\sim 0.5$ arcminute) into ultrahigh-resolution versions (0.04 arcminute).  We make these simulations public at \url{https://lambda.gsfc.nasa.gov/simulation/ultrahigh_resolution_sims.html}, along with the code to recreate them at \url{https://github.com/CMB-HD/hdsims}, for use by the community.

We developed a foreground-cleaning procedure that exploits the localization and frequency dependence of the CIB, radio sources, and tSZ clusters, as well as the ultrahigh-resolution and ultradeep sensitivity of a CMB-HD survey.  This procedure makes use of an iterative matched-filter method and utilizes the 219 and 277~GHz frequency maps to clean point sources and clusters from maps at 90 and 148~GHz, the primary frequencies for cosmological parameter constraints.  

The residual noise power spectra we obtained after our foreground-cleaning procedure is about \num{50\%} higher for the 90 and 148~GHz coadded maps than previously predicted through some guesstimation~\cite{HDparams,subgalacticDM}. This increase in noise power, per $\ell$-mode, results in an increase in cosmological parameters for a $\Lambda$CDM+$N_{\rm{eff}}$+$\sum m_{\nu}$+$T_{\rm{AGN}}+A_{\rm{kSZ}}+n_{\rm{kSZ}}$ model of at most \num{7\%} relative to the idealized forecasts.  We see that the highest increases are for $n_\mathrm{s}$ (\num{7\%}) and $N_\mathrm{eff}$ (\num{3\%}), highlighting the challenges and requirements of these key science targets.  We find that marginalizing over baryonic feedback minimally changes cosmological parameter uncertainties, while marginalizing over the amplitude and shape of the kSZ power spectrum changes cosmological parameter constraints by up to \num{11\%}; in particular, marginalizing over the kSZ parameters shifts \num{$\sigma(N_\mathrm{eff}) =  0.0140$} to \num{$\sigma(N_\mathrm{eff}) =  0.0154$}.\\

We make the foreground-cleaning code developed here public at \url{https://github.com/CMB-HD/hdfgclean} and anticipate that it will be improved upon.  Speeding up the computation time and adding harmonic-space frequency-based deprojection methods are some foreseeable areas for further gains.  This work serves as a first demonstration with simulations that the foreground-removal challenge of a CMB-HD-like survey can be surmounted.

\begin{acknowledgments}
The authors thank Daniel Von Thaden for testing the code presented in this work, which we make public. The authors thank Sebastian Grandis for the suggestion to make Figure~\ref{fig:clustercompleteness}. AM and NS acknowledge support from DOE award number DE-SC0025309 and the Stony Brook OVPR Seed Grant Program.  JA and IB acknowledge support from the REU program supported by the NSF under Grant No. PHY-2243856 and PHY-1852143. IB also acknowledges support from NSF award number 1911370 for the OK-LSAMP program.

The authors would like to thank Stony Brook Research Computing and Cyberinfrastructure, and the Institute for Advanced Computational Science at Stony Brook University for access to the SeaWulf computing system, made possible by grants from the National Science Foundation (award number 1531492 and Major Research Instrumentation award number 2215987), with matching funds from Empire State Development’s Division of Science, Technology and Innovation (NYSTAR) program (contract C210148).


\end{acknowledgments}

\appendix

\section{Computing Power Spectra}
\label{sec:ps}

In the following, we describe how we take a power spectrum of a partial sky map, correcting for the beam, pixel window function, and mask.  We use the pspy package\footnote{\url{pspy.readthedocs.io}} to calculate power spectra. Starting from a CAR map on a patch of the sky that has been convolved with the pixel window function and the instrument beam, we do the following:

\begin{enumerate}
    \item Apodize the edges of the map using a mask that has a \num{$0.5^\circ$} border (for 100 square degree maps) that goes from a value of one at the inner edge smoothly to a value of zero at the outer edge. For the smaller \num{four square degree} maps, we use an apodization width of \num{$0.25^\circ$}. This allows the patch of sky to be periodic at the boundaries.  Given an apodization width, this apodization mask is computed by pspy using Equation 30 of~\cite{Grain2009}. 
    
    \item Deconvolve the pixel window function from the map. (The apodization mask needs to be applied before the pixel-window deconvolution.)

    \item If applicable, apply any additional masks (for example, the mask of leftover missubtracted sources and clusters discussed in Section~\ref{sec:method-clusters}). 
    
    \item Use pspy to take the spherical harmonic transform and calculate the ``raw'' power spectrum at each multipole $\ell$.\footnote{The pspy package also requires the apodization window to be given along with the map since it apodizes the map itself; however, since we already apodized the map before deconvolving the pixel window function, we just pass a map filled with ones instead.} 
    
    \item Compute a mode-decoupling matrix by passing to pspy the mask,\footnote{If another mask has been applied in addition to the apodization mask, then the total mask should be the product of the two masks. When computing spectra for temperature and polarization maps, one can pass pspy different masks for temperature and polarization.} the beam profile, and the bin edges used to bin the final power spectrum.  Throughout this work, we use uniform binning with a bin width of $\Delta \ell = 200$.
    
    \item Use pspy to apply the mode-decoupling matrix to the ``raw'' power spectrum, given the bin edges, to produce a binned power spectrum that has been corrected for the beam and mask. 
\end{enumerate}

\begin{figure}
    \centering
    \includegraphics[width=\columnwidth]{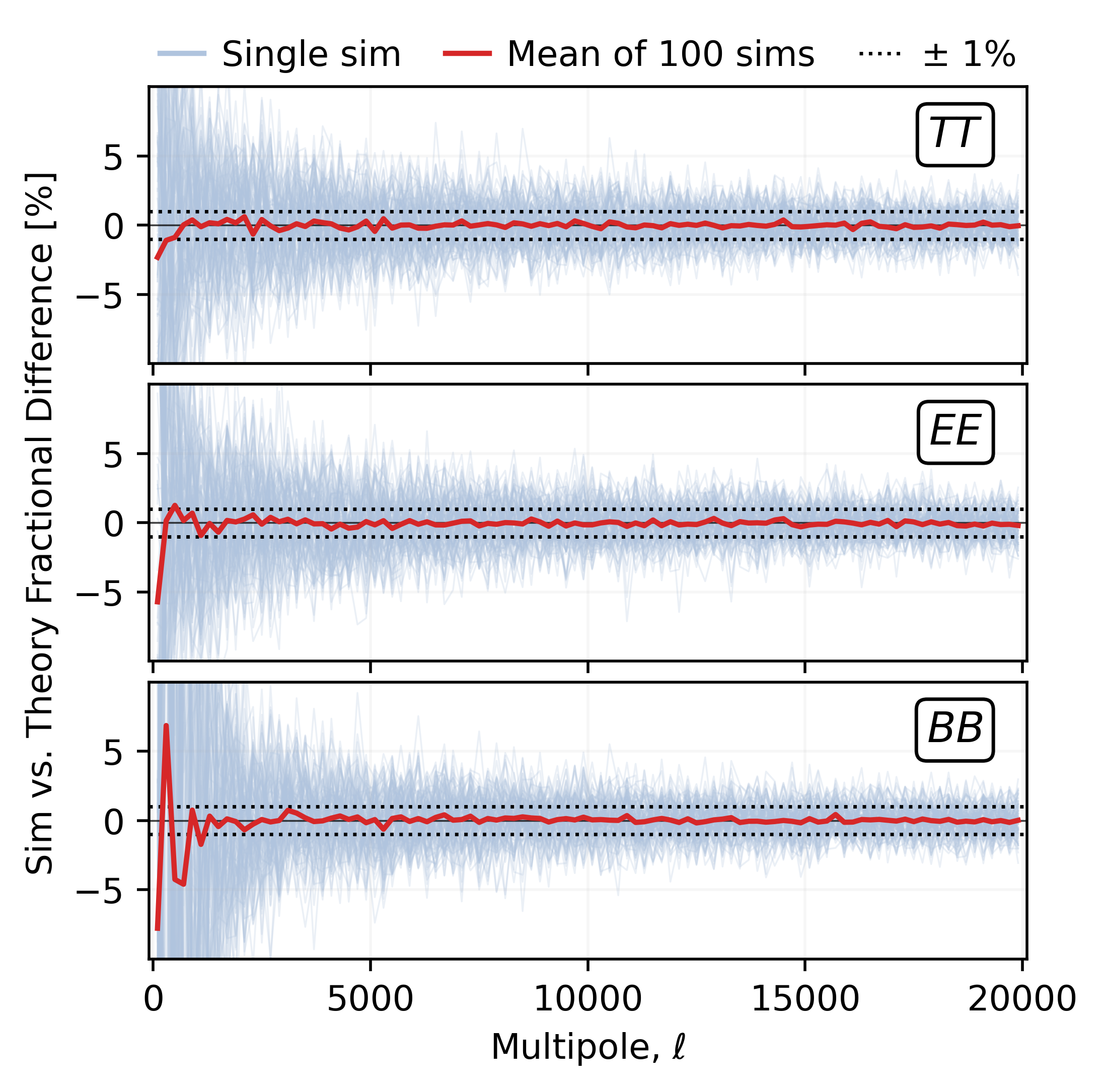}
    \caption{Comparison between the $TT, EE$, and $BB$ power spectra from \num{100} Gaussian realizations of the lensed $T, Q$, and $U$ CMB and the theory power spectra used to generate them. We find that the mean power spectra of the simulations agrees with the theory to within 1\% over most of the range $\ell \in [30, 20,000]$, as indicated by the dotted black lines.  Averaging this difference over all multipoles results in agreement to within 0.1\% for each mean spectrum.  This suggests that our power spectra are unbiased.}
    \label{fig:PowerSpectrumTest}
\end{figure}

To verify the accuracy of this power spectrum procedure, we simulate \num{100} Gaussian realizations of the $T, Q$, and $U$ CMB over a \num{$10^\circ \times 10^\circ$} region of sky using theoretical power spectra. We then calculate the $TT, TE, EE$, and $BB$ power spectra\footnote{Note, pspy automatically calculates $E$ and $B$ spectra given $Q$ and $U$ maps.} of each map following the procedure above and compare the power spectra of the simulations with the theory. The simulations are generated in the same way as the unlensed CMB map in Section~\ref{sec:lensedCMB}, but here we use the lensed, as opposed to the unlensed, CMB theory power spectra from CAMB since we are mainly concerned with the lensed power in this work. The mean power spectra of the simulations agree with the theory spectra to within \num{1\%} over most of the range $\ell \in [30, 20,000]$, as indicated by the dotted black lines in Figure~\ref{fig:PowerSpectrumTest}; averaging this difference over all multipoles results in agreement to within \num{0.1\%} for each mean spectrum, suggesting our power spectra are unbiased.

\section{Calculating Beam Solid Angle}
\label{sec:beam}

\begin{figure}
    \centering
    \includegraphics[width=\linewidth]{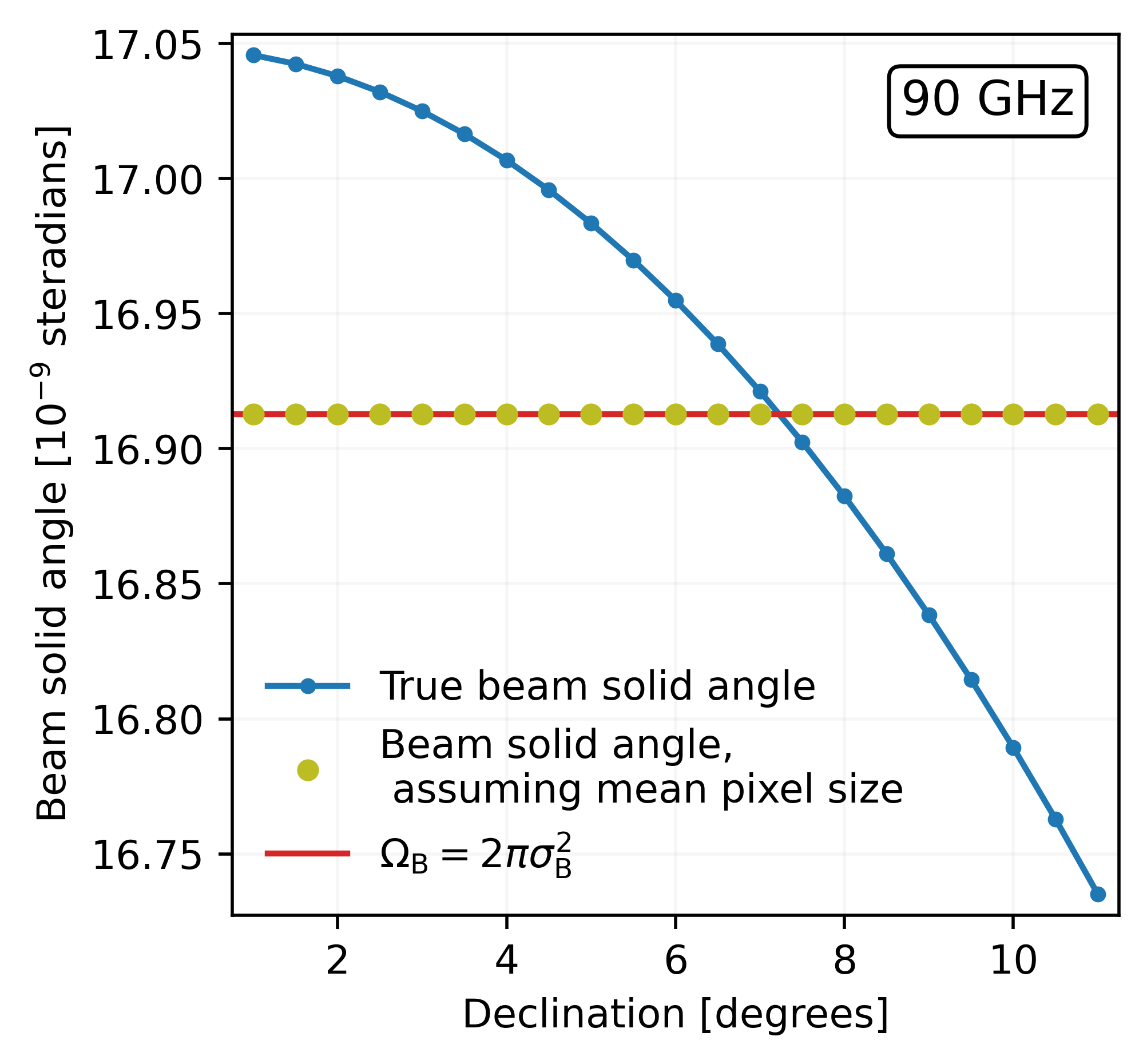}
    \caption{The effective beam solid angle from a Gaussian beam plus the pixelization of the map, as a function of declination. The true beam solid angle is shown in blue, measured in a 100 square degree patch of sky and calculated using Equation~\ref{eq:beam_ratio}; the variation in this beam solid angle is due to the variation of the pixel solid angle with declination. The yellow points show the effective beam solid angle calculated when using the mean pixel solid angle in the map in Equation~\ref{eq:beam_ratio}; this matches the standard analytic calculation of the beam solid angle when neglecting pixelization shown in red, $\Omega_\mathrm{B} = 2 \pi \sigma_\mathrm{B}^2$, where $\sigma_\mathrm{B}$ is the standard deviation of the Gaussian beam.}
    \label{fig:beam}
\end{figure}

As discussed in Section~\ref{sec:filter}, the standard calculation of the beam solid angle needs to be corrected to include the effect of map pixelization.  As mentioned in that Section, we model the beam with a Gaussian profile. The profile, given by Eq.~\ref{eq:beamprofile}, is normalized such that $B(0) = 1$, and integrating over this profile gives the standard beam solid angle, $\int d^2x B(\pmb{x}) = 2 \pi \sigma_\mathrm{B}^2 = \Omega_\mathrm{B}$.  However, since the map is pixelized before being smoothed with a Gaussian beam, we need to calculate $\Omega_\mathrm{B}$ more precisely, or we will under- or over-estimate the flux of sources.

To do this, we note that we measure the flux $S$ of a source from the amplitude of the peak pixel in the filtered map, $T_\mathrm{peak}$, using
\begin{equation} 
    S = T_\mathrm{peak} \Omega_\mathrm{B} \left.\frac{\partial B_\nu(T)}{\partial T}\right|_{T_\mathrm{CMB}}.
\end{equation}
Here, $T_\mathrm{peak}$ has units of $\mu$K, and $\partial B_\nu / \partial T$ is the derivative of the Planck blackbody function at frequency $\nu$, evaluated at CMB temperature $T_\mathrm{CMB}$, and has units of mJy/str/$\mu$K. Multiplying $T_\mathrm{peak}$ by the beam solid angle, $\Omega_\mathrm{B}$, corrects $T_\mathrm{peak}$ for the effect of beam smoothing~\cite{VargasACT2023srcs} and yields the flux of the source $S$.

Since our map is also pixelized, which is effectively smoothing on the pixel scale, we need to correct for this as well.  This can be done by placing a single ``source'' with amplitude $T_0 = 1$~$\mu$K in a single pixel, and convolving this map with the beam.  Then we obtain $T_\mathrm{peak}$, the maximum of the beam-convolved (and ``pixel-convolved'') source, and compute
\begin{equation}
\label{eq:beam_ratio}
    \Omega_\mathrm{B}(\pmb{x}) = \Omega_\mathrm{pix}(\pmb{x}) \frac{1~\mu\mathrm{K}}{T_\mathrm{peak}(\pmb{x})}.
\end{equation}
Here, $\pmb{x}$ is the location of the pixel in the map; different pixels in the map subtend different solid angles, denoted by $\Omega_\mathrm{pix}(\pmb{x})$. When the mean pixel solid angle of the map is used to compute $\Omega_\mathrm{pix}(\pmb{x})$, then $\Omega_\mathrm{B}$ equals $2 \pi \sigma_\mathrm{B}^2$, as shown by the yellow points in Figure~\ref{fig:beam}.  In reality, $\Omega_\mathrm{pix}(\pmb{x})$ varies across the map, 
so we use Eq.~\ref{eq:beam_ratio} to calculate $\Omega_\mathrm{B}(\pmb{x})$ for each pixel.  In Figure~\ref{fig:beam}, we show the true beam solid angle, $\Omega_\mathrm{B}(\pmb{x})$, as blue points and compare it to $2 \pi \sigma_\mathrm{B}^2$ as a function of declination across the map. Neglecting this effect results in an under- or over-estimate of source fluxes by up to about \num{1\%}, as mentioned in Section~\ref{sec:filter}.\footnote{Note that in this section, we are being precise about how knowledge of the beam is used. How well the instrument beam needs to be known is outside the scope of this work.}

\section{CIB Modeling}
\label{sec:CIBmodel}

We model the CIB using the catalog of CIB sources from the S10 catalog. We lower the fluxes of all CIB sources in the S10 catalog by 25\%, as was done in~\cite{vanEngelen:2013rla, SOforecasts}, to better match subsequent observations~\cite{Dunkley:2013vu}. This scaled model matched well both power spectrum and bispectrum measurements of the CIB below $\ell =4000$ from ACT, SPT, and {\it{Planck}}~\cite{vanEngelen:2013rla}.  We refer to this scaled S10 catalog hereafter as the original S10 catalog and contrast it with modified versions of this catalog discussed below.   

\begin{figure}[t]
    \centering
    \includegraphics[width=\columnwidth]{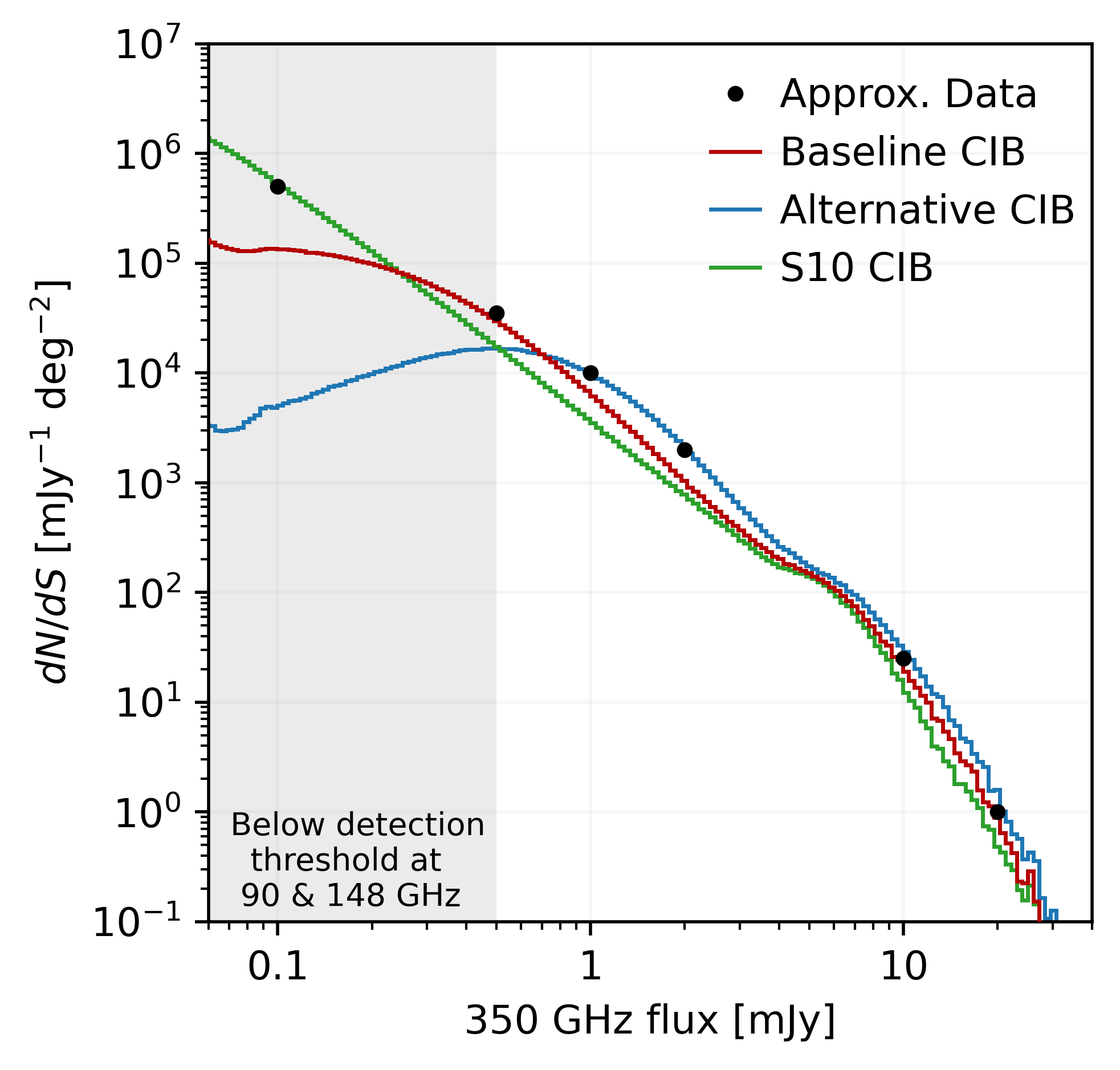}
    \caption{CIB number counts at 350~GHz  within a 100 square degree patch of sky for three CIB models, the original S10 model (green) and two modified models labeled ``baseline'' (red) and ``alternative'' (blue).  The baseline and alternate models are obtained by placing the sources from the S10 catalog in maps with 0.25 and 0.43 arcminute pixel resolution, respectively.  Since multiple sources fall into a single pixel in both these cases, we create new catalogs of these sources, where we let each pixel correspond to one source with the summed flux.  For the S10 model shown in green, we place the S10 catalog sources in a map with 0.04 arcminute resolution, which effectively gives each source its own pixel.  The difference in pixel resolution results in different effective source number counts for a given flux, as shown by the difference in the three curves. We show in shaded gray the source fluxes at 350~GHz that would be below our $4\sigma$ detection threshold at 90 and 148~GHz. From this we see that we would detect and remove more of the CIB for the alternate CIB model than for the S10 model.  We also plot approximate measurements we estimate from Figure 12 of~\protect{\cite{Casey2014}} (black points), which presents a compilation of many CIB number count measurements.   We find the baseline CIB model to be the best overall match to observation, since the alternative model under-estimates lower flux sources, and the S10 model under-estimates higher flux sources.}
    \label{fig:cibmodels}
\end{figure}

\begin{figure}[t]
    \centering
    \includegraphics[width=\columnwidth]{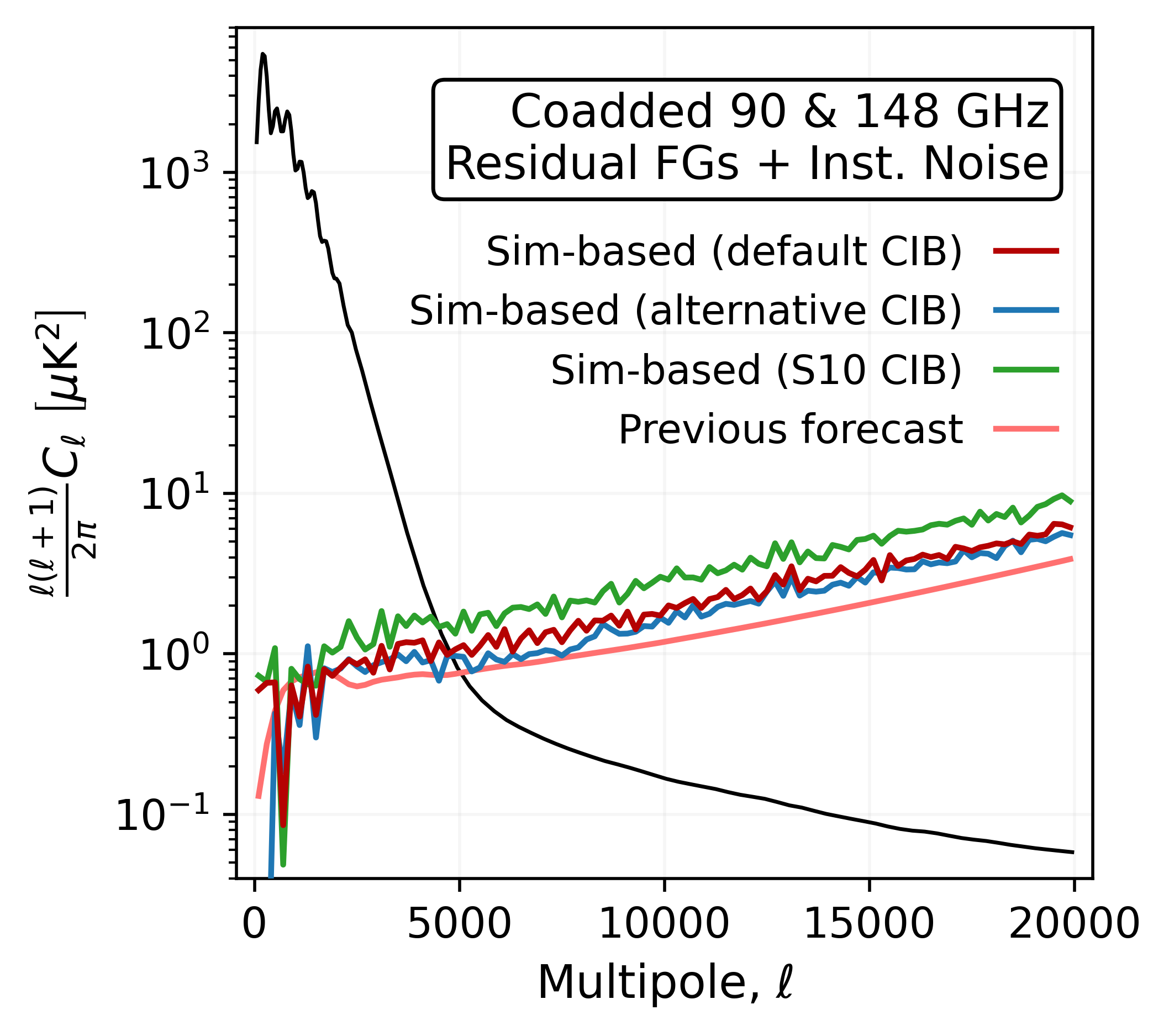}
    \caption{The coadded 90~and 148~GHz power spectrum of the beam-deconvolved instrument noise plus the residual temperature foregrounds after applying our foreground-cleaning method to \num{four square degree} simulations generated with our baseline (dark red), alternative (blue), or S10 (green) CIB model. For comparison, we also show the corresponding previous idealized estimate from~\protect{\cite{han22,HDparams,subgalacticDM}} (light red). The simulation-based power spectrum changes when using different CIB models; we find an increase of 50\%, 30\%, or 125\% relative to the previous estimate for the baseline, alternative, or S10 CIB models, respectively. The baseline CIB model increases cosmological parameter forecasts by at most 7\% for an 11-parameter $\Lambda$CDM+$N_{\rm{eff}}$+$\sum m_{\nu}$+$T_{\rm{AGN}}+A_{\rm{kSZ}}+n_{\rm{kSZ}}$ model compared to previous forecasts; when using the S10 CIB model, the cosmological parameter uncertainties increase by at most 8\% for this 11-parameter model. }
    \label{fig:CoaddedSpectraCIBModels}
\end{figure}

When modeling the CIB using the S10 catalog,  we found that the effective number of sources at a given flux depends on the pixel resolution of our map.  For example, within a $12^\circ \times 12^\circ$ region, the S10 CIB catalog has about 17,800,000 sources, which is roughly 34 sources per square arcminute.  When these sources are placed on a map at the S10 pixel resolution of 0.43 arcminute, there are, on average, about 6 sources per pixel (there are more sources in the catalog than pixels in the map).  The fluxes of all the sources that fall into a given pixel are summed together, resulting in effectively one higher flux source at that location.  If we create a new catalog of these sources, where we let each pixel correspond to one source with the summed flux, then we obtain the CIB number counts shown in Figure~\ref{fig:cibmodels} in blue and labeled ``alternative CIB''.  

If instead we place all the sources in the original S10 CIB catalog, within a $12^\circ \times 12^\circ$ region, on a map with 0.04 arcminute pixel resolution, there are, on average, about 0.05 sources per pixel, i.e.~one source for every 20 pixels (there are fewer sources in the catalog than pixels in the map).  If we create a new catalog of these sources, where we let each pixel correspond to one source, then we obtain the CIB number counts shown in Figure~\ref{fig:cibmodels} in green and labeled ``S10 CIB''.  Since each source essentially has its own pixel when the pixels are 0.04 arcminute, the number counts of this catalog preserve the original S10 CIB model.  

Although the power spectra for the alternative CIB and S10 CIB models are identical and thus match observations, the difference in source number counts at a given flux yields different results when  removing detected sources. We see from Figure~\ref{fig:cibmodels} that the S10 CIB catalog (green curve) has more low flux sources, whereas the alternative CIB model has more bright sources. Since bright sources are easier to detect and remove, assuming the alternative CIB model could yield more optimistic foreground-cleaning results than assuming the S10 CIB model.  

We also make an intermediate model using S10 CIB sources placed on a map with 0.25 arcminute pixel resolution.  The catalog of these sources, where we let each pixel correspond to one source with the summed flux, we call the ``baseline CIB''; we show the number counts of this model in red in Figure~\ref{fig:cibmodels}.

To determine which CIB model to assume in our simulations, we compare them with observations of CIB number counts presented in Figure 12 of the review by~\cite{Casey2014}; Figure 12 in~\cite{Casey2014} presents a compilation of many different number count measurements, and in Figure~\ref{fig:cibmodels} we show our rough estimate of the average of these measurements as black points.\footnote{We do not present an analysis of the uncertainty around the black points in Figure~\ref{fig:cibmodels} as they are used only to provide rough guidance on the reasonableness of CIB models.} 

We see from Figure~\ref{fig:cibmodels} that, in general, the S10 CIB model (green curve) is a good match to observations at lower flux but tends to under-estimate the number of higher-flux sources, especially around 1~mJy at 350 GHz (which is a factor of two above our flux threshold at 90 and 150~GHz).  On the other hand, the alternative CIB model tends to under-estimate the abundance of dimmer CIB sources but is a good match to the number of brighter sources.  The baseline CIB model is a compromise between the two; it only slightly under-estimates the number of low-flux CIB sources and still agrees well with observations at higher fluxes.  Therefore, we choose this as our baseline model used throughout this work, but also run our foreground-cleaning procedure on simulations with the alternative CIB and S10 CIB models to assess the impact. 

For the baseline CIB and alternate CIB models, we add some random scatter to the position of each source in the catalog; without this scatter, the sources would form a grid when they are placed in the higher-resolution (0.04 arcminute) map (having a source location at the center of the lower resolution pixel). The amount of scatter added to each RA and dec is drawn from a Gaussian distribution, with a standard deviation of 20\% of the lower-resolution pixel size (i.e.,~the standard deviation is 0.05 arcmin for the baseline CIB model and 0.086 arcmin for the alternative CIB model.)

To assess the sensitivity of our results to the choice of CIB model, we run our foreground cleaning procedure on simulations generated using the baseline, alternative, and S10 CIB models.  In Figure~\ref{fig:CoaddedSpectraCIBModels}, we show the coadded 90~and 148~GHz power spectrum of the residual foregrounds and instrumental noise from a \num{four square degree} patch of the sky.  We show this for the baseline (darker red), alternative (blue), and S10 (green) CIB models. We also compare these with previous estimates from~\cite{han22, HDparams, subgalacticDM} (lighter red). We find minimal change to the coadded residual spectra when using the alternative versus the baseline CIB model; on average, the former is about \num{30\%} higher than the previous estimate, while the latter is about \num{50\%} higher. As mentioned in Section~\ref{sec:params}, the baseline CIB model increases cosmological parameter forecasts by at most \num{7\%} for an 11-parameter $\Lambda$CDM+$N_{\rm{eff}}$+$\sum m_{\nu}$+$T_{\rm{AGN}}+A_{\rm{kSZ}}+n_{\rm{kSZ}}$ model compared to previous forecasts.  

\begin{table}[t]
    \centering
    \begin{tabular}{l@{\hskip 1em} l@{\hskip 1em}  l l c}
        \toprule
        \toprule
        \multicolumn{1}{l}{} &  \multicolumn{2}{c}{Forecasted  $1\sigma$ Errors} &  & \multicolumn{1}{c}{Ratio of $1\sigma$ Errors} \\
        \cmidrule(){2-3} \cmidrule(){5-5}
        Parameter & No $TT$ & With $TT$ & & With $TT$ / No $TT$ \\
        \midrule
        $\Omega_\mathrm{b} h^2$\dotfill                           & 0.0000267  & 0.0000253  &  & 0.95  \\
        $\Omega_c h^2$\dotfill                                    & 0.000401   & 0.000363   &  & 0.91  \\
        $\ln \left(10^{10} A_\mathrm{s}\right)$\dotfill           & 0.00867    & 0.00844    &  & 0.97  \\
        $n_\mathrm{s}$\dotfill                                    & 0.00202    & 0.00148    &  & 0.73  \\
        $\tau$\dotfill                                            & 0.00461    & 0.00447    &  & 0.97  \\
        $100\theta_\mathrm{MC}$\dotfill                           & 0.0000639  & 0.0000603  &  & 0.94  \\
        $N_\mathrm{eff}$\dotfill                                  & 0.0185     & 0.0140     &  & 0.76  \\
        $\sum m_\nu$~[eV]\dotfill                                 & 0.0291     & 0.0287     &  & 0.99  \\
        $\log_{10}\left(T_\mathrm{AGN}/\mathrm{K}\right)$\dotfill & 0.00467    & 0.00464    &  & 0.99  \\
        \bottomrule
    \end{tabular}
    \caption{Comparison of forecasted $1\sigma$ parameter error bars for a 9-parameter $\Lambda$CDM+$N_{\rm{eff}}$+$\sum m_{\nu}$+$T_{\rm{AGN}}$  model from the combination of mock CMB-HD and DESI data, when the $TT$ power spectrum is excluded and the pol-only $\kappa\kappa$ is used (second column) or when $TT$ is included and the MV $\kappa\kappa$ is used (third column); in both cases, the CMB-HD $TE$, $EE$, and $BB$ power spectra are included. The last column shows the improvement in parameter constraints when including the additional temperature data that is impacted by extragalactic foregrounds; in particular, the error bars on $n_\mathrm{s}$ and $N_\mathrm{eff}$ show the most improvement, with both decreasing by about 25\%.}
    \label{tab:ParamsDataCombinations}
\end{table}

We find that the coadded residual spectra using the S10 CIB model is \num{125\%} higher than previous estimates, since many more of its sources are dim and below the threshold for detection and subtraction; however, we find that cosmological parameter constraints for the 11-parameter model increase by only, at most, \num{8\%} compared to previous forecasts when using the S10 CIB model. This is in part because the polarization spectra also play a role in determining the cosmological parameters and because varying the kSZ parameters absorbs some of the uncertainty from the extragalactic foregrounds (see Figure~\ref{fig:ParamsFromTempPol}, Table~\ref{tab:params_comparison}, and text in Section~\ref{sec:params}).  In Table~\ref{tab:ParamsDataCombinations}, we show the forecasted cosmological parameter constraints for a 9-parameter $\Lambda$CDM+$N_{\rm{eff}}$+$\sum m_{\nu}$+$T_{\rm{AGN}}$ model; we do not vary the kSZ parameters here to ensure a fair comparison since we cannot vary them without including the $TT$ power spectrum. We show the parameter constraints for this model with and without the $TT$ spectrum, and see that the non-$TT$ spectra are already very constraining.  As mentioned above, this partly mitigates the impact of increased extragalactic foregrounds in the $TT$ spectra.  However, the inclusion of $TT$ does improve constraints, in particular on $n_\mathrm{s}$ and $N_{\rm{eff}}$ by about \num{25\%}.  This improvement for $N_{\rm{eff}}$ is important to bring its constraint well below the target of $\Delta N_{\rm{eff}} = 0.027$~\cite{Baumann:2015rya,Baumann:2017gkg}.

\begin{figure*}[t]
    \centering
    \includegraphics[width=\linewidth]{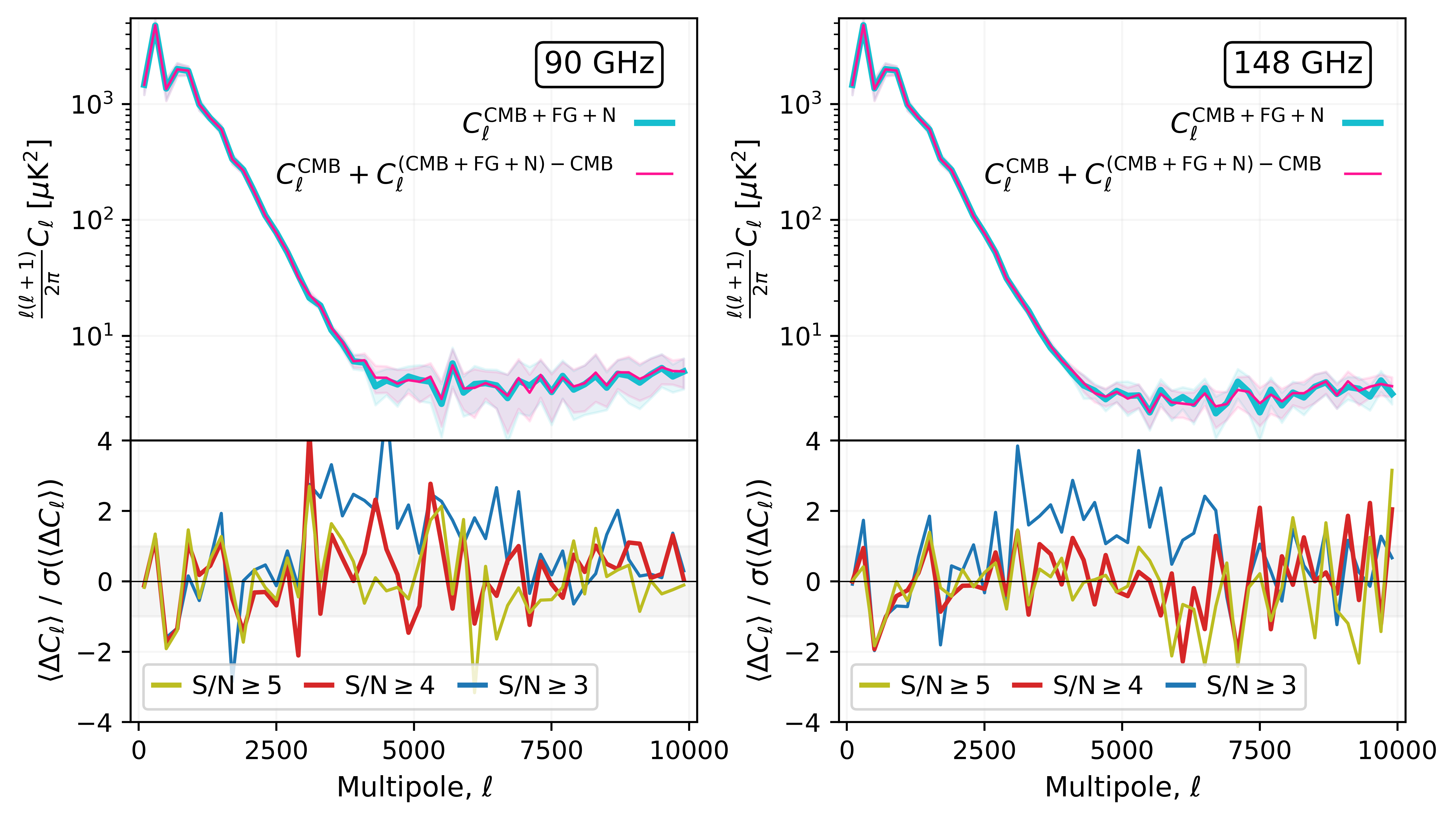}
    \caption{ {\it{Upper panels:}} We show as cyan curves labeled $C_\ell^\mathrm{CMB+FG+N}$ the mean of 16 power spectra after foreground cleaning four square degree maps at 90~GHz (left) and 148~GHz (right) using the methods described in Sections~\ref{sec:method-points} and~\ref{sec:method-clusters}; this spectrum contains the CMB, instrumental noise (N), and the residual foregrounds (FG). We compare these spectra to those shown in pink, which are the mean CMB power spectra of the map realizations, $C_\ell^\mathrm{CMB}$, plus the mean power spectra of the foreground-cleaned maps after subtracting out the CMB realization at the map level, $C_\ell^\mathrm{(CMB+FG+N)-CMB}$. The $\pm 1\sigma$ error on the mean of each mean power spectrum is shown by the shaded region. These curves are obtained using our baseline foreground cleaning method with a minimum SNR of four. If our foreground-cleaning procedure disturbed the underlying CMB, these two spectra would not match.   {\it{Lower panels:}} We show the mean difference between the two power spectra (cyan and pink) as a fraction of the error bar on this mean, using different SNR thresholds for removing point sources and clusters in the foreground cleaning process. For our baseline choice of an SNR threshold of four (red), as well as for an SNR threshold of five (yellow), we do not observe any significant bias from zero in this difference; this suggests that our foreground cleaning method has not altered the CMB signal in the maps. However, we find that this test fails for an SNR threshold of three, and, thus, we adopt a minimum SNR threshold of four throughout this work.}
    \label{fig:SubtractedPowerWithCMB}
\end{figure*}

\section{Did we mess up the CMB when removing sources and clusters?}  \label{sec:CMBtest}

In the following, we discuss our choice for the minimum SNR threshold for point sources and clusters and check that this choice does not bias the underlying CMB map.  One reason such a bias in the CMB map could arise is if our SNR threshold is so low that we start removing hot or cold spots in the CMB map itself or sources/clusters correlated with CMB hot or cold spots.  

We check whether such a bias occurred by comparing the power spectrum after foreground cleaning, $C_\ell^\mathrm{CMB+FG+N}$, to the sum of $C_\ell^\mathrm{CMB}$ (the power spectrum of the CMB realization of the map) and $C_\ell^\mathrm{(CMB+FG+N)-CMB}$ (the power spectrum of the foreground-cleaned map after subtracting the CMB realization at the map level).  Here, ``N'' indicates instrumental noise, and ``FG'' indicates the residual foregrounds.  If our foreground-cleaning procedure disturbed the underlying CMB, then these two spectra would not match. 

We show these spectra in the upper panels of Figure~\ref{fig:SubtractedPowerWithCMB}. The cyan curves show $C_\ell^\mathrm{CMB+FG+N}$, and the pink curves show $C_\ell^\mathrm{CMB}$ + $C_\ell^\mathrm{(CMB+FG+N)-CMB}$, after taking the mean spectra from \num{16} \num{four square degree} maps at 90~GHz (left) and 148~GHz (right); we use the methods described in Sections~\ref{sec:method-points} and~\ref{sec:method-clusters} to remove point sources and clusters from these maps using a minimum SNR threshold of \num{four}. The error on this mean is indicated by the shaded region for each curve, showing the scatter in the power spectra of the individual maps (divided by the square root of the number of maps).  In the lower panels  of Figure~\ref{fig:SubtractedPowerWithCMB}, we show the mean difference between the two power spectra (cyan and pink), $\langle \Delta C_\ell \rangle$, as a fraction of the error on this mean, $\sigma \bigl(\langle \Delta C_\ell \rangle\bigr)$, using different SNR thresholds for point sources and clusters in the foreground cleaning process. 
To quantify if there is a bias across $\ell$ bins, we take the average of the ratio $\langle \Delta C_\ell \rangle / \sigma \bigl(\langle \Delta C_\ell \rangle\bigr)$ over multipoles up to 10,000, and also calculate an error on this average from the bin to bin scatter. For different minimum SNR thresholds, we obtain 
\begin{itemize}
    \item SNR = 5: ~0.07 $\pm$ 0.16 at 90~GHz, 
    
    $\qquad\qquad$ -0.16 $\pm$ 0.16 at 148~GHz;
    
    \item SNR = 4: ~0.27 $\pm$ 0.16 at 90~GHz, 
    
    $\qquad\qquad$ -0.05 $\pm$ 0.15 at 148~GHz;
    
    \item SNR = 3: ~0.97 $\pm$ 0.19 at 90~GHz, 
    
    $\qquad\qquad$ ~0.82 $\pm$ 0.19 at 148~GHz.
\end{itemize}
We calculate this using multipoles up to 10,000 since those multipoles hold most of the constraining power for the cosmological parameters; however, we note that the ultrahigh-resolution of CMB-HD, beyond $\ell=10,000$, is required to clean the foregrounds to the levels achieved here. We find that an SNR threshold of four (red) and five (yellow) yields a difference between these spectra consistent with zero.  This indicates that our foreground cleaning method has not altered the CMB signal in the maps. In contrast, we find that this test fails for an SNR threshold of three, and thus we adopt a minimum SNR threshold of four throughout this work.

\section{Computational and Time Requirements for Simulation Generation and Foreground Cleaning}
\label{sec:computationalNeeds}

Here, we provide a sense of the computational time and requirements to create the ultrahigh-resolution simulations presented here, as well as to perform the foreground cleaning.  In Tables~\ref{tab:SimTimes} and~\ref{tab:FGtimes}, we show the times to generate the simulations and remove the foregrounds for sky regions of different sizes.  To generate ultrahigh-resolution simulations from lower resolution S10 counterparts requires downloading the 130~GB S10 simulation set.  To remove foregrounds from the ultrahigh-resolution simulations requires downloading the \num{66~GB} HD simulation set.

\begin{table}[t]
    \begin{center}
    \begin{tabular}{c@{\hskip 2em} c@{\hskip 2em} c}
      \toprule
      \toprule
       {\bf{Simulation Generation}}  &  \multicolumn{2}{c}{Time (hours)}    \\
       \cmidrule(lr){2-3}
        & $2^\circ\times2^\circ$  &  $10^\circ\times10^\circ$  \\
      \bottomrule
      Download full-sky S10 sims  &  \multicolumn{2}{c}{1.5} \\
      \midrule
      Generate simulations & 1.25 & 2 \\
      Take power spectra  & 0.5 & 1.1 \\
      \midrule
      \textbf{Total minus S10 download} & 1.75 & 3.1 \\
      \textbf{Total time} & 3.25 & 4.6 \\
      \bottomrule
    \end{tabular}    \caption{Listed are the approximate times to generate the ultrahigh-resolution simulation set from a lower-resolution counterpart. We also list times to take power spectra of the entire generated simulation set.  These times  were measured on a ``hbm-long-96core'' queue of the Stony Brook University SeaWulf cluster which has 96 Intel Sapphire Rapids CPUs and 384 GB of memory.} 
    \label{tab:SimTimes}
    \end{center}
\end{table}

\begin{table}[t]
    \begin{center}
    \begin{tabular}{c@{\hskip 2em} c}
      \toprule
      \toprule
        {\bf{Foreground Removal}}  &  Time (hours)    \\
      \bottomrule
      Download HD sims ($10^\circ\times10^\circ$) &  2.5 \\
      \midrule
      Remove CIB and radio sources ($10^\circ\times10^\circ$) & 4  \\
      Remove galaxy clusters ($10^\circ\times10^\circ$) & 1 \\
      \midrule
      \textbf{Total minus HD download} & 5  \\
      \textbf{Total time} & 6.5  \\
      \bottomrule
    \end{tabular}    \caption{Listed are approximate times to remove foregrounds from a 100 square degree region of sky. Here we separate CIB and radio source removal which takes more than four times longer than cluster removal.  These times were measured on the ``short-96core-shared'' queue of the Stony Brook University SeaWulf cluster which used 24 AMD EPYC Milan CPUs with 64 GB of memory.} 
    \label{tab:FGtimes}
    \end{center}
\end{table}

\begin{figure*}[t]
\begin{lstlisting}[caption={CAMB high-accuracy settings used in this work.},label={list:camb}, captionpos=t, language=Python]
import camb
import numpy as np
lmax = 40000 
pars = camb.CAMBparams()
pars.set_cosmology(H0=67.36, ombh2=0.02237, omch2=0.1200, tau=0.0544, num_massive_neutrinos=3, mnu=0.06, nnu=3.044, bbn_predictor="PRIMAT_Yp_DH_ErrorMC_2021.dat")
pars.set_classes(recombination_model="Recfast")
pars.InitPower.set_params(As=np.exp(3.044)*1e-10, ns=0.9649)
pars.set_matter_power(kmax=100, k_per_logint=130)
pars.set_for_lmax(lmax+500, lens_potential_accuracy=30, lens_margin=2050)
pars.set_accuracy(AccuracyBoost=1.1, lSampleBoost=3.0, lAccuracyBoost=3.0, DoLateRadTruncation=False, min_l_logl_sampling=10000)
pars.NonLinear = camb.model.NonLinear_both
pars.NonLinearModel.set_params("mead2016")
\end{lstlisting}
\end{figure*}

\section{CAMB Accuracy Settings } \label{sec:CAMBaccuracy}

The low noise and high resolution of CMB-HD require accurate modeling of the CMB and lensing power spectra to high-$\ell$ (i.e.~$\ell = 20,000$).  A set of CAMB accuracy settings was found in~\cite{HDparams} for CMB-HD lensed and delensed CMB power spectra and CMB lensing spectra that yielded biases on cosmological parameters of less than $0.5\sigma$.  In this work, we make minor additions to the accuracy settings presented in~\cite{HDparams}, primarily to improve the accuracy of the small-scale unlensed CMB power spectra by setting \texttt{min\_l\_logl\_sampling = 10000}; this removed oscillations in the unlensed CMB power spectra at high-$\ell$.  In~\cite{HDparams}, the unlensed CMB was not required; however, in this work, we created simulations starting with a realization of the unlensed CMB and lensed it with the lensing convergence map. In~\cite{HDZack}, it was found that this set of accuracy settings produced CAMB power spectra that are consistent with CLASS~\cite{CLASS} spectra to within \num{0.5\%} over the range $\ell \in [30, 20,000]$ (when using CLASS accuracy settings that also result in cosmological parameter biases less than $0.5\sigma$ for CMB-HD).  We give the full list of CAMB accuracy settings used in this work in Listing~\ref{list:camb} for easy reference.

\begin{figure}[t]
    \centering
    \includegraphics[width=\columnwidth]{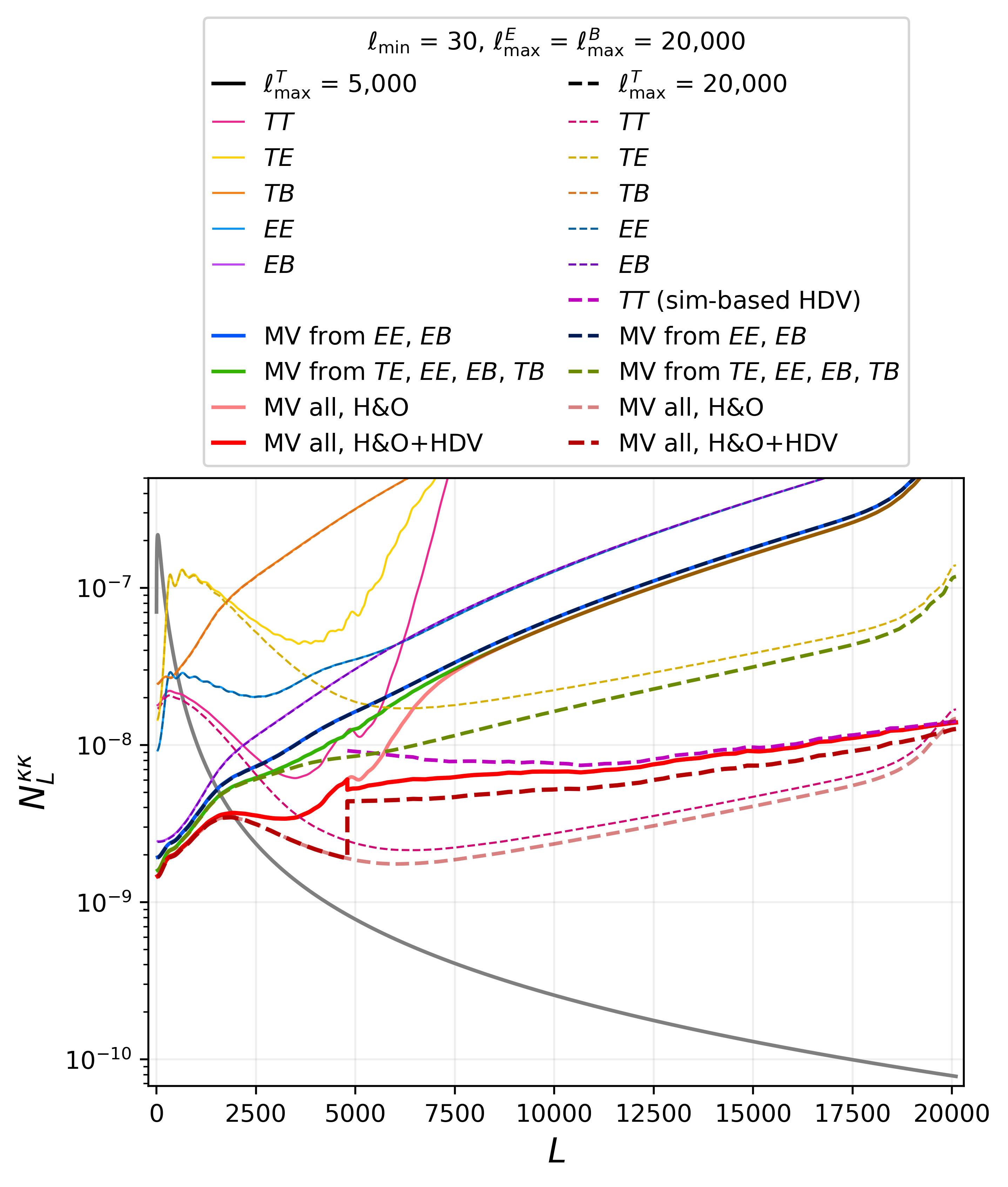}
    \caption{CMB lensing reconstruction noise, $N^{\kappa\kappa}_L$, from different estimators ($TT, TE, TB, EE, EB$ and the minimum variance combination $MV$). We show two cases:~using CMB temperature multipoles up to $\ell_\mathrm{max}^T = 5,000$ to reconstruct the lensing spectra when estimators involve $T$ (solid), and using $\ell_\mathrm{max}^T = 20,000$ (dashed); both cases use $\ell_\mathrm{max} = 20,000$ for $E$ and $B$ maps, and $\ell_\mathrm{min} = 30$ for all maps. We also show $N^{\kappa\kappa}_L$ for $TT$ in the range of $L\in[5000,20,000]$ from a simulation analysis presented in~\protect{\cite{han22}} using a modified lensing estimation technique that aids the mitigation of foregrounds on small scales~\protect{\cite{Hu:2007bt}} (dashed magenta curve); we scale this curve up to reflect the higher noise we find on the CMB temperature power spectrum after foreground cleaning compared to that work (as shown in Figure~\ref{fig:coaddedspectra}). We take as the baseline $N^{\kappa\kappa}_L$ in this work the solid red curve using $\ell_\mathrm{max}^T = 5,000$ for the $TE$ and $TB$ estimators, $\ell_\mathrm{max}^T = 5,000$ for the $TT$ estimator for $L<5000$, $\ell_\mathrm{max} = 20,000$ for the $EE$ and $EB$ estimators, and the scaled sim-based HDV curve from $TT$ for $L>5000$.  We compare cosmological parameter constraints using this MV $\kappa\kappa$ curve to those using a polarization-only $\kappa\kappa$ (blue solid curve) in Table~\ref{tab:params}. We find minimal difference in cosmological parameters, except when also varying the slope and amplitude of the kSZ power spectrum. For the latter case, the lower CMB lensing noise from the MV $\kappa\kappa$ breaks the degeneracy between the kSZ and lensing signal in the CMB $TT$ power spectrum, tightening parameter constraints. }  
    \label{fig:nlkkFromTempPol}
\end{figure}

\section{CMB Lensing Spectra in Different Scenarios} \label{sec:CMBlensing}

The CMB lensing spectra are generated from a four-point function of CMB temperature and polarization maps~\cite{Lewis:2006fu}.  In Figure~\ref{fig:nlkkFromTempPol}, we show the CMB lensing noise curves, $N^{\kappa\kappa}_L$, using different estimators or their combinations; these noise curves (with the exception of the $TT$ curve labeled as sim-based HDV) were calculated using the CLASS delens package~\cite{Hotinli2021}.  In this figure, we also show two scenarios: one where we keep all the CMB temperature information out to $\ell=20,000$ and use that to reconstruct the CMB lensing spectra when temperature data is relevant (dashed curves), and one where we only include CMB temperature information below $\ell_\mathrm{max}^T = 5,000$ (solid curves), where the CMB $TT$ signal dominates over the residual temperature foregrounds (as shown in Figure~\ref{fig:coaddedspectra}). 

A concern is that extragalactic foregrounds in temperature maps can cause biases in the CMB lensing spectra through lensing estimators that include temperature data~\cite{vanEngelen:2013rla}. This holds even for $TE$ and $TB$ estimators, which do not have foreground biases in their respective CMB power spectra but might in CMB lensing spectra via the four-point functions. Thus, choosing $\ell_\mathrm{max}^T = 5,000$ is more conservative when including temperature data.  For the minimum variance lensing spectra used in this work (MV $\kappa\kappa$), we make this choice for the $TE$ and $TB$ estimators;  we also adopt $\ell_\mathrm{max}^T = 5,000$ for the $TT$ estimator for lensing multipoles $L<5000$.

For the $TT$ estimator for lensing multipoles $L>5000$, we explored the foreground bias issue in~\cite{han22}, using simulations with idealized foreground removal to estimate the bias on the CMB lensing spectra.  In that work, we found biases at the level of $1\sigma$ from tSZ and CIB foregrounds for a CMB-HD-like survey.  Since fully repeating that analysis for the new foreground removed temperature maps produced in this paper is beyond the scope of this work, we instead scale the diagonal elements of that covariance matrix by the ratio of $TT$ lensing noise resulting from a quadratic estimator applied to the $TT$ residual noise curves obtained in this work and from the previous idealized estimate (shown as the light and dark red curves in Figure~\ref{fig:coaddedspectra}).  In Figure~\ref{fig:nlkkFromTempPol}, we show this as the dashed magenta curve labeled ``sim-based HDV'' in reference to the specific form of the lensing estimator we used from~\cite{Hu:2007bt}.  

To be conservative, we also show results when using CMB lensing reconstruction from CMB polarization alone (only $EE$ and $EB$ estimators), which are relatively immune to extragalactic foreground contamination.  In Figure~\ref{fig:nlkkFromTempPol}, we show this as the solid blue curve (pol.-only $\kappa\kappa$ in this work), which is significantly higher at high-$L$ than the lensing noise curves that include temperature information.  

We note that there are several techniques that we did not employ, which could further mitigate foreground biases in lensing spectra, such as bias hardened estimators~\cite{ACT:2023ubw}, SCALE lensing estimators~\cite{Chan:2024vbr}, and gradient inversion methods~\cite{Hadzhiyska:2019cle}. Future work will incorporate the foreground removal presented in this work, along with the additional techniques to mitigate extragalactic foregrounds mentioned above, to refine the $N^{\kappa\kappa}_L$ further.

\bibliographystyle{apsrev4-1}
\bibliography{main.bib}

\end{document}